\documentclass[letterpaper,english,aps,prb,floatfix,showpacs,twocolumn,amsfonts,amssymb,superscriptaddress,longbibliography]{revtex4-2}

\usepackage{balance}
\usepackage[T1]{fontenc}
\usepackage[utf8]{inputenc}
\usepackage[english]{babel}
\usepackage{graphics}
\usepackage{amssymb}
\usepackage{amsmath, esint}
\usepackage{overpic}
\usepackage[toc]{appendix}
\usepackage{bbold}

\usepackage{natbib}
\usepackage{xcolor}
\usepackage{soul}
\usepackage{commath}
\usepackage{tikz}
\usepackage{physics}

\usepackage{hyperref}

\newcommand{\be}{\begin{equation}}
\newcommand{\ee}{\end{equation}}
\newcommand{\bspl}{\begin{split}}
\newcommand{\espl}{\end{split}}
\newcommand{\bea}{\begin{eqnarray}}
\newcommand{\eea}{\end{eqnarray}}

\newcommand{\angstrom}{\mbox{\normalfont\AA}}

\newcommand{\bd}{\boldsymbol}

\def\a{\alpha}
\def\b{\beta}

\def\d{\delta}

\def\m{\mu}
\def\l{\lambda}
\def\th{\theta}
\def\t{\tau}

\def\o{\omega}

\def\s{\sigma}

\def\G{\Gamma}
\def\D{\Delta}
\def\O{\Omega}

\def\S{\Sigma}

\def\ra{\rightarrow}
\def\up{\uparrow}

\def\down{\downarrow}

\def\pd{\partial}
\def\nb{\nabla}
\def\bdnb{\bd{\nb}}
\def\bk{{\bf k}}

\def\bA{{\bf A}}

\def\bB{{\bf B}}
\def\bE{{\bf E}}

\def\bJ{{\bf J}}

\def\bx{{\bf x}}

\def\nn{\nonumber}
\def\lb{\label}
\def\pref#1{(\ref{#1})}

\newcount\bozza \bozza=0
\ifnum\bozza=1
\newdimen\shift \shift=-2truecm
\def\lb#1{%
{\label{#1}\rlap{\kern\shift{$\scriptstyle#1$}}}}
\else\def\lb#1{\label{#1}} \fi

\definecolor{darkred}{rgb}{0.55, 0.0, 0.0}
\definecolor{darkpowderblue}{rgb}{0.0, 0.2, 0.6}

\usetikzlibrary{shapes.arrows, fadings}
\tikzfading[name=fadeleft,
  left color=transparent!80, right color=transparent!0]

\begin{document}
\title{Generalized Josephson plasmons in bilayer superconductors}
\author{N. Sellati}
\email{niccolo.sellati@uniroma1.it}
\affiliation{Department of Physics, ``Sapienza'' University of Rome, P.le A. Moro 5, 00185 Rome, Italy}
\author{F. Gabriele}
\affiliation{Department of Physics, ``Sapienza'' University of Rome, P.le A. Moro 5, 00185 Rome, Italy}
\author{C. Castellani}
\affiliation{Department of Physics, ``Sapienza'' University of Rome, P.le A. Moro 5, 00185 Rome, Italy}
\affiliation{Institute for Complex Systems, CNR, UoS Sapienza, 00185 Rome, Italy}
\author{L. Benfatto}
\email{lara.benfatto@roma1.infn.it}
\affiliation{Department of Physics, ``Sapienza'' University of Rome, P.le A. Moro 5, 00185 Rome, Italy}
\affiliation{Institute for Complex Systems, CNR, UoS Sapienza, 00185 Rome, Italy}

\begin{abstract}
Layered superconductors like High-$T_c$ cuprates display out-of-plane plasma oscillations between layers sustained by the weak Josephson coupling among the superconducting sheets, the so-called Josephson plasmons. Bilayer cuprates hosts two of such modes, but due to the anisotropy of the electronic response their description at generic wavevector cannot be separated from that of the in-plane oscillations. In this paper we provide an analytical theoretical framework able to describe the dispersions and the polarizations of the generalized plasma modes of such systems, that has been only partly addressed by previous work in the literature. We then employ it to explain the peculiar characteristics of their linear optical response, by providing a fully microscopic explanation for the appearance of a finite-frequency peak in the real part of the optical conductivity. On a wider perspective, the complete characterization of the Josephson plasma modes provided by our approach represents  a groundwork to address open issues raised by recent experiments with strong THz pulses, able to drive them beyond the linear-response regime. 
\end{abstract}
\date{\today}

\maketitle

\section{Introduction}
Among the various unconventional properties reported for high-temperature superconducting (SC) cuprates, the emergence of a soft plasma edge in the reflectivity measurements for field polarized perpendicularly to the CuO$_2$ planes (say, in the $z$ direction) attracted since the very beginning considerable attention. Indeed, while in the metallic state the weak hopping between planes, along with the strong correlations at play in these systems, make the plasma edge completely damped, below $T_c$ the gap opening removes most of the quasiparticle continuum in the THz range, giving rise to a well-defined $z$-axis reflectivity edge. This feature has been accurately measured by continuous-wave reflectivity measurements long ago in several families of cuprates, hosting one or two layers per unit cell\cite{uchida_prl92,homes_prl93,kim_physicac95,basov_prb94,vandermarel96,basov_prl03}. As usual, the long-wavelength limit of the transverse plasma-polariton, that coincides with the frequency of the plasma edge in reflectivity, identifies also the frequency scale of the longitudinal plasmon, showing that also this mode, connected to fluctuations of the electronic density,   becomes undamped below $T_c$. The advent of time-resolved spectroscopies with short light pulses triggered considerable interest on the fate of these soft plasma modes emerging below $T_c$. This is due to the fact that plasma modes, as connected to the fluctuations of the density, appear also in the spectrum of the SC phase of the complex order parameter, which is its quantum-mechanical conjugate variable\cite{nagaosa,coleman}. A simple way to understand this effect is to recall that the interaction among the phase variables $\theta_n$ in neighbouring planes is described by a Josephson-like model
\begin{align}\lb{jocouplSL}
H_{\mathcal J}=-\mathcal J\sum_n \cos(\theta_n-\theta_{n+1}),
\end{align}
where $n$ is the layer index. Here the coupling constant $\mathcal J$ sets the scale of the out-of-plane stiffness, and then of the SC plasma mode below $T_c$. Since in turn the discrete SC phase gradient $\theta_n-\theta_{n+1}$ is coupled to the electromagnetic (e.m.) gauge field by the minimal-coupling scheme, one can use an intense light pulse to drive SC phase modes beyond the linear regime. Such a possibility has been not only investigated theoretically\cite{nori_review10,nori_natphys06,demler_prb20,gabriele_natcomm21,demler_commphys22},  but it has been clearly demonstrated experimentally in recent years\cite{cavalleri_review,cavalleri_natphys16,cavalleri_science18,averitt_pnas19,cavalleri_prx22,cavalleri_prb22,averitt_prb23,shimano_prb23}.\\
One of the interesting aspects in the description of these soft plasma modes in layered superconductors is that, unless propagation occurs along $z$ or purely in plane, for the frequency and momenta of the THz light one cannot completely separate longitudinal plasmons from transverse plasma polaritons, as it usually happens for isotropic systems at all momenta. This effect, that is already encoded at the level of Maxwell's equations\cite{bulaevskii_prb94,bulaevskii_prb02,machida_prl99,machida_physc00,nori_review10}, leads to the definition of so-called "generalized plasma modes" with mixed longitudinal and transverse character, as it has been highlighted in a recent publication devoted to single-layer superconductors\cite{gabriele_prr22}. As discussed in previous works\cite{bulaevskii_prb94,bulaevskii_prb02,machida_prl99,machida_physc00,nori_review10,demler_commphys22,gabriele_prr22}, the origin of such a mixing fully relies on the anisotropy of the conduction in a layered system, which makes the current response in general not parallel to the applied electric field. This has e.g. the consequence that one can have an induced transverse current even in response to a longitudinal electric field and vice versa, then making it impossible to completely separate longitudinal and transverse e.m. modes for arbitrary direction of the propagating wavevector. Since a transverse current in turn acts as a source of magnetic field, which generates a transverse electric field by Faraday's law, an alternative but yet equivalent way to state the problem is that one must  include retardation  effects of the magnetic field in the response to a longitudinal excitation. Nonetheless, since retardation effects scale as the inverse light velocity (so that they are sometimes also named in this context "relativistic"\cite{demler_prb20,cavalleri_prx22,gabriele_prr22}) for momenta outside the light cone the quantitative effects are negligible, and a longitudinal-transverse decoupling is recovered\cite{gabriele_prr22}.\\
In this paper we analyse the fate of the generalized plasma modes in the case of cuprates systems with two planes per unit cell, as e.g. $\text{YBa}_2\text{Cu}_3\text{O}_{6+x}$ (YBCO). From the technical point of view, we will adopt an effective-action formalism to deal with the SC phase and the e.m. degrees of freedom on the same footing, as discussed recently in Ref.[\onlinecite{gabriele_prr22}] in the case of systems with one plane per unit cell. 
\begin{figure}[t!]
    \centering
    \includegraphics[width=0.5\textwidth,keepaspectratio]{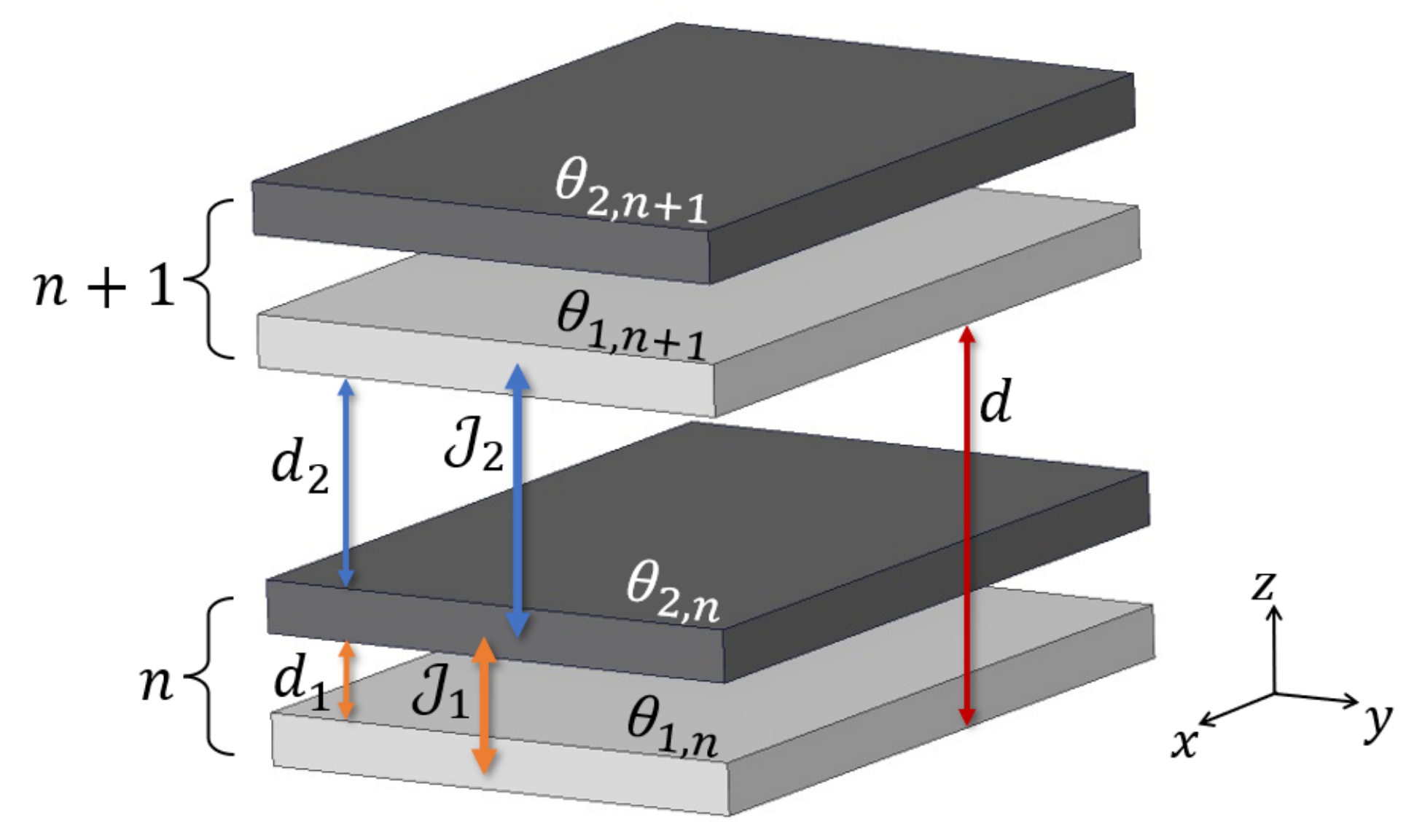}
    \caption{Schematic representation of two subsequent primitive cells of the bilayer lattice. The $n$-th unit cell contains two conducting planes, labeled "1" (light gray) or "2" (dark gray). Subsequent planes have a Josephson-like interaction with constants $\mathcal J_1$ in the intra-bilayer and $\mathcal J_2$ in the inter-bilayer spacing.}
    \lb{unitcell}
\end{figure}
By introducing suitable variables proportional to the physical currents one can indeed generalize the Josephson model \eqref{jocouplSL}, with the twofold advantage to account on the same footing for both retardation effects on the dispersion of the e.m. modes, and non-linear coupling of the SC phase to light. The latter aspect is particularly interesting for future investigation of the non-linear optical response in YBCO, that has been recently explored experimentally by several groups \cite{cavalleri_nmat14,cavalleri_prx22,wangNL_cm22,shimano_prb23}. In this case two plasma edges appear below $T_c$ in the $z$-axis reflectivity, corresponding to the existence of two different inter-layer couplings between planes belonging to the same or to two consecutive unit cells\cite{homes_prl93,kim_physicac95,basov_prb94,vandermarel96}, as sketched in Fig. \ref{unitcell}. As a consequence, in the limit of zero momentum one has three relevant energy scales, a large in-plane plasma frequency $\omega_{xy}$, of the order of 1 eV, and two soft out-of-plane modes $\omega_{z1,z2}$, ranging from few to tens of THz, depending on the doping level\cite{cavalleri_nmat14,vittoria_jap93}. As we shall see, these energy scales define two crossover momenta $|\bk_{c1}|\sim \sqrt{\omega_{z1}^2-\omega_{z2}^2}/c$ and $|\bk_{c2}|\sim \sqrt{\omega_{xy}^2-(\omega_{z1}^2d_1+\omega_{z2}^2d_2)/d}/c$, which account for different manifestations of the mixing among longitudinal and transverse degrees of freedom in the system. Our results not only generalize previous derivations obtained in specific conditions\cite{vandermarel_prb01,vandermarel96}, but they also clarify the nature of the modes, shedding light on the possibility to observe them with different probes. As a direct application, we also derive a general expression for the optical conductivity and we explain the appearance of a  well-defined peak at a frequency $\omega_T^2=(\omega_{z1}^2d_2+\omega_{z2}^2d_1)/d$, that has been indeed reported experimentally\cite{homes_prl93,bernhard_prl11,erb_prl00,tajima_prl09, yamada_prl98,cavalleri_nmat14, wangNL_prx20}. It is worth noting that usually a plasma edge, corresponding to zeros of the dielectric function, does not give rise to a peak in the real part $\sigma_1(\omega)$ of the optical conductivity. A strong absorption peak in $\sigma_1$ arises instead from a resonance in the dielectric function, that is usually unexpected for plasma modes at zero momentum. In previous literature the existence of this peak has been explained by means of a simple but yet very powerful model of capacitive coupling among neighbouring layers\cite{vandermarel96}. Here we derive a similar result within a more formal many-body formalism, which has the advantage to take into account finite-compressibility corrections and to clarify why phase fluctuations should lead to a peak at finite frequency, as opposed to the ordinary single-layer case. Indeed, in the bilayer system the out-of-plane conductivity keeps trace not only of homogeneous phase fluctuations among layers, but also of phase fluctuations with opposite signs in neighbouring layers, that is to some extent the counterpart of the Leggett phase mode\cite{leggett} in multiband superconductors. As we discuss below, the opposite-phase fluctuations give rise to a response at finite frequency which peaks in the limit for vanishing momentum at the frequency scale $\omega_T$.\\
The plan of the paper is the following. In Section \ref{sec2} we introduce the effective-action formalism for the gauge-invariant variables and its connection to the study of SC plasma modes in different contexts. In the introductory subsection \ref{sec2a} we overview the theoretical approach used in the manuscript and we apply it to isotropic superconductors to obtain the well-known dispersion relations of the plasmon and the plasma-polariton. In subsection \ref{sec2b} we employ this structure to anisotropic single-layer superconductors, showing that in anisotropic systems the standard-RPA approach is not sufficient for a complete description of the plasma modes, as already discussed in Ref.\ \cite{gabriele_prr22}. The remainder of the section focuses on bilayer superconductors: in subsection \ref{sec2c} we derive the exact effective action that describes the generalized plasma modes of the system, and use it to compute numerically their dispersions; in subsection \ref{sec2d} we describe their polarizations for different values of the momentum, emphasizing the longitudinal-transverse mixing mechanism and the presence of a purely out-of-plane mode with opposite-phase oscillations; in subsection \ref{sec2e} we evaluate the corrections to the Josephson plasma frequencies given by a finite compressibility in the system. In Section \ref{sec3} we study the linear response of a bilayer superconductor to an external electromagnetic field by evaluating the dielectric function and the optical conductivity of the system with finite-compressibility corrections. Section \ref{sec4} contains the final discussion and conclusions. Further technical details are provided in the Appendices: Appendix \ref{appA} reviews the main steps leading to the Gaussian effective action for a superconductor within the path-integral formalism. Appendix \ref{appB} shows how the Gaussian action for the SC phase and the free e.m. action in a bilayer system follow from a correct discretization of the fields dictated by the Maxwell's equations. In Appendix \ref{appC} we analyse the dispersions of the generalized plasma modes in the non-relativistic regime. \\
\section{Effective-action formalism for plasma modes}\label{sec2}
\subsection{Description of the plasma oscillations via the SC phase in isotropic crystals}\label{sec2a}
Before giving technical details on the derivation of the generalized plasma modes in bilayer superconductors, it can be instructive to briefly outline the strategy for isotropic and single-layer anisotropic systems.
As mentioned in the introduction, the complex order parameter acquires non-zero average value in the SC state below the critical temperature $T_c$, breaking the continuous gauge symmetry. Because of this, a Goldstone mode that is directly linked to the phase fluctuations of the order parameter is expected\cite{nagaosa}. A powerful technique to understand this on a quantum mechanical level relies on the construction of a quantum analogous of the Ginzburg-Landau model: starting from a fermionic model with a BCS-like interaction term one can introduce two effective bosonic fields that play the role of the amplitude and phase of the SC order parameter and apply the Hubbard-Stratonovich procedure to decouple the interaction\cite{nagaosa,coleman,hubbard,stratonovich}. By integrating out the fermions one is left with an effective model that can be expanded up to arbitrary powers in the bosonic fields\cite{nagaosa,aitchison_prb95,depalo_prb99,randeria_prb00,benfatto_prb01,benfatto_prb04,millis_prr20, dassarma_prl90,dassarma_prb91,dassarma_prb95}. By retaining only Gaussian terms in the fluctuations one defines the spectrum of the collective excitations of the system, equivalent to RPA evaluation of the vertex function in the standard diagrammatic language\cite{schrieffer}. This procedure is discussed in details in Appendix \ref{appA}. In this framework, the phase fluctuations $\theta$ at long wavelengths are described by the following imaginary-time action:
\begin{align}\lb{supfluid}
    S_G^{(iso)}[\theta]&=\frac{1}{8}\int d\tau d{\bf x}\bigg[\kappa_0(\partial_\tau \theta)^2 + D_s(\boldsymbol{\nabla}\theta)^2\bigg]=\nn\\
    &=\frac{1}{8}\sum_q\bigg[\kappa_0\Omega_m^2+D_s|\textbf{k}|^2\bigg]|\theta(q)|^2,
\end{align}
where $\kappa_0$ is the bare compressibility, $D_s=n_s/m^*$ is the isotropic superfluid stiffness expressed as the ratio between the superfluid electron density and the effective electron mass, $\tau=it$ is the imaginary time variable and $q=(i\Omega_m,\textbf k)$ is the imaginary-time 4-momentum, with $\Omega_m=2\pi mT$ the bosonic Matsubara frequencies. Even though we will focus here on the $T=0$ case, we will retain the Matsubara formalism that is appropriate for a generalization at finite temperature and allows us for a straightforward derivation of the response function in Sec.\ \ref{sec3}. In the effective-action formalism employed in the present work the energy-momentum dispersions appear as the zeros of the Gaussian action, once the analytical continuation $i\Omega_m\to\omega+i0^+$ has been performed. For neutral superfluid systems, Eq. \eqref{supfluid} identifies the so-called Anderson-Bogoliubov sound mode\cite{anderson_pr58} with dispersion relation $\omega^2=(D_s/\kappa_0)|\textbf{k}|^2$.\\
In a charged superconductor the sound mode is promoted to a plasma mode by adding the effects of the long-range Coulomb interactions $V({\bf k})$ among electrons\cite{anderson_pr58}. Within the effective-action formalism this results is usually achieved\cite{nagaosa,depalo_prb99,randeria_prb00,benfatto_prb01,benfatto_prb04,millis_prr20,dassarma_prl90,dassarma_prb91,dassarma_prb95} by adding a further interaction term in the Hamiltonian describing density-density interactions mediated by  $V({\bf k})$, and decoupling it via an additional Hubbard-Stratonovich field $\rho$ representing the density. Since phase and density are conjugate variables \cite{aitchison_prb95,depalo_prb99,nagaosa} one obtains a direct phase-density coupling in the action, and by integrating out the additional $\rho$ field one recovers the dressing of the compressibility $\kappa_0\to \kappa_0/(1+V({\bf k})\kappa_0)$. As a result the fluctuations of the phase, that reflect density fluctuations, identify a plasma mode as their spectrum acquires a gap\cite{nagaosa,depalo_prb99,randeria_prb00,benfatto_prb01,benfatto_prb04,millis_prr20,anderson_pr58,dassarma_prl90,dassarma_prb91,dassarma_prb95}. However it is instructive for the purpose of this work to employ an alternative derivation for the plasma oscillations of the isotropic superconductors\cite{gabriele_prr22} as it will turn out to be the convenient strategy to be used for anisotropic systems.\\
Starting from Eq. \eqref{supfluid} we introduce an internal e.m. field by means of the minimal coupling substitution\cite{nagaosa,depalo_prb99,randeria_prb00,benfatto_prb04},
\begin{align}\lb{mincoupreal}
    \partial_t\theta&\to\partial_t\theta-2e\phi\nn\\
    \boldsymbol\nabla\theta&\to \boldsymbol\nabla\theta+\frac{2e}{c}\textbf{A},    
\end{align}
which in Matsubara space read
\begin{align}\lb{mincoup}
    \Omega_m\theta(q)&\to\Omega_m\theta(q)-2e\phi(q)\nn\\
    i\textbf{k}\theta(q)&\to i\textbf{k}\theta(q)+\frac{2e}{c}\textbf{A}(q),    
\end{align}
and by including the free e.m. action\cite{nagaosa},
\begin{align}\lb{emaction}
    &S_{\text{e.m.}}[\phi,\textbf A]=\frac{\varepsilon_B}{8\pi}\int d\tau d\textbf{x}\bigg[\frac{1}{\varepsilon_B}(\boldsymbol{\nabla}\times\textbf{A})^2+\nn\\
    &-\bigg(\frac{i\partial_\tau\textbf A}{c}+\boldsymbol{\nabla}\phi\bigg)^2\bigg]=\frac{\varepsilon_B}{8\pi}\sum_q\bigg[\frac{\Omega_m^2}{c^2}|\textbf A(q)|^2+\nn\\
    &-|\textbf k|^2|\phi(q)|^2+\frac{1}{\varepsilon_B}\big|\textbf k \times \textbf A(q)\big|^2+\nn\\
    &+\frac{i\Omega_m}{c}\textbf k \cdot (\phi(q)\textbf A(-q)+ \phi(-q)\textbf A(q))\bigg].
\end{align}
Here $\phi$ and \textbf{A} are the scalar and the vector potential respectively, $-e$ is the charge of the electron, $c$ is the light velocity and $\varepsilon_B$ is the background dielectric constant. Notice that while Eq. \eqref{mincoup} holds for the coupling with both an internal and an external e.m. field, here we only introduce the contribution of the internal fields which we relate to the charge and density fluctuations of the system according to Maxwell's equations. In the total action obtained summing the contributions \eqref{supfluid} and \eqref{emaction} after performing the substitution \eqref{mincoup}, we fix the Weyl gauge, i.e. $\phi=0$, and we then perform the following change of variables:
\begin{align}\lb{GIvarreal}
    \boldsymbol\psi=\boldsymbol\nabla\theta+\frac{2e}{c}\textbf{A},
\end{align}
or equivalently in momentum space,
\begin{align}\lb{GIvar}
    \boldsymbol\psi(q)=i\textbf{k}\theta(q)+\frac{2e}{c}\textbf{A}(q).
\end{align}
By definition these quantities are invariant under the simultaneous gauge transformation\cite{nagaosa,coleman} of the vector potential and of the SC phase by a generic function $\Lambda(q)$:
\begin{align}\lb{gaugetf}
\theta(q)&\to \theta(q) - \frac{2e}{c}\Lambda(q)\nn\\
\textbf{A}(q)&\to\textbf{A}(q)+i\textbf{k}\Lambda(q).
\end{align}
In contrast to the SC phase alone that does not represent a physically observable quantity, the gauge-invariant variables in Eq. \eqref{GIvar} are instead proportional to physical currents. Indeed, analyzing their spectrum is completely equivalent to solving the problem of the electromagnetic wave propagation in the material\cite{gabriele_prr22}. The action then reads:
\begin{align}\lb{GIvarphaseiso}
S^{(iso)}[\theta,\boldsymbol{\psi}]&=\frac{\varepsilon_B}{32\pi e^2}\sum_q\bigg[\bigg(\Omega_m^2+\frac{4\pi e^2}{\varepsilon_B}D_s\bigg)\left|\boldsymbol{\psi}(q)\right|^2+\nn\\
&+\frac{c^2}{\varepsilon_B}\left|\textbf{k}\times\boldsymbol{\psi}(q)\right|^2+ \frac{\Omega_m^2}{\alpha}(1+\alpha|\textbf k|^2)\left|\theta(q)\right|^2+\nn\\
&+i\Omega_m^2\textbf k\cdot \big(\boldsymbol{\psi}(q)\theta(-q)-\boldsymbol{\psi}(-q)\theta(q)\big)\bigg],
\end{align}
where 
\begin{align}\lb{alpha}
\alpha=\frac{\varepsilon_B}{4\pi e^2}\frac{1}{\kappa_0}=\lambda_D^2,
\end{align}
with $\lambda_D$ the Debye screening length. Notice that in the isotropic case here considered only the longitudinal component $\boldsymbol{\psi}_L=(\hat{\textbf{k}}\cdot\boldsymbol{\psi})\hat{\textbf{k}}$ of the gauge-invariant variables couple to the SC phase, while the action of its transverse component $\boldsymbol\psi_T=(\hat{\textbf{k}}\times\boldsymbol{\psi})\times\hat{\textbf{k}}$ is independent. By integrating out the SC phase one is then left with an effective action of the physical fields,
\begin{align}\lb{GIvariso}
S^{(iso)}&[\boldsymbol{\psi}]=\frac{\varepsilon_B}{32\pi e^2}\sum_q\bigg[\bigg(\frac{\Omega_m^2}{1+\alpha|\textbf k|^2}+\omega_P^2\bigg)\left|\boldsymbol{\psi}_L(q)\right|^2+\nn\\
&+\bigg(\Omega_m^2+\omega_P^2+\frac{c^2}{\varepsilon_B}|\textbf{k}|^2\bigg)\left|\boldsymbol{\psi}_T(q)\right|^2\bigg]=\nn\\
&=\frac{1}{32\pi e^2}\sum_q\bigg[\frac{\Omega_m^2}{1+\alpha|\textbf k|^2}\varepsilon_L(\Omega_m,\bk)\left|\boldsymbol{\psi}_L(q)\right|^2+\nn\\
&+\bigg(\Omega_m^2\varepsilon_T(\Omega_m)+\frac{c^2}{\varepsilon_B}|\bk|^2\bigg)\left|\boldsymbol{\psi}_T(q)\right|^2\bigg],
\end{align}
where $\omega_P^2=4\pi e^2 D_s /\varepsilon_B$ is the isotropic plasma frequency and
\begin{align}\lb{dieliso}
&\varepsilon_L(\omega,\bk)=\varepsilon_B\bigg(1-\frac{\omega_P^2\big(1+\alpha|\textbf{k}|^2\big)}{\omega^2}\bigg)\\
&\varepsilon_T(\omega)=\varepsilon_B\bigg(1-\frac{\omega_P^2}{\omega^2}\bigg)
\end{align}
represent the longitudinal and transverse dielectric functions respectively, after the analytical continuation.\\
From Eq. \eqref{GIvariso} one immediately sees that the three components of $\boldsymbol{\psi}$ describe all the e.m. modes in the system\cite{anderson_pr63} given by the poles of the longitudinal and transverse propagators:
\begin{align}\lb{plasmaiso}
&\omega_L^2(\bk)={\omega_P^2(1+\alpha|\textbf{k}|^2)}\nn\\
&\omega_T^2(\bk)={\omega_P^2+\frac{c^2}{\varepsilon_B}|\textbf{k}|^2}.
\end{align}
These results are formally identical to the ones widely known and discussed in literature\cite{nagaosa,depalo_prb99,randeria_prb00,benfatto_prb01,benfatto_prb04,millis_prr20,anderson_pr58}. Nonetheless, a description in terms of the gauge-invariant variables is more convenient in an anisotropic system, in which longitudinal and transverse components are mixed\cite{gabriele_prr22}.

\subsection{Description of the plasma oscillations in single-layer superconductors}\label{sec2b}
A layered superconductor is an example of an anisotropic system in which subsequent SC planes of in-plane lattice constant $a$ and with interlayer distance $d$ interact with a weak Josephson-like coupling\cite{nori_review10,cavalleri_review,keimer_review15,shibauchi_prl94,panagopoulos_prb96,bonn_prl04,fazio_review01} controlled by a constant $\mathcal J$, see Eq. \eqref{jocouplSL} that we report here for convenience: 
\begin{align}\lb{joscouplSL}
H_{\mathcal J}=-\mathcal J\sum_n \cos(\theta_n-\theta_{n+1}),
\end{align}
where $n$ is the primitive cell index. In the following, both for single-layer and for bilayer crystals, we will use the convention by which the  SC sheets are parallel to the $xy$-plane and stacked along the $z$ axis. The SC phase action in Eq. \eqref{supfluid} can be straightforwardly generalized to the anisotropic single-layer case by expanding Eq. \eqref{joscouplSL} to the Gaussian order. This procedure is by all means equivalent to rewriting Eq. \eqref{supfluid} taking into account the anisotropy of the superfluid stiffness\cite{randeria_prb00,benfatto_prb01,millis_prr20}. The Fourier transform is here defined differently from the isotropic case, in such a way that the inter-layer distance becomes explicit in the action. Such a convention will be useful for the generalization to the bilayer case. Denoting the in-plane stiffness by $D_{xy}$ and defining the out-of-plane one as $D_z=4Jd^2$, \textcolor{black}{where $J=\mathcal J/(S d)$ is the density of the Josephson coupling constant $\mathcal J$ along the SC plane of surface $S$}, one then obtains for a single-layer superconductor:
\begin{align}\lb{supfluidSL}
S_G^{(SL)}[\theta]=\frac{d}{8}\sum_q\bigg[\kappa_0\Omega_m^2+D_{xy}k_{xy}^2+D_zq_z^2\bigg]|\theta(q)|^2
\end{align}
where $k_{xy}^2={k_x^2+k_y^2}$ and 
\begin{align}\lb{qz}
q_z=\frac{2}{d}\sin(\frac{k_z d}{2})
\end{align}
accounts for the discrete periodicity along $z$. One can notice that Eq. \eqref{qz} can be recast as the more familiar $q_z^2=2\big(1-\cos(k_zd)\big)/d^2$. The anisotropy of the stiffness is mirrored in the presence of two different plasma frequencies, the in-plane plasma frequency $\omega_{xy}^2=4\pi e^2D_{xy}/\varepsilon_B$ that is typically of the order of the eV, and the Josephson plasma frequency $\omega_{z}^2=4\pi e^2D_{z}/\varepsilon_B$ in the range of THz\cite{uchida_prl92,homes_prl93,kim_physicac95,basov_prb94,vandermarel96}.\\
The procedure making use of the gauge-invariant fields outlined in the isotropic case is useful to treat an anisotropic crystal, as it immediately takes into account all the electromagnetic interactions of the system, not exhausted by the sole Coulomb interaction\cite{gabriele_prr22}. Thus, by repeating the same procedure, one can write the analogous of Eq. \eqref{GIvariso} in terms of the Cartesian components of the gauge-invariant variables:
\begin{align}\lb{GIvarSL}
S^{(SL)}&[\psi]=\frac{\varepsilon_Bd}{32\pi e^2}\sum_q\bigg[\begin{pmatrix}\psi_{x}(q)&\psi_{z}(q)\end{pmatrix}\mathcal{P}_{xz}^{SL}\begin{pmatrix}\psi_x(-q) \\ \psi_{z}(-q)\end{pmatrix}+\nn\\
&+\psi_y(q)\big(\Omega_m^2+\omega_{xy}^2+\frac{c^2}{\varepsilon_B}(k_x^2+q_z^2)\big)\psi_y(-q)\bigg].
\end{align}
Here we chose, without loss of generality, to take the in-plane momentum along the $x$ direction ($k_y=0$). The dynamical matrix associated with the $x$ and $z$ components of the physical variables is
\begin{align}\lb{SLmatrix}
    \mathcal{P}_{xz}^{SL}=\begin{pmatrix}
	\Omega_m^2+\omega_{xy}^2+\frac{c^2}{\varepsilon_B}q_z^2 & -\frac{c^2}        {\varepsilon_B}k_xq_z \\
	-\frac{c^2}{\varepsilon_B}k_xq_z & \Omega_m^2+\omega_{z}^2+\frac{c^2}        {\varepsilon_B}k_x^2
    \end{pmatrix}.
\end{align}
In typical cuprate superconductors one usually finds that $\alpha/d^2$ is small\cite{machida_physc00,konsin_prb98}. As such, in writing Eq. \eqref{SLmatrix} we made the approximation of $\alpha\to 0$, that is equivalent to considering infinite compressibility in the system, although adding corrections due to finite compressibility would be straightforward. In the single-layer superconductors a total of three e.m. modes appear: a decoupled transverse plasma-polariton described by the coefficient of $|\psi_y|^2$, corresponding to an electric field along the $y$ direction and commonly called\cite{alpeggiani_prb13} transverse electric (TE), and two mixed plasma modes, corresponding to a magnetic field along the $y$ direction and called transverse magnetic (TM), whose dispersions are given by zeros of the determinant of $\mathcal{P}_{xz}^{SL}$:
\begin{align}\lb{SLmodes}
&\omega_{\pm}^{2}(\bk)=\frac{1}{2}\bigg(\omega_{xy}^2+\omega_z^2+\frac{c^2}{\varepsilon_B}(k_x^2+q_z^2)\pm\bigg[(\omega_{xy}^2-\omega_z^2)^2+\nn\\
&+\frac{c^4}{\varepsilon_B^2}(k_x^2+q_z^2)^2-2\frac{c^2}{\varepsilon_B}(k_x^2-q_z^2)(\omega_{xy}^2-\omega_z^2)\bigg]^{1/2}\bigg).
\end{align}
%
%
%
As shown in Ref.\ \cite{gabriele_prr22}, where the case $q_z\to k_z$ has been considered, the two modes of Eq. \eqref{SLmatrix} become either purely longitudinal or purely transverse only in the limiting cases $k_x=0$ or $k_z=2\pi l/d$, with $l$ an integer number,
while for any generic direction of $\textbf k$ they display a mixture of longitudinal and transverse character. The dispersions \eqref{SLmodes} are shown in Fig. \ref{SLdisp} as functions of $k_x$ for fixed values of $k_z$, in the usual way the plasmon dispersion corresponding to $\omega_-$ is usually acquired e.g. by RIXS measurements\cite{lee_rixs_nature18,liu_rixs_npjqm20,zhou_prl20}.
\begin{figure}[t!]
    \centering
    \includegraphics[width=0.5\textwidth,keepaspectratio]{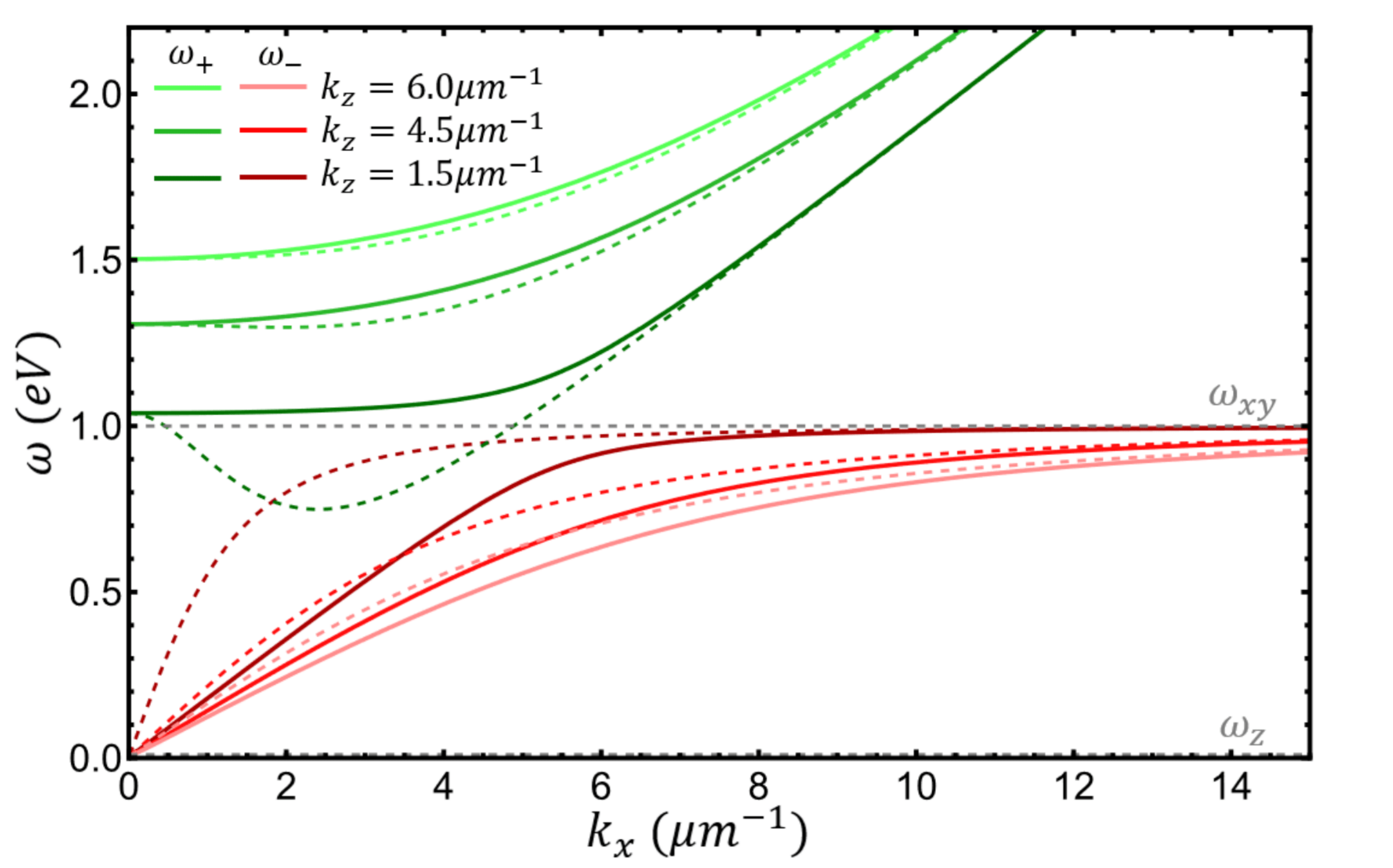}
    \caption{Energy-momentum dispersions of the single-layer superconductor plasma modes $\omega_+$ (green solid lines) and $\omega_-$ (red solid lines) given in Eq. \eqref{SLmodes} as functions of $k_z$ for fixed values of $k_x$. The dispersions are compared with the standard-RPA ones in Eq. \eqref{SLrpamodes}, represented as dashed lines with corresponding colours. Gray dashed lines denote the in-plane and out-of-plane plasma frequencies. Here $\varepsilon_B=1$, $\omega_{xy}=1\text{eV}$, $\omega_z=10\text{meV}$ and $\hbar c=0.187 \text{eV}\mu m$.}
    \lb{SLdisp}
\end{figure}
For the sake of completeness we also show the comparison with the plasmon dispersion $\omega_{-,\text{RPA}}$ of the layered superconductor obtained in the literature within the standard-RPA approach. This consists in including only the RPA dressing of the compressibility by the long-range Coulomb potential, which is equivalent in the present language to  add only the coupling to the scalar potential. For the longitudinal plasmon $\omega_{-,\text{RPA}}$ this is obtained\cite{randeria_prb00,benfatto_prb01,benfatto_prb04,millis_prr20,vandermarel96,dassarma_prl90,dassarma_prb91,dassarma_prb95} by replacing $\kappa_0$ with $\kappa_0/(1+V(k_x,q_z))$ \textcolor{black}{into Eq. \eqref{supfluidSL}}. By doing the analogous approximation for the transverse polariton $\omega_{+,\text{RPA}}$\cite{gabriele_prr22} one gets in the long-wavelength limit the anisotropic generalizations of Eq.s \eqref{plasmaiso} above:
\bea\lb{SLrpamodes}
&\omega_{+,\text{RPA}}^{2}(\bk)=&\frac{\omega_{z}^2k_{x}^2}{(k_x^2+q_z^2)}+\frac{\omega_{xy}^2q_{z}^2}{(k_x^2+q_z^2)}+\frac{c^2}{\varepsilon_B}(k_x^2+q_z^2)\nn\\
&\omega_{-,\text{RPA}}^{2}(\bk)=&\omega_{xy}^2\frac{k_{x}^2}{(k_x^2+q_z^2)}+\omega_{z}^2\frac{q_{z}^2}{(k_x^2+q_z^2)}.
\eea
As shown in Fig.\ \ref{SLdisp} the standard-RPA approach fails at small momenta in describing the correct dispersions, with the velocity of the so-called\cite{lee_rixs_nature18,liu_rixs_npjqm20,zhou_prl20} acoustic plasmon $\omega_{-\text{RPA}}$ diverging as $|\bk|\to0$. Moreover, the crossing among the two $\omega_{\pm,\text{RPA}}$ solutions at finite $k_x$ for intermediate $k_z$ value is an indirect consequence of the fact that the expressions \eqref{SLrpamodes} are non-analytic functions as $|{\bf k}|\rightarrow 0$. Nonetheless, at momenta larger than a scale of the order of 10 $\m$m$^{-1}$ the generalized modes \eqref{SLmodes} approach the RPA results: indeed, in this regime the coupling to the vector potential becomes subleading and accounting only for the effect of Coulomb interactions (i.e. of the scalar potential in the present language) is sufficient for a correct description of the plasma modes, and one indeed recovers the results \eqref{SLrpamodes} usually quoted in the literature in the context e.g. of RIXS measurements\cite{lee_rixs_nature18,liu_rixs_npjqm20,zhou_prl20}. A more detailed discussion of the generalized plasma modes for a single-layer anisotropic superconductor in the small $k_z$ limit, in which $q_z\simeq k_z$, can be found in Ref.[\onlinecite{gabriele_prr22}].

\subsection{Generalization to the bilayer case}\label{sec2c}
In this section we generalize the strategy outlined in the previous two subsections to the case of a bilayer superconductor. The conventions used to describe the out-of-plane layered structure are shown in Fig. \ref{unitcell} for two primitive cells, or "bilayer units". Two SC planes in the same unit cell have an intra-bilayer distance $d_1$, while two subsequent layers belonging to adjacent unit cells have inter-bilayer distance $d_2$, such that $d=d_1+d_2$ identifies the lattice periodicity along the $z$ direction.\\
To correctly describe the phase and the electromagnetic fields some careful steps must be taken. First, we need a discrete notation for all fields along $z$, to account for their different values on the two sheets of a bilayer unit. Secondly, the site of definition of each field and of its derivatives on the bilayer crystal should be chosen coherently with the Maxwell's equations. The problem is not completely trivial, and it is discussed in details in Appendix \ref{appB}.  \\ 
 In order to account for the different nature of the insulating layers in the intra or in the inter-bilayer spacings, we introduce in the system two different Josephson-like interactions\cite{fazio_review01,vandermarel96,vandermarel_prb01}:
\begin{align}\lb{josephcoupl}
&H_{\mathcal J1}=-\mathcal J_1\sum_n\text{cos}(\theta_{1,n}-\theta_{2,n})\nn\\
&H_{\mathcal J2}=-\mathcal J_2\sum_n\text{cos}(\theta_{2,n}-\theta_{1,n+1}),
\end{align}
where the intra-bilayer and the inter-bilayer couplings are respectively controlled by the constants $\mathcal J_1$ and $\mathcal J_2$. The doubling of planes per unit cell has an effect analogous to the folding of the modes that one would observe in a single-layer system described with the "wrong" periodicity. In this last case the modes located at the Brillouin zone boundary would be observed at $k_z=0$ and would be degenerate because $\mathcal J_1=\mathcal J_2$. In the bilayer case however these modes split due to the anisotropy of the Josephson couplings, leading to distinct branches. Such an analogy will be useful in the following to understand the physical origin of some effects. \\
By expanding the cosines and retaining only the second-order terms one can write the Gaussian action of the SC phase, that in real space reads:
\begin{align}\lb{supfluidBL}
&S^{(BL)}_G[\theta]=\frac{1}{8}\int d^2\bx d\tau \sum_n\bigg[\kappa_0\frac{d}{2}[(\partial_\tau\theta_{1,n})^2+(\partial_\tau\theta_{2,n})^2]+\nn\\
&+D_{xy}\frac{d}{2}[(\boldsymbol\nabla_{xy}\theta_{1,n})^2+(\boldsymbol\nabla_{xy}\theta_{2,n})^2]+\nn\\
&+4J_1d_1(\theta_{1,n}-\theta_{2,n})^2+4J_2d_2(\theta_{2,n}-\theta_{1,n+1})^2\bigg]
\end{align}
where $d^2\bx$ is short for $dx\text{ }dy$ and \textcolor{black}{$J_1=\mathcal J_1/(S d_1)$ and $J_2=\mathcal J_2/(S d_2)$ are the densities of the Josephson coupling constants along the SC plane of surface $S$}. The internal electromagnetic field is instead described by the free e.m. action as a generalization of Eq.\ \pref{emaction} to the bilayer case:
\begin{align}\lb{emactionBL}
&S_{\text{e.m.}}^{(BL)}[\phi,\textbf A]=\frac{\varepsilon_B}{8\pi}\int d\tau d^2\bx\sum_n\sum_{\lambda=1,2}\bigg[\frac{d}{2\varepsilon_B}B_{z\lambda,n}^2+\nn\\
&+\frac{d_\lambda}{\varepsilon_B}(B_{x\lambda,n}^2+B_{y\lambda,n}^2)-\frac{d}{2}\left(E_{x\lambda,n}^2+E_{y\lambda,n}^2\right)-d_\lambda E_{z\lambda,n}^2\bigg].
\end{align}
The electric and magnetic fields are defined as
\begin{align}\lb{curl}
\textbf{E}_{\lambda,n}=-\frac{i\partial_\tau \bA_{\lambda,n}}{c}-\boldsymbol\nabla_{\lambda}\phi_{\lambda,n}, \quad
\bB_{\lambda,n}=\boldsymbol{\nabla}_\lambda\times\textbf{A}_{\lambda,n},
\end{align}
where we define $\boldsymbol\nabla_{\lambda}=\begin{pmatrix}\partial_x, & \partial_y, & \Delta_{z\lambda}\end{pmatrix}$ as the gradient operator, with the discrete derivative along the $z$ direction for a generic field $f_\lambda$ that lives on the $\lambda$-th plane given by
\begin{align}\lb{deltaz}
\Delta_{z\lambda}f_{\lambda,n}=
\begin{cases}
\frac{f_{2,n}-f_{1,n}}{d_1}, & \text{if } \lambda=1\\
\frac{f_{1,n+1}-f_{2,n}}{d_2}, & \text{if } \lambda=2.
\end{cases}
\end{align}
The e.m. field is introduced in the SC system described by Eq. \eqref{supfluidBL} by the addition of Eq. \eqref{emactionBL} and by performing the minimal coupling substitution\cite{nagaosa} via the discretization of Eq. \eqref{mincoupreal}, which allows one to immediately define the gauge-invariant fields for a bilayer crystal: in real space, these read
\begin{align}\lb{gaugetfBL}
\boldsymbol\nabla_{xy}\theta_{\lambda,n}&\to\boldsymbol\psi_{xy\lambda,n}=\boldsymbol\nabla_{xy}\theta_{\lambda,n}+\frac{2e}{c}\text{\bA}_{xy\lambda,n}\nn\\
\Delta_{z\lambda}\theta_{\lambda,n}&\to\psi_{z\lambda,n}=\Delta_{z\lambda}\theta_{\lambda,n}+\frac{2e}{c}\text{A}_{z\lambda,n}.
\end{align}
It should be underlined that to keep the gauge-invariant fields consistent with the discretization of the phase and the e.m. fields as discussed in Appendix \ref{appB}, the in-plane components $\boldsymbol\psi_{xy\lambda}$ must be defined on the $\lambda$-th plane while the out-of-plane components $\psi_{z\lambda}$ must be defined on the link between the $\lambda$-th plane and its subsequent, consistent with the physical fact that these quantities are proportional to out-of-plane currents.
\begin{figure*}[t!]
    \centering
    \includegraphics[width=1.0\textwidth,keepaspectratio]{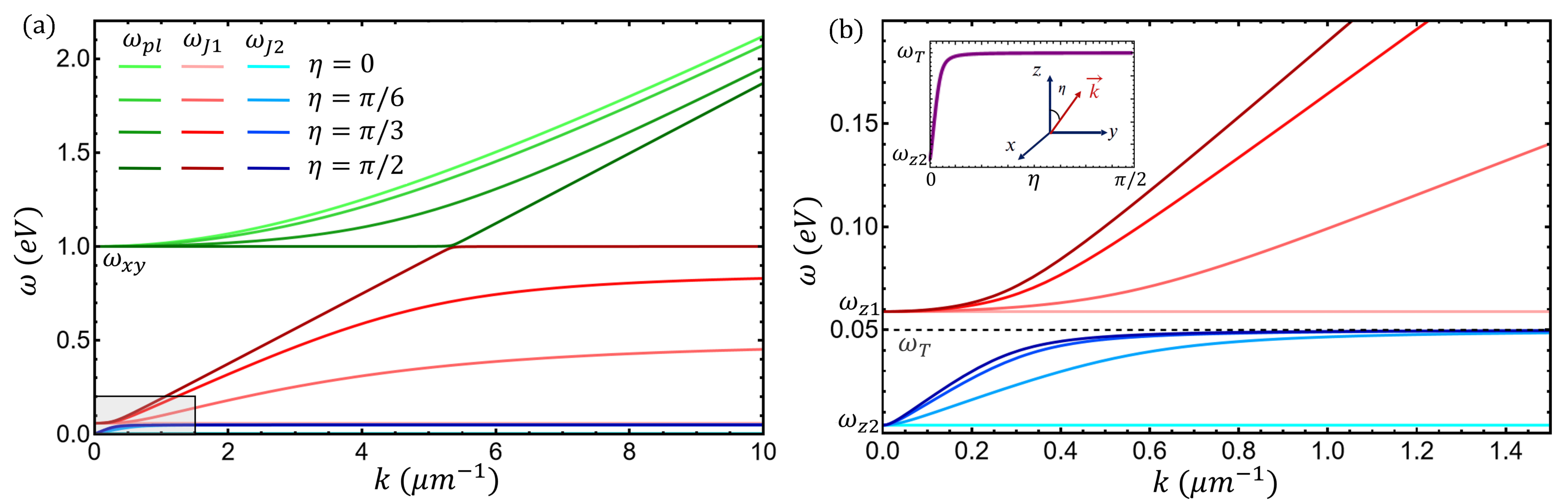}
    \caption{\textbf{(a)} Energy-momentum dispersions of the three lower-in-energy mixed modes $\omega_{pl}$ (green), $\omega_{J1}$ (red) and $\omega_{J2}$ (blue) for selected angles $\eta$ between $\textbf k$ and the $z$ axis. \textbf{(b)} Zoom on the gray-shaded region of panel (a) in which the Josephson modes mix. The transverse plasma frequency $\omega_T$ as defined in Eq. \eqref{omegat} is shown as an horizontal black dashed line. The inset shows the asymptotic value of $\omega_{J2}$ as a function of the angle $\eta$.}
    \lb{dispersion}
\end{figure*}
\begin{figure*}[ht]
    \centering
    \includegraphics[width=1.0\textwidth,keepaspectratio]{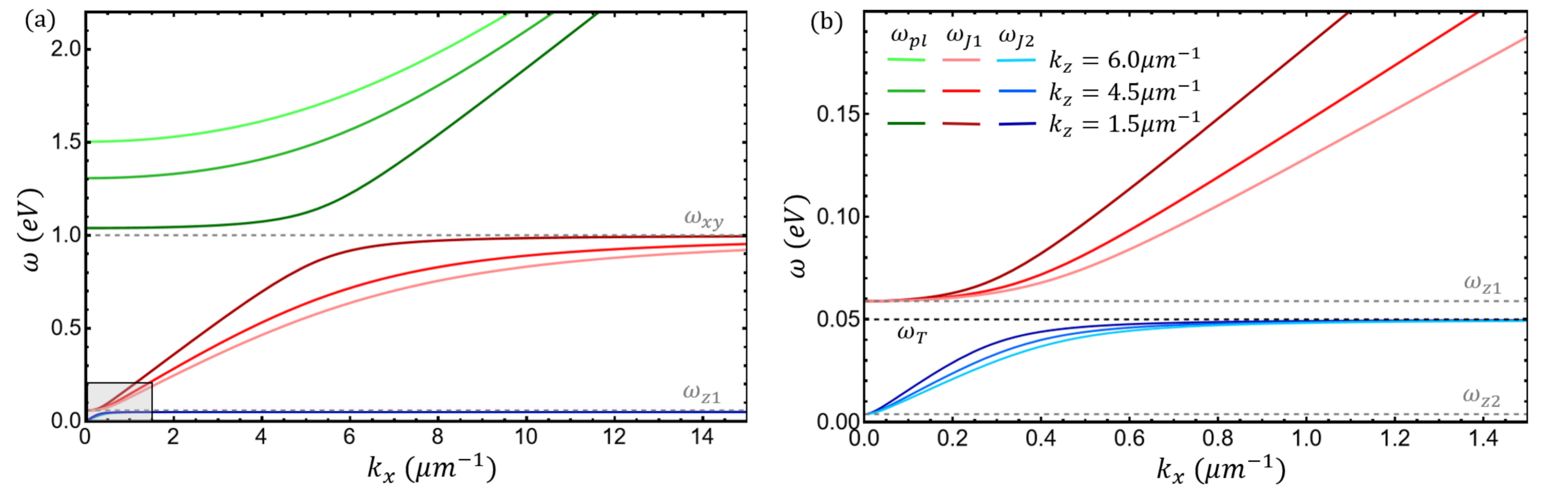}
    \caption{\textbf{(a)} Energy-momentum dispersions of the bilayer superconductor plasma modes $\omega_{pl}$ (green), $\omega_{J1}$ (red) and $\omega_{J2}$ (blue) as functions of $k_x$ for fixed values of $k_z$. Gray dashed lines denote the in-plane plasma frequency and the upper Josephson plasma frequency. \textbf{(b)} Zoom on the gray-shaded region of panel (a) in which the Josephson plasma modes mix. Gray dashed lines denote the upper and lower Josephson plasma frequencies, while the black dashed line denotes the transverse plasma frequency $\omega_T$ as defined in Eq. \eqref{omegat}.}
    \lb{dispersion2}
\end{figure*}
\\
Once the Weyl gauge ($\phi_{\lambda,n}=0$) is chosen,  the system is described by an action $S^{(BL)}[\theta,\boldsymbol\psi]$ which generalises Eq. \eqref{GIvarphaseiso} to the bilayer case. To characterise the plasma modes the SC phase should be integrated out. This calculation is lengthy but straightforward in Fourier space, and the conventions used to define the Fourier transform in the bilayer crystal are discussed in Appendix \ref{appB}. As in the previous subsection, we set the in-plane momentum along the $x$ direction ($k_y=0$) and we here make the approximation of infinite compressibility ($\alpha\to0$). In Fourier space, the action of the gauge-invariant fields once the integration of the SC phases has been carried out can then be written as
\begin{align}\lb{GIvarBL}
S^{(BL)}[\psi]&=\frac{d}{2}\frac{\varepsilon_B}{32\pi e^2}\sum_q\bigg[\boldsymbol\psi_{xz}(q)\mathcal{P}_{xz}^{BL}\boldsymbol\psi_{xz}^T(-q)+\nn\\
&+\boldsymbol\psi_y(q)\mathcal{P}_{y}^{BL}\boldsymbol\psi_y^T(-q)\bigg],
\end{align}
where $\boldsymbol\psi_{xz}=\begin{pmatrix}\psi_{x1}, & \psi_{x2}, &\sqrt{{2d_1}/{d}}\text{ }\psi_{z1}, & \sqrt{{2d_2}/{d}}\text{ }\psi_{z2}\end{pmatrix}$ and $\boldsymbol\psi_{y}=\begin{pmatrix}\psi_{y1}, & \psi_{y2}\end{pmatrix}$. In this basis the coefficient matrix for the $xz$ components of the gauge-invariant fields is 
\begin{align}\lb{BLmatrix}
    \mathcal{P}_{xz}^{BL}=\begin{pmatrix}
	\Omega_m^2\mathbb{1}+\Omega_{xy}^2+\frac{c^2}{\varepsilon_B}\mathcal{Q}_z^{\dagger}\mathcal{Q}_z & -\frac{c^2}{\varepsilon_B}k_x\mathcal{Q}_z^{\dagger}\\
    -\frac{c^2}{\varepsilon_B}k_x\mathcal{Q}_z & \Omega_m^2\mathbb{1}+\Omega_{z}^2+\mathbb{1}\frac{c^2}{\varepsilon_B}k_x^2
    \end{pmatrix},
\end{align}
while the coefficient matrix for the $y$ components is 
\begin{align}\lb{yBLmatrix}
\mathcal{P}_y^{BL}=\left(\Omega_m^2\mathbb1+\Omega_{xy}^2+\mathbb1\frac{c^2}{\varepsilon_B}k_x^2+\frac{c^2}{\varepsilon_B}\mathcal{Q}_z^{\dagger}\mathcal{Q}_z\right).
\end{align}
Where $\mathbb{1}$ is the $2\times2$ identity matrix. The in-plane plasma frequency appears in the matrix $\Omega_{xy}^2=\mathbb{1}\omega_{xy}^2$ while the out-of-plane Josephson plasma frequencies\cite{vandermarel96,vandermarel_prb01,demler_prb20,alpeggiani_prb13}, defined as
\begin{align}\lb{plasmaBL}
\omega_{z\lambda}=\sqrt{\frac{16\pi e^2J_\lambda d_\lambda^2}{\varepsilon_B}},
\end{align}
are inside $\Omega_z^2=\begin{pmatrix}\omega_{z1}^2 & 0 \\ 0 & \omega_{z2}^2\end{pmatrix}$. The matrix $\mathcal{Q}_z$ is defined as
\begin{align}\lb{Qz}
    \mathcal{Q}_{z}=-i\sqrt{\frac{2}{d}}\begin{pmatrix}
        \frac{e^{ik_zd_1/2}}{\sqrt{d_1}} & -\frac{e^{-ik_zd_1/2}}{\sqrt{d_1}}\\
        -\frac{e^{-ik_zd_2/2}}{\sqrt{d_2}} & \frac{e^{ik_zd_2/2}}{\sqrt{d_2}}
    \end{pmatrix},
\end{align}
while $\mathcal{Q}_z^{\dagger}$ is its hermitian conjugate. These two matrices represent the generalization of the out-of-plane momentum $q_z$ in Eq. \eqref{qz} to the bilayer structure, and their product is
\begin{align}\lb{QzQz}
\mathcal{Q}_{z}^\dagger\mathcal{Q}_z=\frac{2}{d}\begin{pmatrix}
\frac{1}{d_1}+\frac{1}{d_2} & -\frac{e^{-ik_zd_1}}{d_1}-\frac{e^{ik_zd_2}}{d_2}\\
 -\frac{e^{ik_zd_1}}{d_1}-\frac{e^{-ik_zd_2}}{d_2}& \frac{1}{d_1}+\frac{1}{d_2}
\end{pmatrix}.
\end{align}
The action in Eq. \eqref{GIvarBL} is the first central result of this work, as it describes the e.m. modes of the bilayer superconductor. Formally it is equivalent to the single-layer one in Eq. \eqref{GIvarSL}, but now it displays a $2\times2$ structure in the $y$ component and a $4\times4$ structure in the $xz$ components. In addition, even though formally $\mathcal{Q}_{z}$ plays in Eq.\ \eqref{GIvarBL} the analogous role of $q_z$ in Eq. \eqref{GIvarSL} for the single-layer case, the analogy is not complete. Indeed, as we will discuss below, $\mathcal{Q}_{z}$ does {\em not} vanish as $k_z=0$, leading to observable and relevant physical consequences in the bilayer system. In general, the $(4\times 4)+(2\times 2)$ structure of the action in Eq. \eqref{GIvarBL} implies that in the bilayer superconductor there are a total of six modes: two decoupled transverse plasma-polaritons (TE) described by $\mathcal{P}_y^{BL}$ and four mixed modes (TM) encoded into $\mathcal{P}_{xz}^{BL}$.\\
The remainder of this section will focus on the behaviour of the energy-momentum dispersions of the mixed TM modes, found numerically as solutions of the characteristic equation obtained by setting the determinant of $\mathcal{P}_{xz}^{BL}$ to zero once the analytical continuation has been performed. The dispersions of the three lower-in-energy mixed modes, which we label $\omega_{pl}$, $\omega_{J1}$ and $\omega_{J2}$ are shown in Fig. \ref{dispersion} for various propagation angles $\eta$ formed by $\bk$ and the $z$ axis. 
In the plots we set $\omega_{xy}=1\text{eV}$ and $\varepsilon_B=1$ for numerical simplicity, while we choose $d_1=3.2\angstrom$, $d_2=8.2\angstrom$ and the Josephson plasma frequencies as $\omega_{z1}=14.2\text{THz}=5.9\cdot 10^{-2}\text{eV}$ and $\omega_{z2}=0.9\text{THz}=3.7\cdot10^{-3}\text{eV}$ to be compatible with those measured in the YBCO cuprate superconductor at doping $x=0.5$ ($T_c=50K$)\cite{cavalleri_prx22,kaiser_prb14}.\\
The limits for $\textbf k\to 0$ of the dispersions are regular and equal to their corresponding plasma frequencies: using $\omega_{xy}>\omega_{z1}>\omega_{z2}$, one immediately sees that
\begin{align}\lb{ktozero}
&\omega_{pl}(\textbf k\to 0)=\omega_{xy}, \nn\\ 
&\omega_{J1}(\textbf k\to 0)=\omega_{z1}, \nn\\ 
&\omega_{J2}(\textbf k\to 0)=\omega_{z2},
\end{align}
regardless of the direction along which such limit is taken. The fourth mode is much higher in energy, with plasma frequency $\sqrt{\omega_{xy}^2+\frac{c^2}{\varepsilon_B}\frac{4}{d_1d_2}}\gg \omega_{xy}$.
As anticipated above, and as it will be discussed further in Sec.\ \ref{sec2d}, this mode can be thought as a folding at $k_z=0$ of the single-layer mode $\omega_+$ at the zone boundary. Since it falls outside the range of frequencies where the model itself can be reasonably applied, it will not be discussed in details in the following. One should note that Eq.s \eqref{ktozero} are only valid in the approximation of infinite compressibility. While this is appropriate for typical cuprate superconductors, in Sec. \ref{sec2e} we will discuss corrections to the $\bk\to0$ limits of the Josephson modes given by a finite compressibility.\\
In the limiting case $\eta=0$ ($k_x=0$) the Josephson modes are non-dispersive at finite $\textbf k$, while $\omega_{pl}$ disperses with the light velocity in the medium $c/\sqrt{\varepsilon_B}$ as expected for a light mode, see Fig. \ref{dispersion}(a,b). In this particular case, the in-plane and out-of-plane modes are decoupled.\\
In any other case the three modes are coupled for finite $\textbf k$ and three different regimes are identified in Fig.\ \ref{dispersion}, separated by two crossover momenta $\textbf k_{c1}$ and $\textbf k_{c2}$. Remarkably, as $\eta=\pi/2$ ($k_z=0$) the upper-right and lower-left blocks of Eq. \eqref{BLmatrix} do not vanish, as the generalized out-of-plane momentum ${\cal Q}_z$ in Eq. \eqref{Qz} does not become the zero matrix, as we anticipated before. This means that the in-plane and out-of-plane components are still coupled even when the momentum is completely along the planes, as opposed to the single-layer case, see Eq. \eqref{SLmatrix} and the discussion below it.\\
For low momenta and for any angle $\eta \neq 0$, the lower Josephson solution $\omega_{J2}$ grows with light velocity, see Fig.\ \ref{dispersion}(b).\\ After the first crossover momentum,
\begin{align}\lb{kc1}
|\textbf k_{c1}|=\frac{\sqrt{\varepsilon_B}}{c}\sqrt{\omega_{z1}^2-\omega_{z2}^2},
\end{align}
it goes towards an asymptotic frequency that depends on the angle $\eta$, see inset of Fig. \ref{dispersion}(b). Its maximum value is taken for $\eta=\pi/2$, where it coincides with a frequency scale named in the previous literature - for reasons that we will clarify below - the "transverse" plasma frequency\cite{vandermarel_prb01,vandermarel96}
\begin{equation}
\lb{omegat}
\omega_T=\sqrt{(\omega_{z1}^2d_2+\omega_{z2}^2d_1)/d}.
\end{equation}
On the contrary, the upper Josephson solution $\omega_{J1}$ grows weakly for small momenta, while it starts dispersing with light velocity above $\textbf k_{c1}$. This behaviour is kept until the second crossover momentum, 
\begin{align}\lb{kc2}
|\textbf k_{c2}|=\frac{\sqrt{\varepsilon_B}}{c}\sqrt{\omega_{xy}^2-(\omega_{z1}^2d_1+\omega_{z2}^2d_2)/d},
\end{align}
above which the solution goes towards an asymptotic frequency that again depends on the angle $\eta$, see Fig.\ \ref{dispersion}(a). As $\eta=\pi/2$, the asymptotic value coincides with the in-plane plasma frequency $\omega_{xy}$. The third solution $\omega_{pl}$ grows weakly below $\textbf k_{c2}$ and starts dispersing with light velocity above it. In this regime, $\omega_{J1}$ and $\omega_{pl}$ follow the analogous behaviour of the single-layer modes $\omega_{-}$ and $\omega_{+}$ of Eq. \eqref{SLmodes}.    \\
To have an idea of the orders of magnitude of the crossover momenta, one can set the light velocity in the medium to $\hbar c\simeq0.187\text{eV}\mu\text{m}$. The lower crossover momentum depends on the difference between the two Josephson plasma frequencies, giving $|\textbf k_{c1}|\sim 0.05-0.5\mu\text{m}^{-1}$ depending on the bilayer system considered. The upper crossover momentum can be estimated by 
considering that in most layered superconductors as e.g. cuprates it is usually $\omega_{xy}\gg\omega_{z1,z2}$, so that with $\omega_{xy}=1\text{eV}$ one has $|\textbf k_{c2}|\simeq 5\mu\text{m}^{-1}$. \\
In Fig.\ \ref{dispersion2} we show the same dispersions as a function of $k_x$ only, for fixed values of $k_z$. In Fig. \ref{dispersion2}(a) one immediately recognizes the close resemblance between the two dispersions $\omega_{J1}$ and $\omega_{pl}$ and their single-layer counterparts $\omega_{-}$ and $\omega_{+}$  shown in Fig.\ \ref{SLdisp}, while the behavior of the two Josephson modes around the first crossover $\textbf k_{c1}$ shown in Fig.\ \ref{dispersion2}(b) is analogous to the one discussed before.\\
The formalism employed in this work allows one to study the modes also in the nonrelativistic regime, where the $\omega_{J2}$ mode changes its behaviour. This is discussed in details in Appendix \ref{appC}.
\begin{figure*}[t!]
    \centering \includegraphics[width=0.95\textwidth,keepaspectratio]{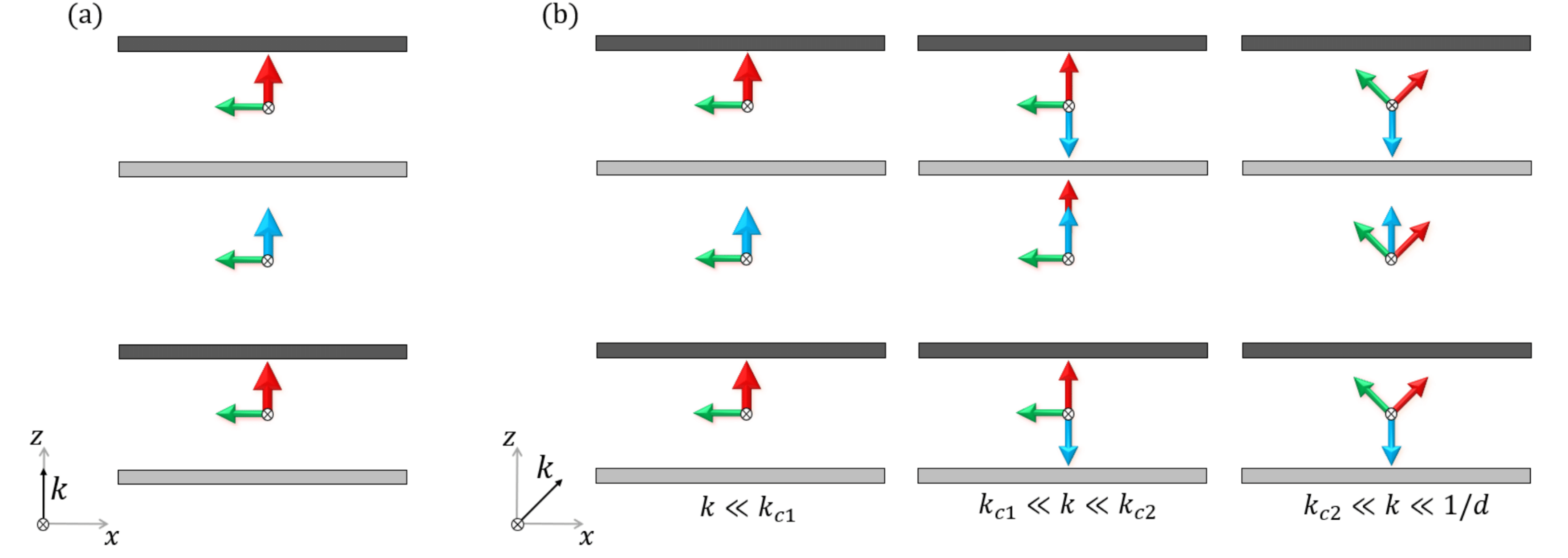}
    \caption{Sketch of the polarizations of the three mixed modes for $\eta=0$ (panel (a)) and for $\eta=\pi/4$ (panel (b) for different momentum regimes) depicted on two bilayer units. The colours are chosen to resemble those of the corresponding dispersions in Fig. \ref{dispersion}: the mode with frequency $\omega_{pl}$ corresponds to the green arrows,  the mode with frequency $\omega_{J1}$ to the red arrows and the mode with frequency $\omega_{J2}$ to the blue arrows. \textcolor{black}{The direction of an arrow denotes the direction of the plasma oscillations while the relative orientation of arrows} with same colour in neighbouring cells denotes whether the oscillations are in-phase or out-of-phase. The width of an arrow is proportional to the magnitude of the corresponding eigenvector component, while their length does not have a physical meaning. For a simple graphical representation we centered the arrows in the spacing between two subsequent layers, although one should always keep in mind that the $x$ components of the eigenvectors always represent oscillations that take place along the SC planes.} 
    \lb{polar}
\end{figure*}
\subsection{Polarizations of the mixed plasma modes}\label{sec2d}

To gain further insight into the nature of the mixed plasma oscillations in a bilayer superconductor it is instructive to have a closer look to their polarizations as functions of $\textbf k$ in the limit of infinite compressibility. These come as normalized eigenvectors to the eigenproblem set by the coefficient matrix $\mathcal{P}_{xz}^{BL}$ in Eq. \eqref{BLmatrix} where one interprets the components $\psi_{x\lambda}$ as the in-plane oscillations on the $\lambda$-th layer and the components $\psi_{z1}$ and $\psi_{z2}$ as the out-of-plane oscillations in the intra-bilayer and in the inter-bilayer respectively. \\
The polarizations of the three lower-in-energy mixed modes are sketched in Fig. \ref{polar} as vectors representing the components of the modes. In the limiting case $\eta=0$, see Fig. \ref{polar}(a), the Josephson modes with constant frequencies $\omega_{z1}$ and $\omega_{z2}$ correspond to oscillations oriented purely along the $z$ direction. In addition, the former describes oscillations living exclusively between the two layers of a same unit cell, while the latter describes oscillations confined to inter-cell layers. The higher-in-energy $\omega_{pl}$ mode corresponds instead to in-plane oscillations that are in phase with respect to the two layers of the unit cell. These "pure" behaviours are respected for any value of the out-of-plane momentum $k_z$.
It is worth noting that we are discussing here polarization eigenvectors in momentum space: as a consequence, while for $k_z\ll \pi/d$ they also represent the oscillation patterns in real space, as $k_z\to \pi/d$ the real-space pattern in neighbouring planes can change with respect to the representation given in Fig.\ \ref{polar} . This is however a trivial effect, and we restrict here for simplicity the discussion to the low-momentum region, in order to visualize in a simple way the distinctive features of the eigenvectors. \\
For any other possible angle $\eta$, shown in Fig.\ \ref{polar}(b), the modes are pure only for $\textbf k\to0$ as they reproduce the scheme of Fig. \ref{polar}(a). For $\textbf k\sim\textbf k_{c1}$ the two Josephson plasma modes mix, the one with frequency $\omega_{J1}$ acquiring an in-phase inter-bilayer component and the one with frequency $\omega_{J2}$ acquiring an opposite-phase intra-bilayer component. Remarkably, the lower Josephson plasma mode holds its opposite-phase oscillations along the $z$ direction as long as its dispersion maintains its saturating behaviour, up to momenta $|\textbf k|\sim1/d$, regardless of the angle $\eta$. Instead, the upper Josephson plasma mode mixes with the $\omega_{pl}$ mode for $\textbf k\sim\textbf k_{c2}$, the former becoming purely longitudinal and the latter becoming purely transverse. This mechanism of longitudinal-transverse mixing between these two in-phase modes happens exactly as it would in a single-layer superconductor as described in Ref.[\onlinecite{gabriele_prr22}]. The fourth higher-in-energy mixed mode displays in-plane oscillations that are in opposite-phase with respect to the two layers of the unit cell for momenta up to $|\textbf k|\sim 1/d$. 
The latter and the low-energy Josephson mode are thus both connected to out-of-phase oscillations in neighbouring layers within the same unit cell. In the limit where $\mathcal J_1=\mathcal J_2$ and $d_1=d_2$ they would then correspond to the modes of the single-layer crystal occurring at the boundaries of the Brillouin zone. As we discussed above, in such picture one can think of these modes as the folded images of the single-layer modes
due to the broken symmetry $\mathcal J_1\neq \mathcal J_2$.

\subsection{Josephson plasma frequencies corrections with finite compressibility}\label{sec2e}
In the previous subsections we made the approximation of $\alpha\to0$, which by Eq. \eqref{alpha} means taking an infinite compressibility or, equivalently, vanishing screening length. This approximation is well-justified in cuprates as the thickness of the SC layers is much larger than the screening length\cite{machida_physc00, konsin_prb98}, and a theory of stacked junctions adopting this approximation\cite{vandermarel96} appears to describe well experimental results on bilayer superconductors\cite{homes_prl93,bernhard_prl11,erb_prl00,tajima_prl09, yamada_prl98,cavalleri_nmat14, wangNL_prx20}. Nonetheless, a consistent interaction between electrons given by a finite compressibility should be considered to estimate the relevance of the corrections to the various physical quantities. This was carried out in Ref. \cite{vandermarel_prb01}: in this subsection we recover the same results using the formalism developed above, to have a better understanding of the physical phenomenon that affects the energy of the Josephson modes for vanishing momentum in bilayer superconductors. Indeed, this case yields some interesting insights that distinguish the bilayer from the single-layer case.\\
By taking a finite value for $\alpha$ one finds that the action of the $x$ and $z$ components of the gauge-invariant variables in Eq. \eqref{GIvarBL} gets corrected as
\begin{align}\lb{BLactAlpha}
&S_{\alpha\neq0}^{(BL)}[\psi_x,\psi_z]=\frac{d}{2}\frac{\varepsilon_B}{32\pi e^2}\sum_q\bigg[\boldsymbol\psi_{xz}(q)\mathcal{P}_{xz}^{BL}\boldsymbol\psi_{xz}^T(-q)+\nn\\
&-\alpha\Omega_m^2\left(\boldsymbol\psi_x(q) k_x+\boldsymbol\psi_z(q)\mathcal{Q}_z\right)\left(\mathbb1+\mathbb1\alpha k_x^2+\alpha \mathcal{Q}_z^\dagger\mathcal{Q}_z\right)^{-1}\nn\\
&\times\left(k_x\boldsymbol\psi_x^T(-q)+\mathcal{Q}_z^\dagger\boldsymbol\psi_z^T(-q)\right)\bigg],
\end{align}
where $\mathcal{P}_{xz}^{BL}$ is defined as in Eq. \eqref{BLmatrix}, $\boldsymbol\psi_x=\begin{pmatrix}\psi_{x1}, & \psi_{x2}\end{pmatrix}$ and $\boldsymbol\psi_z=\begin{pmatrix}\sqrt{2d_1/d}\text{ }\psi_{z1}, & \sqrt{2d_2/d}\text{ }\psi_{z2}\end{pmatrix}$ in agreement with the definitions given above.\\
Although formally this is the same result one would find in the single-layer case\cite{gabriele_prr22}, there is a substantial difference: while in the single-layer crystal the $\alpha$ corrections are purely longitudinal and vanish in the $\bk\to0$ limit, in the bilayer system the corrections have both a longitudinal and a massive component due to the fact that the $\mathcal Q_z$ matrix and its complex conjugate are finite for $k_z\to0$. This implies that the limits for vanishing momentum of the dispersions, i.e. the Josephson plasma frequencies $\omega_{z1}$ and $\omega_{z2}$ defined in Eq. \eqref{plasmaBL}, are corrected with terms of order $\alpha$.\\
To explicitly derive these corrections we here focus only on the $z$ components of the gauge-invariant variables in the limit for $\bk\to0$. As the in-plane and out-of-plane components of the oscillations are decoupled when the in-plane momentum is set to zero due to the vanishing of the off-diagonal elements in the action \eqref{BLactAlpha}, the Josephson plasmons in this limit are described by the $2\times 2$ action
\begin{align}\lb{GIvarAlpha}
S^{(BL)}_{\alpha\neq0}[\psi_z]=\frac{\varepsilon_B}{32\pi e^2}\frac{d}{2}\sum_{i\Omega_m}\bigg[\boldsymbol\psi_{z}(i\Omega_m)\mathcal{P}_{z}^{BL}\boldsymbol\psi_{z}^T(-i\Omega_m)\bigg],
\end{align}
where the coefficient matrix reads
\begin{align}\lb{BLalpha}
\mathcal{P}_z^{BL}=\big[\Omega_z^2+\Omega_m^2\mathbb1\big]-\frac{4}{d}\Omega_m^2\mathcal{C}.
\end{align}
Here, $\mathcal C$ is a matrix with the dimensions of a capacitance defined as
\begin{align}\lb{capac}
\mathcal{C}=\frac{1}{1+4\alpha/(d_1d_2)}\begin{pmatrix}
    \alpha/d_1 & -\alpha/\sqrt{d_1d_2}\\
    -\alpha/\sqrt{d_1d_2} & \alpha/d_2
\end{pmatrix}.
\end{align}
Notice that due to the presence of off-diagonal components in the matrix in Eq. \eqref{capac}, the $z$ components of the gauge-invariant variables are coupled by terms of order $\alpha$. Indeed, performing the analytical continuation $i\Omega_m\to\omega$ and by solving the characteristic equation of $\mathcal{P}_z^{BL}$ one finds the corrected Josephson plasma frequencies, previously reported in Ref.[\onlinecite{vandermarel_prb01}]:
\begin{align}\lb{freqAlpha}
\tilde\omega_{z1,z2}&= \Bigg\{\bigg(\frac{1}{2}+\frac{2\alpha}{d^2}\frac{d}{d_1}\bigg)\omega_{z1}^2+\bigg(\frac{1}{2}+\frac{2\alpha}{d^2}\frac{d}{d_2}\bigg)\omega_{z2}^2+\nn\\
&\pm\Bigg[\bigg[\bigg(\frac{1}{2}+\frac{2\alpha}{d^2}\frac{d}{d_1}\bigg)\omega_{z1}^2-\bigg(\frac{1}{2}+\frac{2\alpha}{d^2}\frac{d}{d_2}\bigg)\omega_{z2}^2\bigg]^2+\nn\\
&+\frac{16\alpha^2}{d^4}\frac{d^2}{d_1d_2}\omega_{z1}^2\omega_{z2}^2\Bigg]^{1/2}\Bigg\}^{1/2}.
\end{align}
To understand the physical phenomenon behind these slight frequency shifts with respect to the original $\omega_{z1,z2}$ one should notice that in Eq. \eqref{BLalpha} all the $\alpha$-dependent terms are in $\mathcal C$, which means that the corrections to the plasma frequencies for $\textbf k\to0$ come from a capacitive coupling between two subsequent layers as expected when a finite compressibility in the planes is taken into account.\\
As we mentioned above, there is no frequency shift for the single-layer Josephson plasma mode if one considered a finite compressibility. The physical reason is that as $k_z\to 0$ the charge distribution is the same in each plane, and then no capacitive coupling between neighbouring planes emerges,  even if a finite compressibility is considered. This is not the case in the bilayer superconductor: indeed, in this case a charge gradient is possible among two layers of the same unit cell even for $k_z\to0$, as the potentials $\phi_1$ and $\phi_2$ are generically different. Such a mechanism is also evidenced by studying the eigenvalues and eigenvectors of the capacitance matrix $\mathcal{C}$: the eigenvector having $\psi_{z2}=\psi_{z1}$ corresponds to an eigenvalue equal to zero, while the eigenvector having $\psi_{z2}=-\psi_{z1}$ corresponds to a non-zero eigenvalue. Since in the single-layer limit in which $\mathcal J_1=\mathcal J_2$ and $d_1=d_2$ the only possible solution of Eq.\ \pref{BLalpha} at $k_z=0$ requires $\psi_{z2}=\psi_{z1}$, one understands why in the single-layer superconductor capacitive effects are irrelevant for vanishing momentum. 
The solution $\psi_{z2}=-\psi_{z1}$ is only acceptable at the Brillouin zone boundary, where the dispersion $\omega_-$ of the single-layer Josephson mode grows linearly\cite{gabriele_prr22} with sound velocity $v_s\propto\sqrt\alpha$. In this framework one can also understand the compressibility corrections to the Josephson frequencies in the bilayer system in Eq. \eqref{freqAlpha} as given by the folding at $k_z=0$ of the single-layer dispersion once the $\mathcal{J}_1=\mathcal{J}_2$ symmetry is broken.
\\ 
The effects of a finite compressibility for finite momenta and in the nonrelativistic regime are discussed in Appendix \ref{appC}.

\section{Linear response to an external e.m. field}\label{sec3}
\subsection{Experimental observations}
In this section we focus on the nontrivial $z$-axis linear optical properties of a bilayer superconductor. As mentioned in the introduction, several experimental papers reported the appearance in the SC state of YBCO of a rather well-defined peak in the real part of the optical conductivity\cite{homes_prl93,bernhard_prl11,erb_prl00,tajima_prl09, yamada_prl98,cavalleri_nmat14, wangNL_prx20} at the transverse plasma frequency $\omega_T$ defined in Eq. \eqref{omegat}. Such an experimental observation has been successfully explained by the so-called Multilayer Model\cite{vandermarel96,vandermarel_prb01} (MLM), which gives a precise recipe on how to reconstruct the dielectric function of the layered system as a series of capacitors represented by each layer. The aim of this section is to derive the results of the MLM within our formalism and discuss its physical implications in light of the characterization of the e.m. modes provided in the previous section.\\
Before giving the technical details, it is worth stressing why the experimental observation of a peak in the real-part of the optical conductivity at a "plasma" frequency appears at first sight rather puzzling. As the discussion in the previous sections highlighted, plasmons are strictly speaking zeros of the dielectric function $\epsilon(\omega)$, see Eq.\ \pref{GIvariso}, which is related to the complex conductivity by the standard relation:  
\begin{align}\lb{epsilon}
\varepsilon(\omega)=\varepsilon_B+ \frac{4\pi i\sigma(\omega)}{\omega}.
\end{align}
In the case of the superconductors the optical conductivity $\sigma(\omega)$, computed via a current-current correlation function, can be indeed expressed via the correlation function for the SC phase fluctuations. However, since $\sigma$ is the response to the local electric field $\bE$, one should consider the irreducible response with respect to the Coulomb interaction\cite{pines,pick_prb70,belitz_prb89}. In other words, $\sigma$ should be related to the phase fluctuations computed {\em without} including\cite{cea_prb14} the RPA dressing of the action via $V(\bk)$. Considering again the simple isotropic case one then finds from Eq.\ \pref{supfluid} that $\sigma(\omega)=-D_s/(i\omega)$, that substituted into Eq.\ \pref{epsilon} leads again to the result in Eq. \pref{dieliso} in the long-wavelength limit. However, the conductivity itself has no features at the plasma frequency $\omega_P$, and its real part is exactly zero in a clean isotropic superconductor. The results in bilayer cuprates show, on the contrary, that $\sigma_1(\omega)$ displays a peak at the frequency $\omega_T$ of Eq. \eqref{omegat}. As we have seen above, this frequency is not connected to an electromagnetic mode at zero momentum, but it is instead connected to the large-momentum limit of the lower $\omega_{J2}$ Josephson plasmon. As we shall see below, the reason behind its appearance in the optical conductivity lies on the fact that the optical response is irreducible with respect to the $\bk\to 0$ Coulomb interaction, but it can be nonetheless affected by the {\em large momentum} electromagnetic interactions, leading to the rather interesting physical effects observed in YBCO.

\subsection{Optical conductivity}

In linear response theory the current density $\textbf{J}(\omega)$ induced by an external monochromatic e.m. field $\textbf{A}^{\text{ext}}(\omega)$ with vanishing momentum can be written as
\begin{align}\lb{linres}
\text{J}_i(\omega)=-\frac{1}{c}\text{K}_{ij}(\omega)\text{A}^{\text{ext}}_j(\omega)
\end{align}
where $\text{K}_{ij}(\omega)$ is the current-current linear response kernel, which in the effective-action formalism in Matsubara space can be evaluated as\cite{nagaosa,coleman}
\begin{align}\lb{kern}
\text{K}_{ij}(i\Omega_m)=\frac{c^2}{d}\frac{\partial^2 S[\textbf A^{\text{ext}}]}{\partial \text A^{\text{ext}}_i(i\Omega_m)\text{ }\partial \text A^{\text{ext}}_j(-i\Omega_m)},
\end{align}
where $S[\textbf A^{\text{ext}}]$ is the effective action obtained after the integration of the internal degrees of freedom of the system. More specifically, within the effective-action scheme employed here we need to integrate out the degrees of freedom of the matter, represented by the SC phase, which is linearly coupled to the gauge field. The out-of-plane lattice constant $d$ appearing in Eq. \eqref{kern} is consistent with our choice for the normalization of the Fourier transforms. In the following we will consider an external uniform electric field polarized along the $z$ direction and incidence parallel to the SC sheets along the $xy$ plane ($k_z=0)$, as it is the case for measurements of the $c$-axis response in Ref.s \cite{homes_prl93,bernhard_prl11,erb_prl00,tajima_prl09, yamada_prl98,cavalleri_nmat14, wangNL_prx20,wangNL_cm22,shimano_prb23}. \\
Our starting point is thus the Gaussian action for the SC phases in Eq. \eqref{supfluidBL} in which we introduce an external vector potential along the $z$ direction by means of the minimal coupling substitution equivalent to Eq.\ \pref{gaugetfBL} above:
\begin{align}\lb{mcAext}
\Delta_{z\lambda}\theta_{\lambda,n}\to\Delta_{z\lambda}\theta_{\lambda,n}+\frac{2e}{c}\text{A}^{\text{ext}}_{z,n}.
\end{align}
Let's at first suppose that, as in the isotropic case, one should not dress the internal degrees of freedom with the Coulomb interaction, i.e. in the language of the internal e.m. fields used in the previous section one should not consider an internal scalar potential coupled to the SC phase. In this case, the optical conductivity could be easily derived by shifting to Fourier space according to the rules discussed in Appendix \ref{appB} and defining the variables 
\begin{align}\lb{thetapm}
&\theta_{+}(q)=\theta_2(q)+\theta_1(q),\nn\\
&\theta_{-}(q)=\theta_2(q)-\theta_1(q),
\end{align}
so that the total action reads
\begin{align}\lb{sthetapm}
&S[\theta_{\pm},\text A_{z}^{\text{ext}}]=\frac{1}{8}\sum_q\bigg[\frac{d}{4}\big(\kappa_0\Omega_m^2+D_{xy}k_{xy}^2\big)\big|\theta_{+}(q)\big|^2\big)+\nn\\
&+\bigg(\frac{d}{4}\kappa_0\Omega_m^2+\frac{d}{4}D_{xy}k_{xy}^2+4J_1d_1+4J_2d_2\bigg)\big|\theta_-(q)\big|^2\bigg]+\nn\\
&+\frac{\varepsilon_B}{8\pi c^2}\sum_q\bigg[\big(\omega_{z1}^2d_1+\omega_{z2}^2d_2\big)|\text A_{z}^{\text{ext}}(q)|^2\bigg]+\nn\\
&+\frac{\varepsilon_B}{16\pi e c}\sum_q\bigg[\big(\omega_{z1}^2-\omega_{z2}^2\big)\big(\theta_{-}(q)\text A_z^{\text{ext}}(-q)+h.c.\big)\bigg].
\end{align}
Notice that because the external field is polarized along $z$ it only couples to the phase gradient in the $z$ direction, that is represented, in the present discrete notation, by the $\theta_-$ variable. It is then straightforward to show, after the integration of the internal degree of freedom $\theta_-$ and taking the limit for $k_{xy}\to0$, that one is left with
\begin{align}\lb{linresAct}
S[\text A_z^{\text{ext}}]=\frac{d}{2c^2}\sum_{i\Omega_m}\text A_z^{\text{ext}}(i\Omega_m)\text{K}_{zz}(i\Omega_m)\text A_z^{\text{ext}}(-i\Omega_m),
\end{align}
where the current-current linear response kernel is
\begin{align}\lb{defsigma}
\text K_{zz}&(i\Omega_m)=\frac{\varepsilon_B}{4\pi d}\bigg[\big(\omega_{z1}^2d_1+\omega_{z2}^2d_2\big)+\nn\\
-&\frac{\varepsilon_B}{32\pi e^2}(\omega_{z1}^2-\omega_{z2}^2)^2\langle\theta_-(i\Omega_m)\theta_-(-i\Omega_m)\rangle\bigg]=\nn\\
=&\frac{\varepsilon_B}{4\pi d}\bigg[\big(\omega_{z1}^2d_1+\omega_{z2}^2d_2\big)-\frac{4\alpha}{d}\frac{(\omega_{z1}^2-\omega_{z2}^2)^2}{(\frac{4\alpha}{d_1d_2}\omega_T^2-(i\Omega_m)^2)}\bigg].
\end{align}
Consequently one is able to write the conductivity after the analytical continuation $i\Omega_m\to \omega+i0^+$ as
\begin{align}\lb{cond}
\sigma(\omega)&=\sigma_1(\omega)+i\sigma_2(\omega)=\frac{i}{\omega+i0^+}\text{K}_{zz}(\omega+i0^+)=\nn\\
&=\left[\pi K_{zz1}(\omega)\delta(\omega)-\frac{K_{zz2}(\omega)}{\omega}\right]+\frac{iK_{zz1}(\omega)}{\omega}.
\end{align}
As one can see in the square brackets, in the bilayer system the real part of the conductivity at $k_z=0$ is given by two terms. The first one is a delta-like response at $\omega=0$ given by the real part $K_{zz1}$ of the response kernel. The second term of Eq.\ \pref{cond} given by the imaginary part $K_{zz2}$ represents instead a delta-like response at finite-frequency, absent in the single-layer case, controlled by the relative intra-cell phase fluctuations described by the variable $\theta_-$. However, even though this second contribution admits a peak, it is not at $\omega_T$. More importantly, such a correction vanishes in the limit of infinite compressibility, that is the appropriate one for cuprates, as we discussed above.\\
So far we did not include any long-range effect, with the idea that for isotropic and anisotropic single-layer superconductors one should not consider the internal Coulomb interactions, as doing so in these systems would mean taking into account reducible diagrams\cite{pines,pick_prb70,belitz_prb89, cea_prb14}, as we discussed in the previous subsection. However, in bilayer superconductors one finds that an internal scalar potential does not only describe the long-range Coulomb interactions but also an intra-bilayer interaction at $|\bk|=0$ that should be taken into account.\\
We thus introduce again the internal scalar potential by means of the first minimal coupling substitution in Eq. \pref{mincoup}.
One sees that, in analogy with the definitions \eqref{thetapm} of the phase variables, there are two possible combinations of the scalar potentials:  
\begin{align}\lb{phipm}
&\phi_{+}(q)=\phi_2(q)+\phi_1(q),\nn\\
&\phi_{-}(q)=\phi_2(q)-\phi_1(q).
\end{align}
We also introduce the free e.m. action in a bilayer system as in Eq. \eqref{emactionBL} expressing it by means of the $\phi_\pm$ variables,
\begin{align}\lb{emphi}
S_{\text{e.m.}}^{(BL)}[\phi_{\pm}]&=-\frac{\varepsilon_B}{8\pi}\sum_q\bigg[\frac{d}{4}k_{xy}^2|\phi_+(q)|^2+\nn\\
&+\left(\frac{d}{d_1d_2}+\frac{d}{4}k_{xy}^2\right)|\phi_-(q)|^2\bigg].
\end{align}
From Eq.\ \pref{emphi} one immediately understands that 
the $\phi_-$ combination describes an intra-cell potential gradient that corresponds to short-range Coulomb interactions. As such its fluctuations must be included, in full analogy with the usual procedure in the case of ab-initio DFT calculations\cite{pick_prb70}.  
 One can better understand the picture behind this procedure by looking at a single-layer superconductor with broken translation symmetry because of different Josephson couplings between the planes, $\mathcal J_1\neq \mathcal J_2$. In this system, the phase $\theta_-$ can be interpreted as a fold of $\theta_+$ at the Brillouin zone boundary, i.e. $\theta_-(k_z=0)$ plays a role analogous to $\theta_+(k_z=\pi/(d/2))$, that corresponds indeed to oscillations with opposite phases in neighbouring planes, similarly to what we discussed in the previous sections for the gauge-invariant variables. The bilayer structure has then the effect to couple phase fluctuations at the zone boundary to the $k_z=0$ response.\\
 Thus the relevant action for computing the average value of the phase modes is built by adding to Eq. \eqref{sthetapm} the free e.m. action for $\phi_-(q)$ and the action resulting from the minimal coupling substitution: 
\begin{align}\lb{thetaphi}
S[\theta_{\pm}&,\phi_-,\text A^{\text{ext}}_z]=S[\theta_{\pm},\text A_z^{\text{ext}}]+\nn\\
&-\frac{\varepsilon_B}{8\pi}\sum_q\bigg[\bigg(\frac{d}{d_1d_2}+\frac{\pi e^2d}{\varepsilon_B}\kappa_0+\frac{d}{4}k_{xy}^2\bigg)|\phi_-(q)|^2\bigg]+\nn\\
&+\frac{e}{4}\sum_q\bigg[\frac{d}{4}\kappa_0\Omega_m\big(\theta_-(q)\phi_-(-q)-h.c.\big)\bigg],
\end{align}
%
while $\phi_+$ fluctuations should not be included as they describe the long-range Coulomb interaction. Notice that the scalar potential $\phi_-$ only couples to the $\theta_-$ combination, so that only this degree of freedom is dressed by the finite-range Coulomb interaction. It may be argued that the procedure here employed does not consider an internal vector potential, which is instead crucial in order to correctly characterise the plasma modes as discussed in the previous section. Nevertheless, one can check that all the short-range interactions described by couplings between $\theta_-$ and the components of the vector potential can be set to zero by a convenient gauge choice.\\
By integrating out the short-range Coulomb interactions $\phi_-$ and then the internal degree of freedom $\theta_-$ one finds that the second term of Eq.\ \pref{defsigma} gets corrected, and the linear response kernel reads
\begin{align}\lb{kernzz}
\text{K}_{zz}(i\Omega_m)&=\frac{\varepsilon_B}{4\pi d}\bigg[\big(\omega_{z1}^2d_1+\omega_{z2}^2d_2\big)+\nn\\
&-\left(1+\frac{4\alpha}{d_1d_2}\right)\frac{d_1d_2}{d}\frac{(\omega_{z1}^2-\omega_{z2}^2)^2}{\big(\tilde\omega_T^2-(i\Omega_m)^2)}\bigg],
\end{align}
where $\tilde\omega_T^2$ generalizes the transverse plasma frequency in Eq. \eqref{omegat} to a finite compressibility, 
\begin{equation}\lb{omegattilde}
\tilde\omega_T^2=\omega_T^2\left(1+\frac{4\alpha}{d_1d_2}\right).
\end{equation}
\begin{figure}[t!]
    \centering
    \includegraphics[width=0.5\textwidth,keepaspectratio]{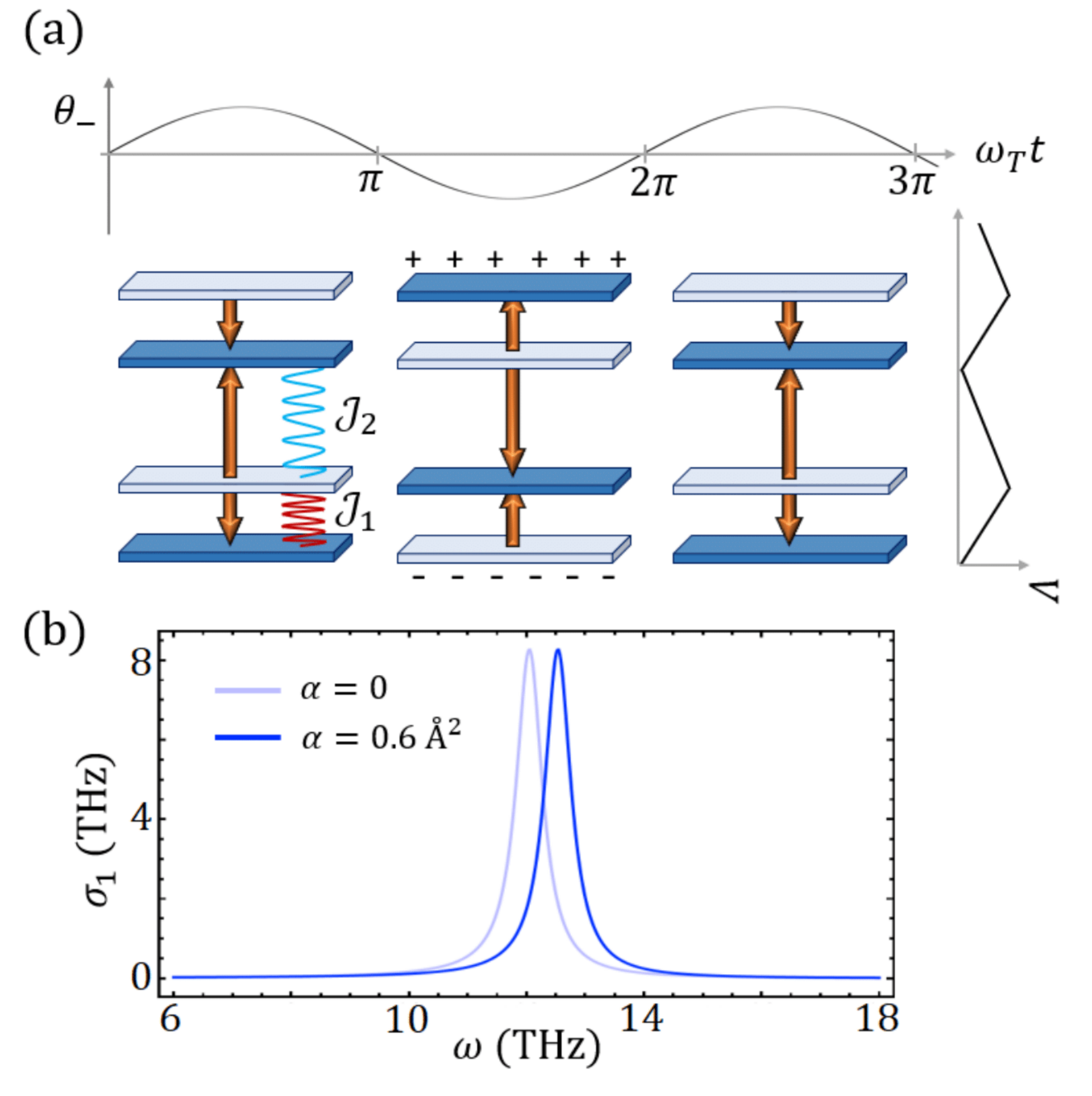}
    \caption{\textbf{(a)} Sketch of the mode $\theta_-$ at frequency $\omega_T$ excited by the external field, as a function of the time. The arrows represent the currents flowing from a layer to another: their widths represent the magnitude of the currents while their lengths do not have a physical meaning and are chosen only for graphic purposes. The layers with net positive charge are shown in blue, while the layers with net negative charge are shown in gray. On the right side we plot the voltage distribution as a function of $z$. \textbf{(b)} Real part of the conductivity shown as a function of the frequency without (gray) and with (blue) compressibility corrections ($\alpha=0.6\angstrom^2$) as written in Eq. \eqref{cond1}. In the plots we chose $\omega_{z1}=14.2\text{THz}$, $\omega_{z2}=0.9\text{THz}$, $d_1=3.2\angstrom$ and $d_2=8.2\angstrom$, so that $\omega_T=12.1\text{THz}$ and $\tilde\omega_T=13.2\text{THz}$. A finite damping parameter $\Gamma=0.3\omega_{z2}$ is also introduced.}
    \lb{recond}
\end{figure}\\
In typical bilayer cuprates as e.g. YBCO one finds that $\left|\tilde\omega_T-\omega_T\right|\sim1\text{THz}$. One can then write the conductivity as in Eq. \eqref{cond}, where now the real part reads
\begin{align}\lb{cond1}
&\sigma_1(\omega)=\frac{\varepsilon_B}{4d}\bigg[\big(\omega_{z1}^2d_1+\omega_{z2}^2d_2\big)\delta(\omega)+\left(1+\frac{4\alpha}{d_1d_2}\right)\times\nn\\
&\times\frac{d_1d_2}{d}\frac{(\omega_{z1}^2-\omega_{z2}^2)^2}{2\tilde\omega_T^2}\big[\delta(\omega-\tilde\omega_T)+\delta(\omega+\tilde\omega_T)-2\delta(\omega)\big]\bigg]
\end{align}
and the imaginary part reads
\begin{align}\lb{cond2}
\sigma_2(\omega)&=\frac{\varepsilon_B}{4\pi d}\frac{1}{\omega}\bigg[\big(\omega_{z1}^2d_1+\omega_{z2}^2d_2\big)+\nn\\
&-\left(1+\frac{4\alpha}{d_1d_2}\right)\frac{d_1d_2}{d}\frac{(\omega_{z1}^2-\omega_{z2}^2)^2}{\tilde\omega_T^2-\omega^2}\bigg].
\end{align}
Notice that in the infinite-compressibility limit $\alpha\to 0$ the finite-frequency correction stays finite and one recovers the result deduced in Ref. \cite{vandermarel96}. The phase oscillations described by $\theta_-$ are sketched in Fig. \ref{recond}(a), while the real part of the conductivity $\sigma_1(\omega)$ is shown in Fig. \ref{recond}(b) with and without the compressibility corrections. A finite damping parameter $\Gamma$ is also introduced in the finite-frequency part of Eq. \eqref{cond1} to have a direct representation of the optical experiments\cite{homes_prl93,bernhard_prl11,erb_prl00,tajima_prl09, yamada_prl98,cavalleri_nmat14, wangNL_prx20}. Indeed, the optical conductivity displays a peak at the transverse plasma frequency $\tilde\omega_T$ defined in Eq.\ \pref{omegattilde}, weighted by the difference between the two squared Josephson plasma frequencies defined in Eq. \eqref{plasmaBL}.\\
The procedure here employed makes clear that the peak appears exactly at the transverse plasma frequency because it is a manifestation at $|\bk|=0$ of the lower Josephson plasma mode $\omega_{J2}$ at the border of the Brillouin zone. Indeed, its weight given by the difference of the plasma frequencies squared is due to the fact that this low-energy mode is linked to opposite-phase out-of-plane currents, as discussed in section \ref{sec2d}.\\
With the imaginary part of the conductivity $\sigma_2(\omega)$ from Eq. \eqref{cond} one can evaluate the total out-of-plane superfluid stiffness of a bilayer superconductor: in the infinite compressibility limit this reads
\begin{align}\lb{stiff}
e^2D_z^{(BL)}=\omega\sigma_2(\omega)\big|_{\omega\to0}=\frac{\varepsilon_B}{4\pi}\frac{\omega_{z1}^2\omega_{z2}^2}{\omega_T^2}.
\end{align}
This can be rewritten as a combination of the intra-bilayer and inter-bilayer superfluid stiffnesses defined as $D_{z\lambda}=\omega_{z\lambda}^2\varepsilon_B/(4\pi e^2)$, 
\begin{align}\lb{stiff2}
D_z^{(BL)}=d\frac{D_{z1}D_{z2}}{D_{z1}d_2+D_{z2}d_1}.
\end{align}
Due to the fact that in typical cuprate superconductors $D_{z1}\gg D_{z2}$, from Eq. \eqref{stiff2} one can see that the total out-of-plane stiffness is dominated by the weaker inter-bilayer stiffness $D_{z2}$. One easily understands this result in the context of a 1D superconducting chain in which the total stiffness is given by $D_{s}=[\sum_i(1/D_i)]^{-1}$, $D_i$ being the stiffness of the $i$-th link, and thus it is always dominated by the weakest link in the chain\cite{seibold_prb15}.\\
From Eq.\ \pref{kernzz} one can also obtain the dielectric function of the system, which takes the form
\begin{align}\lb{eps}
\varepsilon(\omega)=\varepsilon_B\frac{(\omega^2-\tilde\omega_{z1}^2)(\omega^2-\tilde\omega_{z2}^2)}{\omega^2(\omega^2-\tilde\omega_T^2)},
\end{align}
with $\tilde\omega_{z\lambda}$ defined as in Eq.s \eqref{freqAlpha}. This is the result of the MLM quoted in Ref\ \cite{vandermarel_prb01}.\\
As suggested by Eq. \eqref{GIvariso} in the isotropic case, the dielectric function appears also as the coefficient to the transverse component of the gauge-invariant variables. This is also valid in the present case. We start from the action in Eq. \eqref{GIvarAlpha} written in the $\bk\to0$ limit relevant for this section, and perform the following change of variables:
\begin{align}\lb{psi+-}
\psi_{+}&=\frac{1}{d}(\psi_{z2}d_2+\psi_{z1}d_1)\nn\\
\psi_{-}&=\frac{1}{d}(\psi_{z2}d_2-\psi_{z1}d_1).
\end{align}
Notice that by the definition of $\psi_{z\lambda}$ in Eq. \eqref{gaugetfBL}, the combination $\psi_+$ is independent of the SC phase and reads:
\begin{align}
\psi_+=\frac{2e}{c}\frac{1}{d}(A_{z1}d_1+A_{z2}d_2).
\end{align}
As such, this particular combination of $\psi_{z1}$ and $\psi_{z2}$ is the uniform ($k_z=0$) transverse component of the gauge field. One can then rewrite Eq.\ \pref{GIvarAlpha} in terms of the $\psi_{\pm}$ variables, and integrate out the  $\psi_-$ combination, which plays the analogous role of $\theta_-$ dressed by short-range Coulomb interaction in the previous derivation. It is then straightforward to see that one is  left with
\begin{align}
S^{(BL)}_{\alpha\neq0}[\psi_+]=\frac{d}{32\pi e^2}\sum_{i\Omega_m}\bigg[\Omega_m^2\varepsilon(i\Omega_m)|\psi_+(i\Omega_m)|^2\bigg],
\end{align}
where $\varepsilon(i\Omega_m)$ goes into the dielectric function in Eq. \eqref{eps} once the analytical continuation has been performed. Thus, also in the language of the gauge-invariant variables one is able to recover the crucial role in bilayer superconductors of the coupling between the short-range Coulomb interactions and the $k_z=0$ response, encoded in Eq.\ \eqref{GIvarAlpha} by the finite coupling among $\psi_+$ and $\psi_-$.

\section{Conclusions}\label{sec4}
In the present manuscript we provided a detailed analysis of the e.m. modes in a model system for a bilayer superconductor, i.e. a layered superconductor with two layers per unit cell, characterised by different intra-cell and inter-cell Josephson couplings among the SC sheets. Such a model provides an excellent description of the optical response of YBCO cuprates, one of the most studied families of high-$T_c$ superconductors. In particular, while the linear c-axis optical response of YBCO has been experimentally investigated long ago\cite{homes_prl93,bernhard_prl11,erb_prl00,tajima_prl09, yamada_prl98}, its non-linear out-of-plane THz response attracted renewed interest in recent years thanks to the promise to use intense light pulses to control the nonlinear driving of the soft, undamped Josephson plasmon emerging below $T_c$\cite{cavalleri_nmat14,cavalleri_prx22,wangNL_cm22,shimano_prb23}. In this paper we addressed two main issues: (i) the derivation of the energy-momentum dispersion for both polariton and plasmons  at arbitrary wavelength; (ii) the derivation of the linear optical response along the $c$ axis. The issue (i) is motivated by the observation that in a layered system the usual decoupling among longitudinal and transverse e.m. degrees of freedom, that holds in isotropic systems at all length scales, is only quantitatively valid at momenta larger enough than the light cone. To state the problem differently, the density and current fluctuations get intrinsically mixed at low momenta, leading to hybrid light-matter modes that preserve simultaneously both longitudinal and transverse character. By using an effective-action approach where the matter and the internal e.m. degrees of freedom are treated on the same footing, we showed that the dispersions of the generalized plasma modes can be obtained by the zeros of the matrix of the physical gauge-invariant variables, given by the compact and analytical expression in Eq.\ \pref{GIvarBL}. One has three modes in the relevant range of energies, two of them starting from the frequency scales connected to intra-cell and inter-cell Josephson couplings among layers, and a third one starting from the larger in-plane plasma frequency. Even though the existence of multiple modes and their numerical dispersions in selected regimes were already discussed previously in the literature\cite{vandermarel_prb01,demler_prb20,alpeggiani_prb13}, the main advantage of our approach is to reduce the numerical complexity of the derivation to a simple eigenvalue problem of an analytical matrix. This also simplifies considerably the analysis of the polarization dependence of the modes in the various regime for the momenta. We then identified two crossover scales: above the lower one $|\bk_{c1}|\sim \sqrt{\omega_{z1}^2-\omega_{z2}^2}/c$ one finds the mixing among the two lower Josephson plasmons, with one mode evolving analogously to the lower e.m. mode of the single-layer case, and the latter evolving towards a low-energy mode around the frequency $\omega_T=\sqrt{(\omega_{z1}^2d_2+\omega_{z2}^2d_1)/d}$, \textcolor{black}{which represents opposite-phase plasma oscillations between the planes, polarized along the c-axis} for a wide range of momenta up to $|\bk|\sim1/d$. Above a second crossover scale $|\bk_{c2}|\sim \sqrt{\omega_{xy}^2-(\omega_{z1}^2d_1+\omega_{z2}^2d_2)/d}/c$ the two upper modes have the analogous evolution of the two modes of  the single-layer case\cite{gabriele_prr22}, and progressively approach the pure transverse/longitudinal modes usually predicted within a standard-RPA approach.\\
To make closer connection with previous work focusing on the $c$-axis linear response, we also computed the linear optical conductivity. We showed that the theoretically predicted\cite{vandermarel_prb01,vandermarel96} and experimentally observed \cite{homes_prl93,bernhard_prl11,erb_prl00,tajima_prl09, yamada_prl98,cavalleri_nmat14, wangNL_prx20} peak in the optical conductivity at the scale $\omega_T$ can be understood as an effect of an unusual finite-frequency correction to the optical response due to out-of-phase fluctuations of the SC phase in neighbouring layers within the same unit cell. This interpretation explains also the rather unexpected observation of a plasmon-like peak in the optical conductivity. Indeed, the general expectation is that since the conductivity is the current response to the local electric field, screening effects due to Coulomb interactions should not be included. In other words, in the usual diagrammatic language the optical conductivity is obtained as a current response function irreducible with respect  to the Coulomb interaction\cite{pines,pick_prb70,belitz_prb89}. As such, the conductivity should not carry signatures of the plasma modes, that appear instead in the dielectric function describing the screening. However, in the bilayer case the beating mode connected to relative phase fluctuations among neighbouring planes within the same unit cell intrinsically couples to Coulomb interactions at {\em large} momenta, i.e. at short length scales, and as such must be included in the physical response, as it is usually done in DFT calculations in lattice systems\cite{pick_prb70}. Following this procedure we then reproduced the observed experimental peak at $\omega_T$ and we also computed its corrections for finite compressibility, that can be relevant to locate it for different doping levels. \\
Besides such a direct application to the computation of the linear response, the results of the present manuscript, including the methodological ones, provide a framework to address several open issues still under discussion for what concerns recent experiments using strong THz fields in YBCO\cite{cavalleri_nmat14,cavalleri_prx22,wangNL_cm22,shimano_prb23}. Indeed, a precise characterization of the modes and their polarizations is the crucial prerequisite in order to understand the possible mechanisms responsible for their contribution to the non-linear optical response. So far, both a coupling to an infrared phonon  mode\cite{cavalleri_prx22} and a direct non-linear coupling of plasmons to light\cite{gabriele_natcomm21} have been proposed as possible pathways for non-linear driving of plasma waves in YBCO.  How these proposals can be justified at a full microscopic level is still an open question, that certainly deserves future investigation.

\vspace{1cm} {\bf Acknowledgments}
We acknowledge financial support by EU under project MORE-TEM ERC-SYN (grant agreement No 951215), and by Sapienza University  under project Ateneo 2021 (RM12117A4A7FD11B) and Ateneo 2022 (RP1221816662A977).

\appendix

\section{Phase-only effective action in the path-integral formalism}\label{appA}
Let us start from a grand-canonical hamiltonian for a generic single-band superconductor:
\begin{align}\lb{ham}
\hat{H}-\mu\hat{N}=\sum_{\bk,\s} \xi_{\bk}\hat{c}^\dagger_{\bk\s}\hat{c}_{\bk\s}+\hat{H}_I,
\end{align}
where $\sigma$ is the spin index, $\xi_{\bk}$ is the band dispersion with respect to the chemical potential $\mu$, $\hat c_{\bk,\sigma}^\dagger$ and $\hat c_{\bk,\sigma}$ are the electron creation and annihilation operators respectively. The interacting hamiltonian $\hat{H}_I$ reads\cite{randeria_prb00}:
\begin{align}\lb{hamint}
\hat{H}_I=-\frac{U}{N}\sum_{\bk'}
\hat{\Phi}_\D^\dagger(\bk')\hat{\Phi}_\D(\bk')
\end{align}
where $\hat{\Phi}_\D(\bk')=\sum_{\bk}\gamma(\bk)\hat{c}_{-\bk-\bk'/2,\down}\hat{c}_{\bk-\bk'/2,\up}$, with $\gamma(\bk)=\cos(k_xa)-\cos(k_ya)$ accounting for the $d$-wave symmetry of the order parameter, $U>0$ is the SC coupling constant and $N$ denotes the number of lattice sites. In order to compute thermal averages over the hamiltonian \pref{ham} we use the path integral formulation. Within such framework the imaginary-time action for fermions\cite{nagaosa} can be written as
\begin{align}\lb{act}
S&[c,\overline{c}]=S_0+S_I=\nn\\
&=\int_0^\b d\t\left[\sum_{\bk\s}\overline{c}_{\bk\s}\left(\pd_\t+\xi_{\bk}\right)c_{\bk\s}+H_I(\bk,\t)\right],
\end{align}
where $\t=i t$ is the imaginary time variable summed from $0$ to $\b=\frac{1}{T}$ and $\overline c$ and $c$ are the Grassmann variables associated to the creation and annihilation operators respectively. To obtain the effective action in terms of the order-parameter collective degrees of freedom, the interacting action is decoupled in the particle-particle channel by means of the Hubbard-Stratonovich (HS) transformation by introducing the auxiliary complex field $\D$:
\begin{align}
\D(\bx,\tau)=(\D_0+\d\D(\bx,\tau))e^{i\th (\bx,\tau)} 
\end{align}
where $\D_0$ is the mean-field expectation value of the amplitude associated to the SC energy gap, $\d\D$ and $\th$ are amplitude and phase fluctuations. By making an appropriate gauge transformation on the Grassmann fields $c$ and $\overline{c}$ it is possible to make the dependence on the phase $\th$ explicit in the action. Then we introduce the Nambu spinors $\Psi_\bk^\dagger=\left(c_{\bk\up}^{\dagger}, c_{\bk\down} \right)$, by which one can define the BCS Green Function as
\begin{align}
\hat{\mathcal{G}}_0(\bk,i\o_\nu)&= -\int_0 ^\b d\t\langle
\mathcal{T}\big(\hat{\Psi}_{\bk}(\t)\hat{\Psi}^\dagger_{\bk}(0)
\big)\rangle e^{i\o_\nu \t}=\nn\\
&=\frac{i\o_\nu\hat{\t}_0+\xi_\bk\hat{\t}_3-\D_0\gamma(\bk)\hat{\t}_1}{(i\o_\nu)^2-E_\bk^2}.
\end{align}
Here $\omega_\nu=(2\nu+1)\pi T$ are the Matsubara fermionic frequencies, $E_\bk =\sqrt{\xi_\bk^2+(\D_0\gamma(\bk))^2}$ the quasiparticles energy and $\hat{\t}_i$ the Pauli matrices.\\
With these transformations on Eq. \eqref{act}, one finds that the HS transform of $S_I$ is independent of the phase fluctuations, while the free contribution now reads:
\begin{align}
\tilde{S}_0=S_0+\int d\bx d\t \overline{\Psi}(\bx,\t)
\hat{\Sigma}(\bx,\t)\Psi(\bx,\t).
\end{align}
$\hat{\Sigma}$ is the self-energy, which depends, in principle, on both amplitude and phase fluctuations. Nonetheless, as long as one is interested in the low-temperature dynamics of phase fluctuations in layered cuprates, amplitude fluctuations can be neglected\cite{benfatto_prb04}. The self energy then reads:
\begin{align}\lb{self}
\hat{\Sigma}=\left[ \frac{i}{2}\pd_{\t}\th+\frac{1}{8m^*}\left(\bdnb\th\right)^2\right]\hat{\t}_3+\left[ \frac{i}{4m^*} \bdnb\th \cdot \overset\leftrightarrow\nb\right]\hat{\t}_0,
\end{align}
where $\overset\leftrightarrow\nb=\overset\rightarrow\nb-\overset\leftarrow\nabla$, \textcolor{black}{with $\overset\rightarrow{\nabla}$ ($\overset\leftarrow{\nabla}$) the gradient operator acting on the function on its right (left)}. Notice that, according to the Goldstone theorem, the phase $\th$ appears in the self-energy only trough its time and spatial derivatives, i.e. there are no mass terms for $\th$.\\
Since the action is quadratic in the fermionic variables, we can now integrate them out. Ignoring the amplitude mean-field expectation value $\Delta_0$, such procedure leads to the following effective action for the phase fluctuations:
\begin{align}\lb{seffexp}
S_{\text{eff}}[\th]=\text{Tr}\sum_{n=1}^{+\infty} \frac{\big(\hat{\mathcal{G}}_0\hat{\Sigma}\big)^n}{n}
\end{align}
where the trace is computed over both spin and momentum degrees of freedom. In order to study the phase dynamics we can compute this effective action at Gaussian level, truncating the sum for $n\leq2$:
\begin{align}\lb{gaussapp}
S_G[\th]=\frac{1}{8}\sum_q
\bigg[&-\O_m^2\chi^{\rho\rho}(q)+\bk^\a \bk^\b\chi^{jj}_{\a\b}(q)+\nn\\
&-2i\O_m\bk^\a\chi^{\rho j}_{\a}(q)
\bigg]|\th(q)|^2 \nn\\
\end{align}
where $q=(i\O_m,\bk)$ is the imaginary-time 4-momentum with $\Omega_m=2\pi m T$ the bosonic Matsubara frequencies, and 
\begin{align}
\chi^{\rho\rho}&(q)=
\frac{T}{N}\sum_{q'}\text{Tr}\Big[
\hat{\mathcal{G}}_0(q'+q)\hat{\t}_3
\hat{\mathcal{G}}_0(q')\hat{\t}_3\Big]\nn\\
\chi^{\rho j}_\a&(q)=
\frac{T}{N}\sum_{q'}\frac{\bk'_\a+\frac{\bk_\a}{2}}{m^*}
\text{Tr}\Big[\hat{\mathcal{G}}_0(q'+q)\hat{\t}_0
\hat{\mathcal{G}}_0(q')\hat{\t}_3\Big]\nn\\
\chi^{jj}_{\a\b}&(q)=\frac{n}{m^*}\d_{\a\b}+\nn\\
&+\frac{T}{N}\sum_{q'}
\frac{\bk'_\a+\frac{\bk_\a}{2}}{m^*}
\frac{\bk'_\b+\frac{\bk_\b}{2}}{m^*}
 \text{Tr}\Big[\hat{\mathcal{G}}_0(q'+q)\hat{\t}_0
\hat{\mathcal{G}}_0(q')\hat{\t}_0\Big]
\end{align}
are the BCS response functions, which contain all the information on the microscopic fermionic degrees of freedom. Again, if one is interested in the low-temperature phase-dynamics, one can evaluate the BCS bubbles in the static limit $i\O_m=0$, $\bk\to0$: within such approximation Eq. \eqref{gaussapp} goes to the superfluid action in Eq. \eqref{supfluid} of the main text. Notice the symmetry of the SC order parameter only enters the problem via the $\gamma(\bk)$ factor which modulates the SC gap $\Delta_0\gamma(\bk)$. As a consequence, the structure \pref{gaussapp} is general\cite{randeria_prb00,benfatto_prb01}, and the main dependence on the symmetry of the SC gap appears in the temperature dependence of the current-current correlation function, leading to a temperature dependence of the BCS superfluid stiffness $D_s$ that is linear at low $T$ in the $d$-wave case, in contrast to the exponential suppression for the fully gapped  $s$-wave case.
\\

\section{Fields discretization in bilayer crystals}\label{appB}
In this appendix we derive the action associated with the free contribution of the electromagnetic (e.m.) fields and fix the discretization required for the SC phase action. 
To achieve this, a possible procedure would be to work with continuous Maxwell's equation and discrete density and current defined on an anisotropic lattice, which can then be reduced to a layered structure by taking the continuum limit in the $x$ and $y$ directions. This would require to express the free e.m. fields in terms of a sum on the momenta appropriate to the lattice so to have the correct periodicity and momentum conservation. For instance, the bare Coulomb interaction for a single-layer system would be described by the well-known propagator\cite{fetter} 
\begin{align}\lb{coulfet}
V_C^{SL}=\frac{2\pi e^2 d}{k_x}\frac{\sinh(k_x d)}{ \cosh(k_x d) - \cos( k_z d )}.
\end{align}
For a bilayer crystal this procedure would be quite heavy. In this work we adopt an alternative procedure, in which one goes back from the discretized Maxwell's equations to the action that generates them via the variational principle. For instance, within this approach the single-layer Coulomb interaction is given by 
\begin{align}\lb{coulsl}
V_C^{SL}=\frac{4\pi e^2}{ k_x^2 + 4/d^2 \sin^2(k_z d/2) }.
\end{align}
Comparing this with Eq. \eqref{coulfet}, one understands that the two approaches are equivalent at leading order in $\bk d$ and even if the latter is not as common as the standard anisotropic discretization, it can be generalized to include the needed fields on the bilayer lattice.\\
Let us consider a rectangular lattice structure. With no loss of generality, $x$ denotes the in-plane direction with lattice constant $a$, while $z$ is the out-of-plane coordinate with intra and inter-bilayer spacings $d_1$ and $d_2$ respectively. For the sake of simplicity, here we do not consider explicitly the $y$ dimension of the lattice, although the $y$ direction must be considered in order to correctly define the field components. Such discretization defines in a single unit cell two distinct rectangular regions of area $a\cdot d_\lambda$ called plaquettes. The lattice and the quantities defined on it are shown in Fig. \ref{grid}. To recover the results of the main text one should take the limit $a\to 0$ at the end.\\
As a first step, we define the scalar potential $\phi$ and the components of the vector potential $\text A_x$ and $\text A_z$ on such a lattice properly. A consistent choice is to define $\phi$ on the lattice sites and $\text A_{i}$ on the links between two neighbouring sites along the $i$ direction. As a consequence, the magnetic field is along the $y$ direction and lies at the center of the $\lambda$-th plaquette,
\begin{align}\lb{Bpot}
B_{y\l,n}=
\D_{z\l} \text A_{x\l,n}-
\D_x \text A_{z\l,n}.
\end{align}
where the out-of-plane discrete derivative in this case is defined as in Eq. \eqref{deltaz} in the main text and the in-plane discrete derivative $\D_x$ acts on a generic function $f_{\l,n}$ according to:
\begin{align}\lb{xder}
\D_x f_{\l,n}(x)=
\frac{f_{\l,n}(x+a)-
f_{\l,n}(x)}{a}.
\end{align}
In the continuum limit $a\ra 0$ relevant for the main text, $\D_x$ simply reduces to the in-plane partial-derivative operator $\pd_x$. On the other hand, the two components of electric field in imaginary-time formalism are
%
\begin{align}\lb{Epotx}
E_{x\l,n}=-{\D}_{x}\phi_{\l,n}
-\frac{i}{c}\frac{\pd \text A_{x\l,n}}{\pd \tau},
\end{align}
defined on the link along the $x$ direction between the sites $x$ and $x+a$, and
\begin{align}\lb{Epotz}
E_{z\l,n}=-{\D}_{z,\l}\phi_{\l,n}
-\frac{i}{c}\frac{\pd \text A_{z\l,n}}{\pd \tau},
\end{align}
defined on the link along the $z$ direction between two subsequent layers.
\begin{figure}[t!]
    \centering
    \includegraphics[width=0.5\textwidth,keepaspectratio]{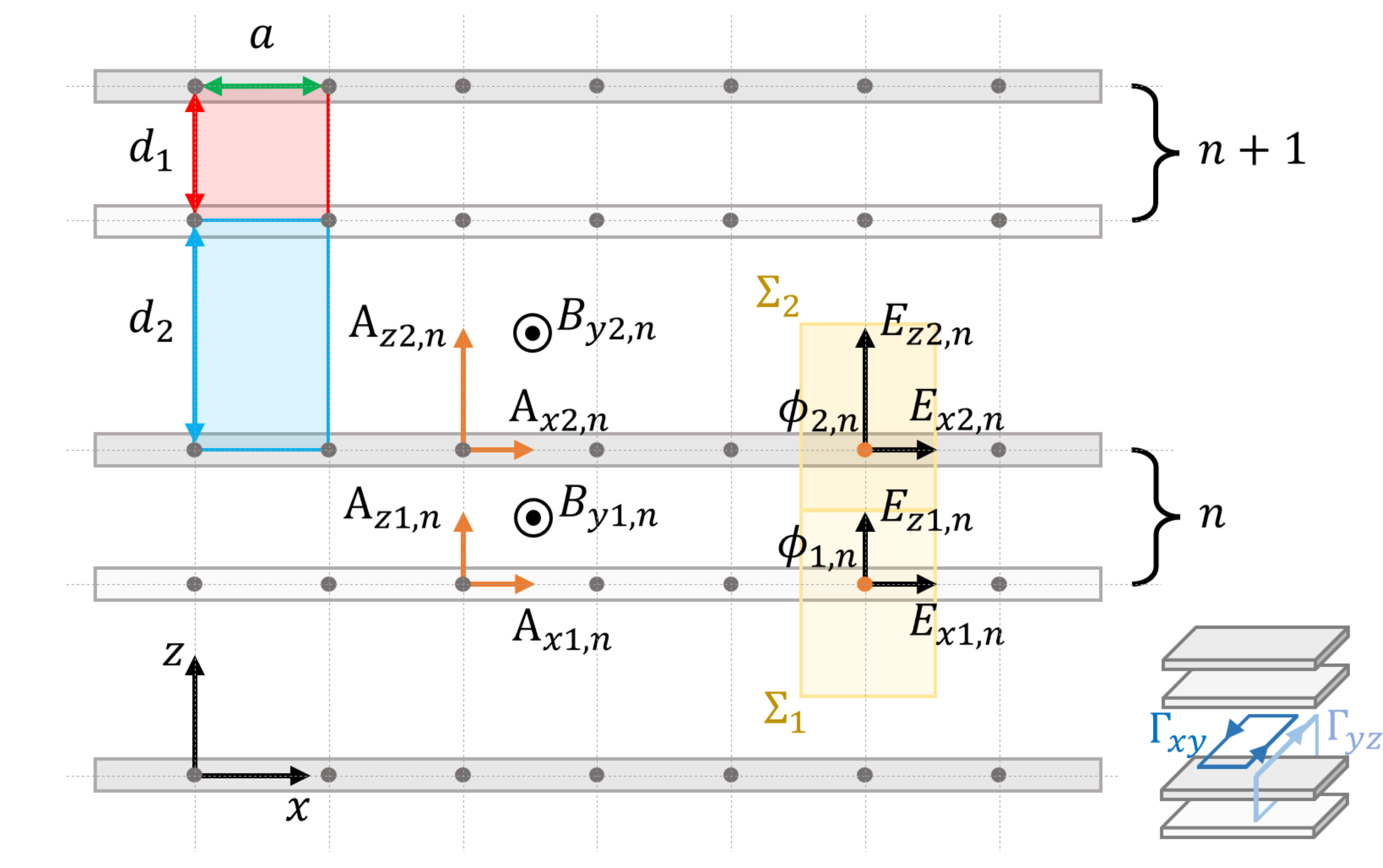}
    \caption{Sketch of the bilayer lattice structure. The spacings between the sites (gray dots) define two "plaquettes" of area $a\cdot d_1$ (red) or $a\cdot d_2$ (blue). The e.m. scalar potential is defined on the sites, as shown by the orange dots in a sample unit cell, while the components of the vector potential are defined on the links between two sites, as shown by orange arrows. This choice fixes the positions of the magnetic and electric fields according to the definitions \eqref{Bpot}, \eqref{Epotx} and \eqref{Epotz}. Here we also depict the two Gaussian surfaces along the $xz$-plane, $\Sigma_1$ and $\Sigma_2$ (perimeter of the yellow areas) in a sample unit cell and the two Amperian loops relevant for the $y$ component of the magnetic field, $\Gamma_{xy}$ and $\Gamma_{yz}$ (blue rectangles) on the 3D model on the bottom-right corner.}
    \lb{grid}
\end{figure}\\
To compute the electrostatic action, one needs to discretise Gauss's law for a bilayer lattice. In the most general case, its integral formulation reads
%
\begin{align}\lb{gaulaw}
\oint\limits_{\S(\O)} \bE(\bx)\cdot\hat{\mathbf{n}}(\bx)
d S=\frac{4\pi}{\varepsilon_B}Q_{enc}
\end{align}
In Eq. \pref{gaulaw}, $Q_{enc}=\int\limits_{\O}d \bx \rho(\bx)$ is the total electric charge contained into a generic volume $\O$, $\S(\O)$ is the closed Gaussian surface enclosing both $\O$ and the charge, and $\hat{\mathbf{n}}(\bx)$ is the versor normal to the surface at the point $\bx$. In order to apply Eq. \pref{gaulaw} to our system, we have to choose two Gaussian surfaces properly, for label-1 and for label-2 sites. 
Within our 2D lattice, the volume $\Omega$ is reduced to a surface and a generic Gaussian surface is equivalent to a closed path in the $xz$ plane. We thus choose rectangular paths, such that each side of the rectangle crosses perpendicularly one component of the electric field. A good choice consists in two rectangular Gaussian surfaces, both having in-plane and out-of-plane dimensions $a$ and $d/2$ respectively, aligned so that each of them encloses a single lattice site. 
The total fluxes over these two surfaces $\S_1$ and $\S_2$ are then given by:
\begin{align}\lb{S1}
&\oint \limits_{\S_1} \bE(\bx)\cdot\hat{\mathbf{n}}(\bx)
d S=\nn\\
&=\frac{d}{2}\left[
E_{x1,n}\left(x+a\right)-
E_{x1,n}\left(x\right)
\right]
+a\left[
E_{z2,n}-
E_{z1,n}
\right]
\end{align}
\begin{align}\lb{S2}
&\oint \limits_{\S_2} \bE(\bx)\cdot\hat{\mathbf{n}}(\bx)
d S=\nn\\
&=\frac{d}{2}\left[
E_{x2,n}\left(x+a\right)-
E_{x2,n}\left(x\right)
\right]
+
a\left[
E_{z1,n+1}-
E_{z2,n}
\right]
\end{align}
By using Eq.s \eqref{S1} and \eqref{S2} in Eq. \eqref{gaulaw} with $Q_{enc}^{(1)}=q_{1,n}$ and $Q_{enc}^{(2)}=q_{2,n}$, we obtain the equation of motion for the electric field:
\begin{align}\lb{E}
\frac{d}{2}\left(\D_{x}E_{x\l,n}+
\D_{z\l}E_{z\l,n}\right)
=\frac{4\pi}{\varepsilon_B}\frac{q_{\l,n}}{a},
\end{align}
where the derivative along $z$ acts now on a quantity defined on the link between two out-of-plane sites as:
\begin{align}\lb{deltaz2}
\D_{z\l}{}f_{\l,n}=
\begin{cases}
\frac{f_{2,n}-
f_{1,n}}{d/2},\text{  }\l=1
\\
\frac{f_{1,n+1}-
f_{2,n}}{d/2},\text{  }\l=2
\end{cases}
\end{align}
Eq. \pref{E} can also be seen as the equation of motion given by the imaginary-time action
\begin{align}\lb{SE}
S_E^{(BL)}&=\sum_\l^{1,2}
\frac{d}{2}
\sum_n \int d^2\bx d\tau\text{  }
\rho_{\l,n}\phi_{\l,n}+\nn\\
&-\frac{\varepsilon_B}{8\pi}
 \sum_\l^{1,2}\sum_n \int d^2\bx d\tau
\left[
\frac{d}{2}E_{x\l,n}^2+
d_\l E_{z\l,n}^2
\right],
\end{align}
where the limit $a\to0$ relevant for the main text is taken, and $\rho_{\l,n}=\lim_{a\to 0}q_{\l,n}/a$ is the 2D charge density. \\
The magnetic contribution, in full analogy with the derivation for the electrostatic term above, requires the discretization of Amp\`ere's law, whose integral form reads
\begin{align}\lb{amplaw}
\oint \limits_{\G(\S)} \bB(\bx)\cdot d\mathbf{l}=\frac{4\pi}{c}\text I_{enc}
\end{align}
where $\text I_{enc}=\iint\limits_{\S}\bJ(\bx)\cdot\hat{\mathbf{n}}(\bx) dS$ denotes the electric current flowing through a generic surface $\S$ bounded by the Amperian loop $\G(\S)$ and $d\mathbf{l}$ is the infinitesimal length element parallel to the curve at the point $\bx$. Taking vanishing $y$ dimension, one needs to define the two components of the linear current density as $I_{x\l,n}$ and $I_{z\l,n}$ on the links. The former is enclosed in a rectangular Amperian loop $\Gamma_{yz}$ along the $yz$ plane with vanishing $y$ dimension so that the equation of motion for the magnetic field along this path reads
\begin{align}\lb{B1}
\frac{d}{2} \D_{z\l} B_{y\l,n}=-\frac{4\pi}{c} I_{x\l,n},
\end{align}
where $\Delta_{z\lambda}$ acts on $B_{y\l,n}$ according to Eq. \eqref{deltaz2}; the latter is enclosed in a rectangular Amperian loop $\Gamma_{xy}$ along the $xy$ plane, and the equation of motion for the magnetic field in the limit of vanishing $y$ dimension reads in this case
\begin{align}\lb{B2}
a \D_x B_{y\l,n}=\frac{4\pi}{c} I_{z\l,n}.
\end{align}
%
%
Eq.s \eqref{B1} and \eqref{B2} follow from the variational principle associated with the action
\begin{align}\lb{SB}
S_B^{(BL)}=\sum_\l^{1,2}\bigg[
\frac{d}{2}
&\sum_n \int d^2\bx d\tau\text{  }
\text J_{x\l,n}\frac{\text{A}_{x\l,n}}{c}+\nn\\
+d_\lambda
&\sum_n \int d^2\bx d\tau\text{  }
\text J_{z\l,n}\frac{\text{A}_{z\l,n}}{c}+\nn\\
+\frac{d_\l}{8\pi}  
&\sum_{n}  
\int d^2\bx d\tau
B_{y\l,n}^2\bigg].
\end{align}
where $\text J_{i\l,n}=\lim_{a\to0}I_{i\l,n}/a$ and we send $a\to0$.\\
The light-matter interaction and the free e.m. dynamics are thus described by the action obtained by the sum of Eq.s \eqref{SE} and \eqref{SB}, which reads\cite{homann_prl20,homann_22,homann_prb21}
\begin{widetext}
\begin{align}\lb{Sblapp}
&S^{(BL)}[\phi,\textbf A, \rho, \textbf J]=S_E^{(BL)}+S_B^{(BL)}=\sum_{\lambda}^{1,2}\sum_{n}\int d^2\bx d\tau\text{  }
\left[\frac{d}{2}\rho_{\l,n}\phi_{\l,n}+\frac{d}{2}\text J_{x\l,n}\frac{\text{A}_{x\l,n}}{c}+d_\lambda
\text J_{z\l,n}\frac{\text{A}_{z\l,n}}{c}\right]+\nn\\
&+\frac{\varepsilon_B}{8\pi}
\sum_\l^{1,2}
\sum_n \int d^2\bx d\tau 
\left[\frac{d_\l}{\varepsilon_B}\left(\D_{z\lambda}\text A_{x\l,n}-
\pd_x \text A_{z\l,n}\right)^2
-\frac{d}{2}
\left(\pd_x\phi_{\l,n}+
\frac{i}{c}\pd_\tau \text A_{x\l,n}\right)^2
-d_\l
\left(\D_{z\lambda}\phi_{\l,n}+
\frac{i}{c}\pd_\tau \text A_{z\l,n}\right)^2
\right]
\end{align}
\end{widetext}
where we made explicit the magnetic and electric fields according to Eq.s \eqref{Bpot}, \eqref{Epotx} and \eqref{Epotz}. In the language of the SC phase field $\theta$, defined for consistency on the sites, one can identify the charge density and the current density as
\begin{align}\lb{phasecharge}
&\rho_{\lambda,n}=\frac{e}{4}\kappa_0(i\partial_\tau\theta_{\lambda,n}-2e\phi_{\lambda,n})\nn\\
&\text J_x = \frac{e}{4}D_{xy}\left(\nabla_{x}\theta_{\lambda,n}+\frac{2e}{c}\text A_{x\lambda,n}\right)\nn\\
&\text J_z = eJ_\lambda d_{\lambda}^2\left(\Delta_{z\lambda}\theta_{\lambda,n}+\frac{2e}{c}\text A_{z\lambda,n}\right) .
\end{align}
While the calculations above were performed with a magnetic field along $y$, Eq. \eqref{Sblapp} can be generalized for a magnetic field along a generic direction. As such, the first row of Eq. \eqref{Sblapp} sets the discretization constants for the SC phase action as written in Eq. \eqref{supfluidBL} in the main text, while the second row turns into the free e.m. contribution as given by Eq. \eqref{emactionBL}.\\
We conclude this appendix by establishing the rules for the Fourier transform along the $z$ axis used in the main text. First, we fix the origin of the frame of reference on the link between two subsequent layers of an arbitrary unit cell. On the $n$-th unit cell, a generic field $f_{\lambda,n}^{s,l}$ defined either on-site ($s$) or on the link ($l$), transforms as
\begin{align}\lb{fourBL}
f_{\lambda,n}^{s,l}=\sum_{k_z}e^{ik_z(nd+h_\lambda^{s,l})}f_{\lambda}(k_z),
\end{align}
where 
\begin{align}\lb{hs}
h_\lambda^s=\begin{cases}
-d_1/2,\text{   }&\lambda=1\\
d_1/2,\text{   }&\lambda=2
\end{cases},
\end{align}
and
\begin{align}\lb{hl}
h_\lambda^l=\begin{cases}
0,\text{   }&\lambda=1\\
d/2,\text{   }&\lambda=2
\end{cases}.
\end{align}
\section{Dispersions of the plasma modes in the nonrelativistic regime}\label{appC}
\begin{figure}[t!]
    \centering    \includegraphics[width=0.5\textwidth,keepaspectratio]{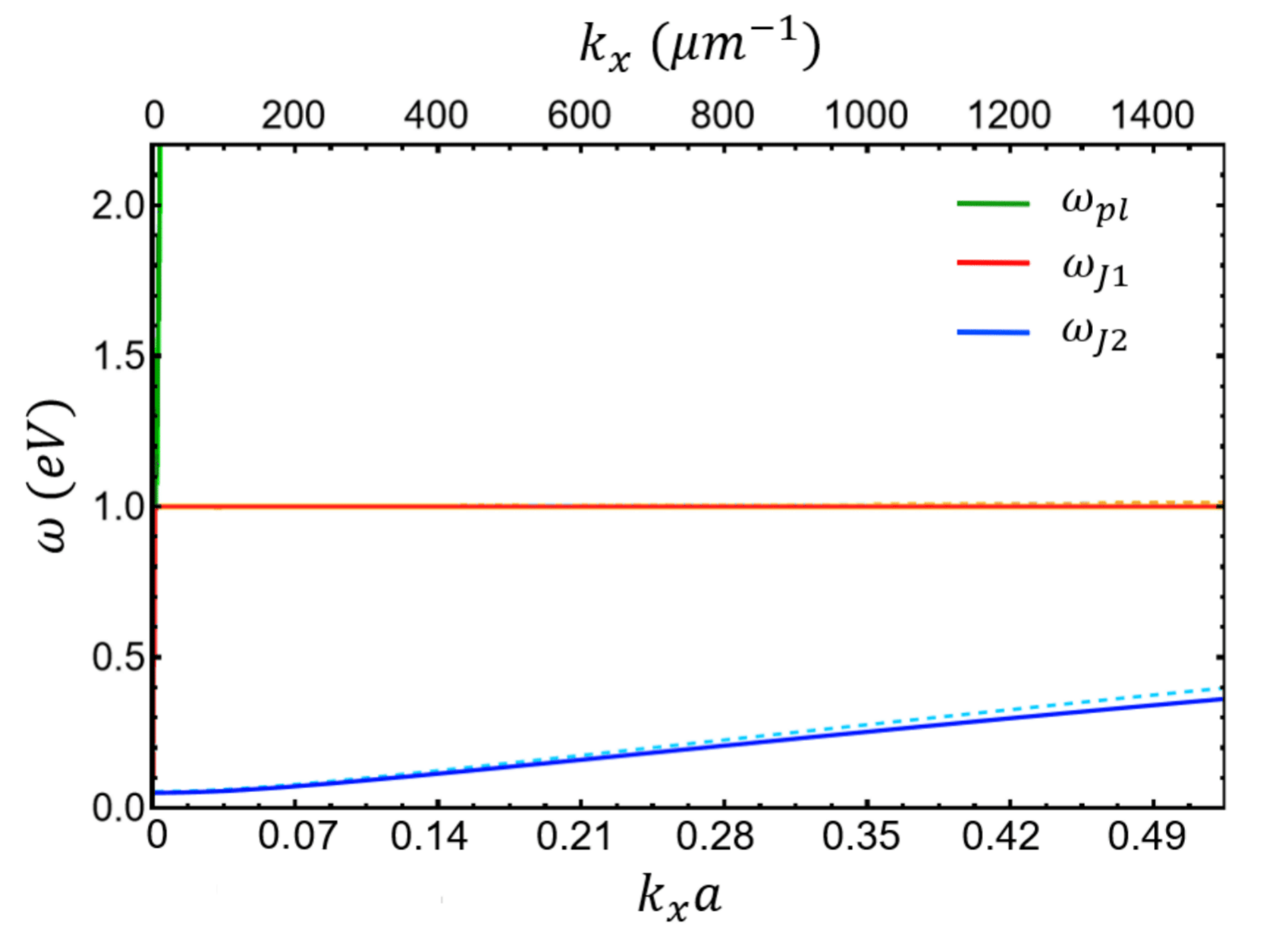}
    \caption{Energy-momentum dispersions as functions of $k_x$ for $k_z=0$ ($\eta=\pi/2$) of the three mixed modes $\omega_{pl}$ (green), $\omega_{J1}$ (red) and $\omega_{J2}$ (blue) in the nonrelativistic regime. Dashed lines show the compressibility corrections given by a finite $\alpha=0.6\angstrom^2$. The in-plane lattice constant in this plot is $a=3.5\angstrom$.}
    \lb{nonrel}
\end{figure}
The formalism employed in this paper allows one to study the plasma modes of a bilayer superconductor in every region of the energy-momentum space. A more in-depth study of the $\omega_{J2}$ low-energy Josephson plasmon reveals a nontrivial behaviour in the nonrelativistic regime. As discussed in Sec. \ref{sec2c} and Sec. \ref{sec2d} in the main text, for $|\bk|\gg|\bk_{c1}|$ this mode saturates towards a constant value of the frequency while sustaining opposite-phase out-of-plane oscillations regardless of the direction of propagation. However, for momenta approaching $1/d$ the dispersion actually starts growing linearly, as shown with solid lines in Fig. \ref{nonrel}. This effect could not be captured by previous studies\cite{bulaevskii_prb94,vandermarel_prb01,koyama_prb96,koyama02,cavalleri_prl16,cavalleri_prb17,alpeggiani_prb13} focused on the optical properties of the Josephson plasmons that used the fact that $\omega_{xy}\gg\omega_{z1,z2}$ to ignore the in-plane dynamics. Moreover, this dispersive behaviour cannot be associated with the Bogoliubov sound dispersion\cite{demler_prb20}, as it appears even for $\alpha=0$ and the dispersion velocity appears to be two orders of magnitude greater than the typical sound velocities in metals. Instead, since this effect is only visible as $k_x\neq 0$ and the velocity increases for $\eta$ approaching $\pi/2$, we associate this behaviour to the non-negligible in-plane dynamics of the charges in the nonrelativistic regime. This is supported by the fact that the polarization of the Josephson mode acquires slowly-increasing opposite-phase in-plane components as its dispersion grows.
\\
A finite compressibility ($\alpha\neq 0$), apart from the small corrections to the plasma frequencies at $|\bk|=0$ discussed in section \ref{sec2e}, affects significantly the dispersions of the modes only in the nonrelativistic regime, as shown in Fig. \ref{nonrel} with dashed lines. In particular, a finite $\alpha$ causes a weak linear behaviour in the $\omega_{J1}$ Josephson mode, with sound velocity $v_s\propto\sqrt{\alpha}$. The same compressibility effect takes place similarly in the $\omega_{J2}$ mode, adding up to its stronger linear behaviour in the nonrelativistic regime discussed above. The effects of a finite compressibility are instead not appreciable on the $\omega_{pl}$ mode, as the sound velocity is much smaller than the light velocity with which the mode is growing, $v_s\ll c$.

\bibliography{bibl.bib} 

\begin{thebibliography}{74}%
\makeatletter
\providecommand \@ifxundefined [1]{%
 \@ifx{#1\undefined}
}%
\providecommand \@ifnum [1]{%
 \ifnum #1\expandafter \@firstoftwo
 \else \expandafter \@secondoftwo
 \fi
}%
\providecommand \@ifx [1]{%
 \ifx #1\expandafter \@firstoftwo
 \else \expandafter \@secondoftwo
 \fi
}%
\providecommand \natexlab [1]{#1}%
\providecommand \enquote  [1]{``#1''}%
\providecommand \bibnamefont  [1]{#1}%
\providecommand \bibfnamefont [1]{#1}%
\providecommand \citenamefont [1]{#1}%
\providecommand \href@noop [0]{\@secondoftwo}%
\providecommand \href [0]{\begingroup \@sanitize@url \@href}%
\providecommand \@href[1]{\@@startlink{#1}\@@href}%
\providecommand \@@href[1]{\endgroup#1\@@endlink}%
\providecommand \@sanitize@url [0]{\catcode `\\12\catcode `\$12\catcode
  `\&12\catcode `\#12\catcode `\^12\catcode `\_12\catcode `\%12\relax}%
\providecommand \@@startlink[1]{}%
\providecommand \@@endlink[0]{}%
\providecommand \url  [0]{\begingroup\@sanitize@url \@url }%
\providecommand \@url [1]{\endgroup\@href {#1}{\urlprefix }}%
\providecommand \urlprefix  [0]{URL }%
\providecommand \Eprint [0]{\href }%
\providecommand \doibase [0]{http://dx.doi.org/}%
\providecommand \selectlanguage [0]{\@gobble}%
\providecommand \bibinfo  [0]{\@secondoftwo}%
\providecommand \bibfield  [0]{\@secondoftwo}%
\providecommand \translation [1]{[#1]}%
\providecommand \BibitemOpen [0]{}%
\providecommand \bibitemStop [0]{}%
\providecommand \bibitemNoStop [0]{.\EOS\space}%
\providecommand \EOS [0]{\spacefactor3000\relax}%
\providecommand \BibitemShut  [1]{\csname bibitem#1\endcsname}%
\let\auto@bib@innerbib\@empty
\bibitem [{\citenamefont {Tamasaku}\ \emph {et~al.}(1992)\citenamefont
  {Tamasaku}, \citenamefont {Nakamura},\ and\ \citenamefont
  {Uchida}}]{uchida_prl92}%
  \BibitemOpen
  \bibfield  {author} {\bibinfo {author} {\bibfnamefont {K.}~\bibnamefont
  {Tamasaku}}, \bibinfo {author} {\bibfnamefont {Y.}~\bibnamefont {Nakamura}},
  \ and\ \bibinfo {author} {\bibfnamefont {S.}~\bibnamefont {Uchida}},\ }\href
  {\doibase 10.1103/PhysRevLett.69.1455} {\bibfield  {journal} {\bibinfo
  {journal} {Phys. Rev. Lett.}\ }\textbf {\bibinfo {volume} {69}},\ \bibinfo
  {pages} {1455} (\bibinfo {year} {1992})}\BibitemShut {NoStop}%
\bibitem [{\citenamefont {Homes}\ \emph {et~al.}(1993)\citenamefont {Homes},
  \citenamefont {Timusk}, \citenamefont {Liang}, \citenamefont {Bonn},\ and\
  \citenamefont {Hardy}}]{homes_prl93}%
  \BibitemOpen
  \bibfield  {author} {\bibinfo {author} {\bibfnamefont {C.~C.}\ \bibnamefont
  {Homes}}, \bibinfo {author} {\bibfnamefont {T.}~\bibnamefont {Timusk}},
  \bibinfo {author} {\bibfnamefont {R.}~\bibnamefont {Liang}}, \bibinfo
  {author} {\bibfnamefont {D.~A.}\ \bibnamefont {Bonn}}, \ and\ \bibinfo
  {author} {\bibfnamefont {W.~N.}\ \bibnamefont {Hardy}},\ }\href {\doibase
  10.1103/PhysRevLett.71.1645} {\bibfield  {journal} {\bibinfo  {journal}
  {Phys. Rev. Lett.}\ }\textbf {\bibinfo {volume} {71}},\ \bibinfo {pages}
  {1645} (\bibinfo {year} {1993})}\BibitemShut {NoStop}%
\bibitem [{\citenamefont {Kim}\ \emph {et~al.}(1995)\citenamefont {Kim},
  \citenamefont {Somal}, \citenamefont {Czyzyk}, \citenamefont {{van der
  Marel}}, \citenamefont {Wittlin}, \citenamefont {Gerrits}, \citenamefont
  {Duijn}, \citenamefont {Hien},\ and\ \citenamefont
  {Menovsky}}]{kim_physicac95}%
  \BibitemOpen
  \bibfield  {author} {\bibinfo {author} {\bibfnamefont {J.~H.}\ \bibnamefont
  {Kim}}, \bibinfo {author} {\bibfnamefont {H.}~\bibnamefont {Somal}}, \bibinfo
  {author} {\bibfnamefont {M.}~\bibnamefont {Czyzyk}}, \bibinfo {author}
  {\bibfnamefont {D.}~\bibnamefont {{van der Marel}}}, \bibinfo {author}
  {\bibfnamefont {A.}~\bibnamefont {Wittlin}}, \bibinfo {author} {\bibfnamefont
  {A.}~\bibnamefont {Gerrits}}, \bibinfo {author} {\bibfnamefont
  {V.}~\bibnamefont {Duijn}}, \bibinfo {author} {\bibfnamefont
  {N.}~\bibnamefont {Hien}}, \ and\ \bibinfo {author} {\bibfnamefont
  {A.}~\bibnamefont {Menovsky}},\ }\href {\doibase
  https://doi.org/10.1016/0921-4534(95)00198-0} {\bibfield  {journal} {\bibinfo
   {journal} {Physica C: Superconductivity}\ }\textbf {\bibinfo {volume}
  {247}},\ \bibinfo {pages} {297 } (\bibinfo {year} {1995})}\BibitemShut
  {NoStop}%
\bibitem [{\citenamefont {Basov}\ \emph {et~al.}(1994)\citenamefont {Basov},
  \citenamefont {Timusk}, \citenamefont {Dabrowski},\ and\ \citenamefont
  {Jorgensen}}]{basov_prb94}%
  \BibitemOpen
  \bibfield  {author} {\bibinfo {author} {\bibfnamefont {D.~N.}\ \bibnamefont
  {Basov}}, \bibinfo {author} {\bibfnamefont {T.}~\bibnamefont {Timusk}},
  \bibinfo {author} {\bibfnamefont {B.}~\bibnamefont {Dabrowski}}, \ and\
  \bibinfo {author} {\bibfnamefont {J.~D.}\ \bibnamefont {Jorgensen}},\ }\href
  {\doibase 10.1103/PhysRevB.50.3511} {\bibfield  {journal} {\bibinfo
  {journal} {Phys. Rev. B}\ }\textbf {\bibinfo {volume} {50}},\ \bibinfo
  {pages} {3511} (\bibinfo {year} {1994})}\BibitemShut {NoStop}%
\bibitem [{\citenamefont {van~der Marel}\ and\ \citenamefont
  {Tsvetkov}(1996)}]{vandermarel96}%
  \BibitemOpen
  \bibfield  {author} {\bibinfo {author} {\bibfnamefont {D.}~\bibnamefont
  {van~der Marel}}\ and\ \bibinfo {author} {\bibfnamefont {A.}~\bibnamefont
  {Tsvetkov}},\ }\href {\doibase https://doi.org/10.1007/BF02548125} {\bibfield
   {journal} {\bibinfo  {journal} {Czech. J. of Phys.}\ }\textbf {\bibinfo
  {volume} {46}},\ \bibinfo {pages} {3165} (\bibinfo {year}
  {1996})}\BibitemShut {NoStop}%
\bibitem [{\citenamefont {Dordevic}\ \emph {et~al.}(2003)\citenamefont
  {Dordevic}, \citenamefont {Komiya}, \citenamefont {Ando},\ and\ \citenamefont
  {Basov}}]{basov_prl03}%
  \BibitemOpen
  \bibfield  {author} {\bibinfo {author} {\bibfnamefont {S.~V.}\ \bibnamefont
  {Dordevic}}, \bibinfo {author} {\bibfnamefont {S.}~\bibnamefont {Komiya}},
  \bibinfo {author} {\bibfnamefont {Y.}~\bibnamefont {Ando}}, \ and\ \bibinfo
  {author} {\bibfnamefont {D.~N.}\ \bibnamefont {Basov}},\ }\href {\doibase
  10.1103/PhysRevLett.91.167401} {\bibfield  {journal} {\bibinfo  {journal}
  {Phys. Rev. Lett.}\ }\textbf {\bibinfo {volume} {91}},\ \bibinfo {pages}
  {167401} (\bibinfo {year} {2003})}\BibitemShut {NoStop}%
\bibitem [{\citenamefont {Nagaosa}\ and\ \citenamefont
  {Heusler}(1999)}]{nagaosa}%
  \BibitemOpen
  \bibfield  {author} {\bibinfo {author} {\bibfnamefont {N.}~\bibnamefont
  {Nagaosa}}\ and\ \bibinfo {author} {\bibfnamefont {S.}~\bibnamefont
  {Heusler}},\ }\href {https://books.google.it/books?id=C9uAXYIlFhMC} {\emph
  {\bibinfo {title} {Quantum Field Theory in Condensed Matter Physics}}},\
  Texts and monographs in physics\ (\bibinfo  {publisher} {Springer, New York,
  NY},\ \bibinfo {year} {1999})\BibitemShut {NoStop}%
\bibitem [{\citenamefont {Coleman}(2015)}]{coleman}%
  \BibitemOpen
  \bibfield  {author} {\bibinfo {author} {\bibfnamefont {P.}~\bibnamefont
  {Coleman}},\ }\href {\doibase 10.1017/CBO9781139020916} {\emph {\bibinfo
  {title} {Introduction to Many-Body Physics}}}\ (\bibinfo  {publisher}
  {Cambridge University Press},\ \bibinfo {year} {2015})\BibitemShut {NoStop}%
\bibitem [{\citenamefont {Savel'ev}\ \emph {et~al.}(2010)\citenamefont
  {Savel'ev}, \citenamefont {Yampol'skii}, \citenamefont {Rakhmanov},\ and\
  \citenamefont {Nori}}]{nori_review10}%
  \BibitemOpen
  \bibfield  {author} {\bibinfo {author} {\bibfnamefont {S.}~\bibnamefont
  {Savel'ev}}, \bibinfo {author} {\bibfnamefont {V.~A.}\ \bibnamefont
  {Yampol'skii}}, \bibinfo {author} {\bibfnamefont {A.~L.}\ \bibnamefont
  {Rakhmanov}}, \ and\ \bibinfo {author} {\bibfnamefont {F.}~\bibnamefont
  {Nori}},\ }\href {\doibase 10.1088/0034-4885/73/2/026501} {\bibfield
  {journal} {\bibinfo  {journal} {Reports on Progress in Physics}\ }\textbf
  {\bibinfo {volume} {73}},\ \bibinfo {pages} {026501} (\bibinfo {year}
  {2010})}\BibitemShut {NoStop}%
\bibitem [{\citenamefont {Savel'ev}\ \emph {et~al.}(2006)\citenamefont
  {Savel'ev}, \citenamefont {Rakhmanov}, \citenamefont {Yampol'skii},\ and\
  \citenamefont {Nori}}]{nori_natphys06}%
  \BibitemOpen
  \bibfield  {author} {\bibinfo {author} {\bibfnamefont {S.}~\bibnamefont
  {Savel'ev}}, \bibinfo {author} {\bibfnamefont {A.~L.}\ \bibnamefont
  {Rakhmanov}}, \bibinfo {author} {\bibfnamefont {V.~A.}\ \bibnamefont
  {Yampol'skii}}, \ and\ \bibinfo {author} {\bibfnamefont {F.}~\bibnamefont
  {Nori}},\ }\href {\doibase 10.1038/nphys358} {\bibfield  {journal} {\bibinfo
  {journal} {Nature Physics}\ }\textbf {\bibinfo {volume} {2}},\ \bibinfo
  {pages} {521} (\bibinfo {year} {2006})}\BibitemShut {NoStop}%
\bibitem [{\citenamefont {Michael}\ \emph {et~al.}(2020)\citenamefont
  {Michael}, \citenamefont {von Hoegen}, \citenamefont {Fechner}, \citenamefont
  {F\"orst}, \citenamefont {Cavalleri},\ and\ \citenamefont
  {Demler}}]{demler_prb20}%
  \BibitemOpen
  \bibfield  {author} {\bibinfo {author} {\bibfnamefont {M.~H.}\ \bibnamefont
  {Michael}}, \bibinfo {author} {\bibfnamefont {A.}~\bibnamefont {von Hoegen}},
  \bibinfo {author} {\bibfnamefont {M.}~\bibnamefont {Fechner}}, \bibinfo
  {author} {\bibfnamefont {M.}~\bibnamefont {F\"orst}}, \bibinfo {author}
  {\bibfnamefont {A.}~\bibnamefont {Cavalleri}}, \ and\ \bibinfo {author}
  {\bibfnamefont {E.}~\bibnamefont {Demler}},\ }\href {\doibase
  10.1103/PhysRevB.102.174505} {\bibfield  {journal} {\bibinfo  {journal}
  {Phys. Rev. B}\ }\textbf {\bibinfo {volume} {102}},\ \bibinfo {pages}
  {174505} (\bibinfo {year} {2020})}\BibitemShut {NoStop}%
\bibitem [{\citenamefont {Gabriele}\ \emph {et~al.}(2021)\citenamefont
  {Gabriele}, \citenamefont {Udina},\ and\ \citenamefont
  {Benfatto}}]{gabriele_natcomm21}%
  \BibitemOpen
  \bibfield  {author} {\bibinfo {author} {\bibfnamefont {F.}~\bibnamefont
  {Gabriele}}, \bibinfo {author} {\bibfnamefont {M.}~\bibnamefont {Udina}}, \
  and\ \bibinfo {author} {\bibfnamefont {L.}~\bibnamefont {Benfatto}},\ }\href
  {\doibase 10.1038/s41467-021-21041-6} {\bibfield  {journal} {\bibinfo
  {journal} {Nature Communications}\ }\textbf {\bibinfo {volume} {12}},\
  \bibinfo {pages} {752} (\bibinfo {year} {2021})}\BibitemShut {NoStop}%
\bibitem [{\citenamefont {Dolgirev}\ \emph {et~al.}(2022)\citenamefont
  {Dolgirev}, \citenamefont {Zong}, \citenamefont {Michael}, \citenamefont
  {Curtis}, \citenamefont {Podolsky}, \citenamefont {Cavalleri},\ and\
  \citenamefont {Demler}}]{demler_commphys22}%
  \BibitemOpen
  \bibfield  {author} {\bibinfo {author} {\bibfnamefont {P.~E.}\ \bibnamefont
  {Dolgirev}}, \bibinfo {author} {\bibfnamefont {A.}~\bibnamefont {Zong}},
  \bibinfo {author} {\bibfnamefont {M.~H.}\ \bibnamefont {Michael}}, \bibinfo
  {author} {\bibfnamefont {J.~B.}\ \bibnamefont {Curtis}}, \bibinfo {author}
  {\bibfnamefont {D.}~\bibnamefont {Podolsky}}, \bibinfo {author}
  {\bibfnamefont {A.}~\bibnamefont {Cavalleri}}, \ and\ \bibinfo {author}
  {\bibfnamefont {E.}~\bibnamefont {Demler}},\ }\href {\doibase
  10.1038/s42005-022-01007-w} {\bibfield  {journal} {\bibinfo  {journal}
  {Communications Physics}\ }\textbf {\bibinfo {volume} {5}},\ \bibinfo {pages}
  {234} (\bibinfo {year} {2022})}\BibitemShut {NoStop}%
\bibitem [{\citenamefont {Laplace}\ and\ \citenamefont
  {Cavalleri}(2016)}]{cavalleri_review}%
  \BibitemOpen
  \bibfield  {author} {\bibinfo {author} {\bibfnamefont {Y.}~\bibnamefont
  {Laplace}}\ and\ \bibinfo {author} {\bibfnamefont {A.}~\bibnamefont
  {Cavalleri}},\ }\href {\doibase 10.1080/23746149.2016.1212671} {\bibfield
  {journal} {\bibinfo  {journal} {Advances in Physics: X}\ }\textbf {\bibinfo
  {volume} {1}},\ \bibinfo {pages} {387} (\bibinfo {year} {2016})}\BibitemShut
  {NoStop}%
\bibitem [{\citenamefont {Rajasekaran}\ \emph {et~al.}(2016)\citenamefont
  {Rajasekaran}, \citenamefont {Casandruc}, \citenamefont {Laplace},
  \citenamefont {Nicoletti}, \citenamefont {Gu}, \citenamefont {Clark},
  \citenamefont {Jaksch},\ and\ \citenamefont
  {Cavalleri}}]{cavalleri_natphys16}%
  \BibitemOpen
  \bibfield  {author} {\bibinfo {author} {\bibfnamefont {S.}~\bibnamefont
  {Rajasekaran}}, \bibinfo {author} {\bibfnamefont {E.}~\bibnamefont
  {Casandruc}}, \bibinfo {author} {\bibfnamefont {Y.}~\bibnamefont {Laplace}},
  \bibinfo {author} {\bibfnamefont {D.}~\bibnamefont {Nicoletti}}, \bibinfo
  {author} {\bibfnamefont {G.~D.}\ \bibnamefont {Gu}}, \bibinfo {author}
  {\bibfnamefont {S.~R.}\ \bibnamefont {Clark}}, \bibinfo {author}
  {\bibfnamefont {D.}~\bibnamefont {Jaksch}}, \ and\ \bibinfo {author}
  {\bibfnamefont {A.}~\bibnamefont {Cavalleri}},\ }\href
  {https://doi.org/10.1038/nphys3819} {\bibfield  {journal} {\bibinfo
  {journal} {Nature Physics}\ }\textbf {\bibinfo {volume} {12}},\ \bibinfo
  {pages} {1012} (\bibinfo {year} {2016})}\BibitemShut {NoStop}%
\bibitem [{\citenamefont {Rajasekaran}\ \emph {et~al.}(2018)\citenamefont
  {Rajasekaran}, \citenamefont {Okamoto}, \citenamefont {Mathey}, \citenamefont
  {Fechner}, \citenamefont {Thampy}, \citenamefont {Gu},\ and\ \citenamefont
  {Cavalleri}}]{cavalleri_science18}%
  \BibitemOpen
  \bibfield  {author} {\bibinfo {author} {\bibfnamefont {S.}~\bibnamefont
  {Rajasekaran}}, \bibinfo {author} {\bibfnamefont {J.}~\bibnamefont
  {Okamoto}}, \bibinfo {author} {\bibfnamefont {L.}~\bibnamefont {Mathey}},
  \bibinfo {author} {\bibfnamefont {M.}~\bibnamefont {Fechner}}, \bibinfo
  {author} {\bibfnamefont {V.}~\bibnamefont {Thampy}}, \bibinfo {author}
  {\bibfnamefont {G.~D.}\ \bibnamefont {Gu}}, \ and\ \bibinfo {author}
  {\bibfnamefont {A.}~\bibnamefont {Cavalleri}},\ }\href {\doibase
  10.1126/science.aan3438} {\bibfield  {journal} {\bibinfo  {journal}
  {Science}\ }\textbf {\bibinfo {volume} {359}},\ \bibinfo {pages} {575}
  (\bibinfo {year} {2018})}\BibitemShut {NoStop}%
\bibitem [{\citenamefont {Cremin}\ \emph {et~al.}(2019)\citenamefont {Cremin},
  \citenamefont {Zhang}, \citenamefont {Homes}, \citenamefont {Gu},
  \citenamefont {Sun}, \citenamefont {Fogler}, \citenamefont {Millis},
  \citenamefont {Basov},\ and\ \citenamefont {Averitt}}]{averitt_pnas19}%
  \BibitemOpen
  \bibfield  {author} {\bibinfo {author} {\bibfnamefont {K.~A.}\ \bibnamefont
  {Cremin}}, \bibinfo {author} {\bibfnamefont {J.}~\bibnamefont {Zhang}},
  \bibinfo {author} {\bibfnamefont {C.~C.}\ \bibnamefont {Homes}}, \bibinfo
  {author} {\bibfnamefont {G.~D.}\ \bibnamefont {Gu}}, \bibinfo {author}
  {\bibfnamefont {Z.}~\bibnamefont {Sun}}, \bibinfo {author} {\bibfnamefont
  {M.~M.}\ \bibnamefont {Fogler}}, \bibinfo {author} {\bibfnamefont {A.~J.}\
  \bibnamefont {Millis}}, \bibinfo {author} {\bibfnamefont {D.~N.}\
  \bibnamefont {Basov}}, \ and\ \bibinfo {author} {\bibfnamefont {R.~D.}\
  \bibnamefont {Averitt}},\ }\href {\doibase 10.1073/pnas.1908368116}
  {\bibfield  {journal} {\bibinfo  {journal} {Proceedings of the National
  Academy of Sciences}\ }\textbf {\bibinfo {volume} {116}},\ \bibinfo {pages}
  {19875} (\bibinfo {year} {2019})}\BibitemShut {NoStop}%
\bibitem [{\citenamefont {von Hoegen}\ \emph {et~al.}(2022)\citenamefont {von
  Hoegen}, \citenamefont {Fechner}, \citenamefont {F\"orst}, \citenamefont
  {Taherian}, \citenamefont {Rowe}, \citenamefont {Ribak}, \citenamefont
  {Porras}, \citenamefont {Keimer}, \citenamefont {Michael}, \citenamefont
  {Demler},\ and\ \citenamefont {Cavalleri}}]{cavalleri_prx22}%
  \BibitemOpen
  \bibfield  {author} {\bibinfo {author} {\bibfnamefont {A.}~\bibnamefont {von
  Hoegen}}, \bibinfo {author} {\bibfnamefont {M.}~\bibnamefont {Fechner}},
  \bibinfo {author} {\bibfnamefont {M.}~\bibnamefont {F\"orst}}, \bibinfo
  {author} {\bibfnamefont {N.}~\bibnamefont {Taherian}}, \bibinfo {author}
  {\bibfnamefont {E.}~\bibnamefont {Rowe}}, \bibinfo {author} {\bibfnamefont
  {A.}~\bibnamefont {Ribak}}, \bibinfo {author} {\bibfnamefont
  {J.}~\bibnamefont {Porras}}, \bibinfo {author} {\bibfnamefont
  {B.}~\bibnamefont {Keimer}}, \bibinfo {author} {\bibfnamefont
  {M.}~\bibnamefont {Michael}}, \bibinfo {author} {\bibfnamefont
  {E.}~\bibnamefont {Demler}}, \ and\ \bibinfo {author} {\bibfnamefont
  {A.}~\bibnamefont {Cavalleri}},\ }\href {\doibase 10.1103/PhysRevX.12.031008}
  {\bibfield  {journal} {\bibinfo  {journal} {Phys. Rev. X}\ }\textbf {\bibinfo
  {volume} {12}},\ \bibinfo {pages} {031008} (\bibinfo {year}
  {2022})}\BibitemShut {NoStop}%
\bibitem [{\citenamefont {Fu}\ \emph {et~al.}(2022)\citenamefont {Fu},
  \citenamefont {Nicoletti}, \citenamefont {Fechner}, \citenamefont {Buzzi},
  \citenamefont {Gu},\ and\ \citenamefont {Cavalleri}}]{cavalleri_prb22}%
  \BibitemOpen
  \bibfield  {author} {\bibinfo {author} {\bibfnamefont {D.}~\bibnamefont
  {Fu}}, \bibinfo {author} {\bibfnamefont {D.}~\bibnamefont {Nicoletti}},
  \bibinfo {author} {\bibfnamefont {M.}~\bibnamefont {Fechner}}, \bibinfo
  {author} {\bibfnamefont {M.}~\bibnamefont {Buzzi}}, \bibinfo {author}
  {\bibfnamefont {G.~D.}\ \bibnamefont {Gu}}, \ and\ \bibinfo {author}
  {\bibfnamefont {A.}~\bibnamefont {Cavalleri}},\ }\href {\doibase
  10.1103/PhysRevB.105.L020502} {\bibfield  {journal} {\bibinfo  {journal}
  {Phys. Rev. B}\ }\textbf {\bibinfo {volume} {105}},\ \bibinfo {pages}
  {L020502} (\bibinfo {year} {2022})}\BibitemShut {NoStop}%
\bibitem [{\citenamefont {Kaj}\ \emph {et~al.}(2023)\citenamefont {Kaj},
  \citenamefont {Cremin}, \citenamefont {Hammock}, \citenamefont {Schalch},
  \citenamefont {Basov},\ and\ \citenamefont {Averitt}}]{averitt_prb23}%
  \BibitemOpen
  \bibfield  {author} {\bibinfo {author} {\bibfnamefont {K.}~\bibnamefont
  {Kaj}}, \bibinfo {author} {\bibfnamefont {K.~A.}\ \bibnamefont {Cremin}},
  \bibinfo {author} {\bibfnamefont {I.}~\bibnamefont {Hammock}}, \bibinfo
  {author} {\bibfnamefont {J.}~\bibnamefont {Schalch}}, \bibinfo {author}
  {\bibfnamefont {D.~N.}\ \bibnamefont {Basov}}, \ and\ \bibinfo {author}
  {\bibfnamefont {R.~D.}\ \bibnamefont {Averitt}},\ }\href {\doibase
  10.1103/PhysRevB.107.L140504} {\bibfield  {journal} {\bibinfo  {journal}
  {Phys. Rev. B}\ }\textbf {\bibinfo {volume} {107}},\ \bibinfo {pages}
  {L140504} (\bibinfo {year} {2023})}\BibitemShut {NoStop}%
\bibitem [{\citenamefont {Katsumi}\ \emph {et~al.}(2023)\citenamefont
  {Katsumi}, \citenamefont {Nishida}, \citenamefont {Kaiser}, \citenamefont
  {Miyasaka}, \citenamefont {Tajima},\ and\ \citenamefont
  {Shimano}}]{shimano_prb23}%
  \BibitemOpen
  \bibfield  {author} {\bibinfo {author} {\bibfnamefont {K.}~\bibnamefont
  {Katsumi}}, \bibinfo {author} {\bibfnamefont {M.}~\bibnamefont {Nishida}},
  \bibinfo {author} {\bibfnamefont {S.}~\bibnamefont {Kaiser}}, \bibinfo
  {author} {\bibfnamefont {S.}~\bibnamefont {Miyasaka}}, \bibinfo {author}
  {\bibfnamefont {S.}~\bibnamefont {Tajima}}, \ and\ \bibinfo {author}
  {\bibfnamefont {R.}~\bibnamefont {Shimano}},\ }\href {\doibase
  10.1103/PhysRevB.107.214506} {\bibfield  {journal} {\bibinfo  {journal}
  {Phys. Rev. B}\ }\textbf {\bibinfo {volume} {107}},\ \bibinfo {pages}
  {214506} (\bibinfo {year} {2023})}\BibitemShut {NoStop}%
\bibitem [{\citenamefont {Bulaevskii}\ \emph {et~al.}(1994)\citenamefont
  {Bulaevskii}, \citenamefont {Zamora}, \citenamefont {Baeriswyl},
  \citenamefont {Beck},\ and\ \citenamefont {Clem}}]{bulaevskii_prb94}%
  \BibitemOpen
  \bibfield  {author} {\bibinfo {author} {\bibfnamefont {L.~N.}\ \bibnamefont
  {Bulaevskii}}, \bibinfo {author} {\bibfnamefont {M.}~\bibnamefont {Zamora}},
  \bibinfo {author} {\bibfnamefont {D.}~\bibnamefont {Baeriswyl}}, \bibinfo
  {author} {\bibfnamefont {H.}~\bibnamefont {Beck}}, \ and\ \bibinfo {author}
  {\bibfnamefont {J.~R.}\ \bibnamefont {Clem}},\ }\href {\doibase
  10.1103/PhysRevB.50.12831} {\bibfield  {journal} {\bibinfo  {journal} {Phys.
  Rev. B}\ }\textbf {\bibinfo {volume} {50}},\ \bibinfo {pages} {12831}
  (\bibinfo {year} {1994})}\BibitemShut {NoStop}%
\bibitem [{\citenamefont {Helm}\ and\ \citenamefont
  {Bulaevskii}(2002)}]{bulaevskii_prb02}%
  \BibitemOpen
  \bibfield  {author} {\bibinfo {author} {\bibfnamefont {C.}~\bibnamefont
  {Helm}}\ and\ \bibinfo {author} {\bibfnamefont {L.~N.}\ \bibnamefont
  {Bulaevskii}},\ }\href {\doibase 10.1103/PhysRevB.66.094514} {\bibfield
  {journal} {\bibinfo  {journal} {Phys. Rev. B}\ }\textbf {\bibinfo {volume}
  {66}},\ \bibinfo {pages} {094514} (\bibinfo {year} {2002})}\BibitemShut
  {NoStop}%
\bibitem [{\citenamefont {Machida}\ \emph {et~al.}(1999)\citenamefont
  {Machida}, \citenamefont {Koyama},\ and\ \citenamefont
  {Tachiki}}]{machida_prl99}%
  \BibitemOpen
  \bibfield  {author} {\bibinfo {author} {\bibfnamefont {M.}~\bibnamefont
  {Machida}}, \bibinfo {author} {\bibfnamefont {T.}~\bibnamefont {Koyama}}, \
  and\ \bibinfo {author} {\bibfnamefont {M.}~\bibnamefont {Tachiki}},\ }\href
  {\doibase 10.1103/PhysRevLett.83.4618} {\bibfield  {journal} {\bibinfo
  {journal} {Phys. Rev. Lett.}\ }\textbf {\bibinfo {volume} {83}},\ \bibinfo
  {pages} {4618} (\bibinfo {year} {1999})}\BibitemShut {NoStop}%
\bibitem [{\citenamefont {Machida}\ \emph {et~al.}(2000)\citenamefont
  {Machida}, \citenamefont {Koyama}, \citenamefont {Tanaka},\ and\
  \citenamefont {Tachiki}}]{machida_physc00}%
  \BibitemOpen
  \bibfield  {author} {\bibinfo {author} {\bibfnamefont {M.}~\bibnamefont
  {Machida}}, \bibinfo {author} {\bibfnamefont {T.}~\bibnamefont {Koyama}},
  \bibinfo {author} {\bibfnamefont {A.}~\bibnamefont {Tanaka}}, \ and\ \bibinfo
  {author} {\bibfnamefont {M.}~\bibnamefont {Tachiki}},\ }\href {\doibase
  https://doi.org/10.1016/S0921-4534(99)00612-7} {\bibfield  {journal}
  {\bibinfo  {journal} {Physica C: Superconductivity}\ }\textbf {\bibinfo
  {volume} {331}},\ \bibinfo {pages} {85 } (\bibinfo {year}
  {2000})}\BibitemShut {NoStop}%
\bibitem [{\citenamefont {Gabriele}\ \emph {et~al.}(2022)\citenamefont
  {Gabriele}, \citenamefont {Castellani},\ and\ \citenamefont
  {Benfatto}}]{gabriele_prr22}%
  \BibitemOpen
  \bibfield  {author} {\bibinfo {author} {\bibfnamefont {F.}~\bibnamefont
  {Gabriele}}, \bibinfo {author} {\bibfnamefont {C.}~\bibnamefont
  {Castellani}}, \ and\ \bibinfo {author} {\bibfnamefont {L.}~\bibnamefont
  {Benfatto}},\ }\href {\doibase 10.1103/PhysRevResearch.4.023112} {\bibfield
  {journal} {\bibinfo  {journal} {Phys. Rev. Res.}\ }\textbf {\bibinfo {volume}
  {4}},\ \bibinfo {pages} {023112} (\bibinfo {year} {2022})}\BibitemShut
  {NoStop}%
\bibitem [{\citenamefont {Hu}\ \emph {et~al.}(2014)\citenamefont {Hu},
  \citenamefont {Kaiser}, \citenamefont {Nicoletti}, \citenamefont {Hunt},
  \citenamefont {Gierz}, \citenamefont {Hoffmann}, \citenamefont {Le~Tacon},
  \citenamefont {Loew}, \citenamefont {Keimer},\ and\ \citenamefont
  {Cavalleri}}]{cavalleri_nmat14}%
  \BibitemOpen
  \bibfield  {author} {\bibinfo {author} {\bibfnamefont {H.}~\bibnamefont
  {Hu}}, \bibinfo {author} {\bibfnamefont {S.}~\bibnamefont {Kaiser}}, \bibinfo
  {author} {\bibfnamefont {D.}~\bibnamefont {Nicoletti}}, \bibinfo {author}
  {\bibfnamefont {C.~R.}\ \bibnamefont {Hunt}}, \bibinfo {author}
  {\bibfnamefont {I.}~\bibnamefont {Gierz}}, \bibinfo {author} {\bibfnamefont
  {M.~C.}\ \bibnamefont {Hoffmann}}, \bibinfo {author} {\bibfnamefont
  {M.}~\bibnamefont {Le~Tacon}}, \bibinfo {author} {\bibfnamefont
  {T.}~\bibnamefont {Loew}}, \bibinfo {author} {\bibfnamefont {B.}~\bibnamefont
  {Keimer}}, \ and\ \bibinfo {author} {\bibfnamefont {A.}~\bibnamefont
  {Cavalleri}},\ }\href {\doibase 10.1038/nmat3963} {\bibfield  {journal}
  {\bibinfo  {journal} {Nature Materials}\ }\textbf {\bibinfo {volume} {13}},\
  \bibinfo {pages} {705} (\bibinfo {year} {2014})}\BibitemShut {NoStop}%
\bibitem [{\citenamefont {Yuan}\ \emph {et~al.}(2022)\citenamefont {Yuan},
  \citenamefont {Shi}, \citenamefont {Yue}, \citenamefont {Li}, \citenamefont
  {Wang}, \citenamefont {Xu}, \citenamefont {Xu}, \citenamefont {Wang},
  \citenamefont {Gan}, \citenamefont {Chen}, \citenamefont {Lin}, \citenamefont
  {Wang}, \citenamefont {Jin}, \citenamefont {Wang}, \citenamefont {Luo},
  \citenamefont {Zhang}, \citenamefont {Wu}, \citenamefont {Liu}, \citenamefont
  {Hu}, \citenamefont {Li}, \citenamefont {Zhou}, \citenamefont {Wu},
  \citenamefont {Dong},\ and\ \citenamefont {Wang}}]{wangNL_cm22}%
  \BibitemOpen
  \bibfield  {author} {\bibinfo {author} {\bibfnamefont {J.~Y.}\ \bibnamefont
  {Yuan}}, \bibinfo {author} {\bibfnamefont {L.~Y.}\ \bibnamefont {Shi}},
  \bibinfo {author} {\bibfnamefont {L.}~\bibnamefont {Yue}}, \bibinfo {author}
  {\bibfnamefont {B.~H.}\ \bibnamefont {Li}}, \bibinfo {author} {\bibfnamefont
  {Z.~X.}\ \bibnamefont {Wang}}, \bibinfo {author} {\bibfnamefont {S.~X.}\
  \bibnamefont {Xu}}, \bibinfo {author} {\bibfnamefont {T.~Q.}\ \bibnamefont
  {Xu}}, \bibinfo {author} {\bibfnamefont {Y.}~\bibnamefont {Wang}}, \bibinfo
  {author} {\bibfnamefont {Z.~Z.}\ \bibnamefont {Gan}}, \bibinfo {author}
  {\bibfnamefont {F.~C.}\ \bibnamefont {Chen}}, \bibinfo {author}
  {\bibfnamefont {Z.~F.}\ \bibnamefont {Lin}}, \bibinfo {author} {\bibfnamefont
  {X.}~\bibnamefont {Wang}}, \bibinfo {author} {\bibfnamefont {K.}~\bibnamefont
  {Jin}}, \bibinfo {author} {\bibfnamefont {X.~B.}\ \bibnamefont {Wang}},
  \bibinfo {author} {\bibfnamefont {J.~L.}\ \bibnamefont {Luo}}, \bibinfo
  {author} {\bibfnamefont {S.~J.}\ \bibnamefont {Zhang}}, \bibinfo {author}
  {\bibfnamefont {Q.}~\bibnamefont {Wu}}, \bibinfo {author} {\bibfnamefont
  {Q.~M.}\ \bibnamefont {Liu}}, \bibinfo {author} {\bibfnamefont {T.~C.}\
  \bibnamefont {Hu}}, \bibinfo {author} {\bibfnamefont {R.~S.}\ \bibnamefont
  {Li}}, \bibinfo {author} {\bibfnamefont {X.~Y.}\ \bibnamefont {Zhou}},
  \bibinfo {author} {\bibfnamefont {D.}~\bibnamefont {Wu}}, \bibinfo {author}
  {\bibfnamefont {T.}~\bibnamefont {Dong}}, \ and\ \bibinfo {author}
  {\bibfnamefont {N.~L.}\ \bibnamefont {Wang}},\ }\href@noop {} {\enquote
  {\bibinfo {title} {Revealing strong coupling of collective modes between
  superconductivity and pseudogap in cuprate superconductor by terahertz third
  harmonic generation},}\ } (\bibinfo {year} {2022}),\ \Eprint
  {http://arxiv.org/abs/2211.06961} {arXiv:2211.06961 [cond-mat.supr-con]}
  \BibitemShut {NoStop}%
\bibitem [{\citenamefont {Jiang}\ \emph {et~al.}(1993)\citenamefont {Jiang},
  \citenamefont {Yuan}, \citenamefont {How}, \citenamefont {Widom},
  \citenamefont {Vittoria},\ and\ \citenamefont {Drehman}}]{vittoria_jap93}%
  \BibitemOpen
  \bibfield  {author} {\bibinfo {author} {\bibfnamefont {H.}~\bibnamefont
  {Jiang}}, \bibinfo {author} {\bibfnamefont {T.}~\bibnamefont {Yuan}},
  \bibinfo {author} {\bibfnamefont {H.}~\bibnamefont {How}}, \bibinfo {author}
  {\bibfnamefont {A.}~\bibnamefont {Widom}}, \bibinfo {author} {\bibfnamefont
  {C.}~\bibnamefont {Vittoria}}, \ and\ \bibinfo {author} {\bibfnamefont
  {A.}~\bibnamefont {Drehman}},\ }\href@noop {} {\bibfield  {journal} {\bibinfo
   {journal} {Journal of applied physics}\ }\textbf {\bibinfo {volume} {73}},\
  \bibinfo {pages} {5865} (\bibinfo {year} {1993})}\BibitemShut {NoStop}%
\bibitem [{\citenamefont {van~der Marel}\ and\ \citenamefont
  {Tsvetkov}(2001)}]{vandermarel_prb01}%
  \BibitemOpen
  \bibfield  {author} {\bibinfo {author} {\bibfnamefont {D.}~\bibnamefont
  {van~der Marel}}\ and\ \bibinfo {author} {\bibfnamefont {A.~A.}\ \bibnamefont
  {Tsvetkov}},\ }\href {\doibase 10.1103/PhysRevB.64.024530} {\bibfield
  {journal} {\bibinfo  {journal} {Phys. Rev. B}\ }\textbf {\bibinfo {volume}
  {64}},\ \bibinfo {pages} {024530} (\bibinfo {year} {2001})}\BibitemShut
  {NoStop}%
\bibitem [{\citenamefont {Dubroka}\ \emph {et~al.}(2011)\citenamefont
  {Dubroka}, \citenamefont {R\"ossle}, \citenamefont {Kim}, \citenamefont
  {Malik}, \citenamefont {Munzar}, \citenamefont {Basov}, \citenamefont
  {Schafgans}, \citenamefont {Moon}, \citenamefont {Lin}, \citenamefont {Haug},
  \citenamefont {Hinkov}, \citenamefont {Keimer}, \citenamefont {Wolf},
  \citenamefont {Storey}, \citenamefont {Tallon},\ and\ \citenamefont
  {Bernhard}}]{bernhard_prl11}%
  \BibitemOpen
  \bibfield  {author} {\bibinfo {author} {\bibfnamefont {A.}~\bibnamefont
  {Dubroka}}, \bibinfo {author} {\bibfnamefont {M.}~\bibnamefont {R\"ossle}},
  \bibinfo {author} {\bibfnamefont {K.~W.}\ \bibnamefont {Kim}}, \bibinfo
  {author} {\bibfnamefont {V.~K.}\ \bibnamefont {Malik}}, \bibinfo {author}
  {\bibfnamefont {D.}~\bibnamefont {Munzar}}, \bibinfo {author} {\bibfnamefont
  {D.~N.}\ \bibnamefont {Basov}}, \bibinfo {author} {\bibfnamefont {A.~A.}\
  \bibnamefont {Schafgans}}, \bibinfo {author} {\bibfnamefont {S.~J.}\
  \bibnamefont {Moon}}, \bibinfo {author} {\bibfnamefont {C.~T.}\ \bibnamefont
  {Lin}}, \bibinfo {author} {\bibfnamefont {D.}~\bibnamefont {Haug}}, \bibinfo
  {author} {\bibfnamefont {V.}~\bibnamefont {Hinkov}}, \bibinfo {author}
  {\bibfnamefont {B.}~\bibnamefont {Keimer}}, \bibinfo {author} {\bibfnamefont
  {T.}~\bibnamefont {Wolf}}, \bibinfo {author} {\bibfnamefont {J.~G.}\
  \bibnamefont {Storey}}, \bibinfo {author} {\bibfnamefont {J.~L.}\
  \bibnamefont {Tallon}}, \ and\ \bibinfo {author} {\bibfnamefont
  {C.}~\bibnamefont {Bernhard}},\ }\href {\doibase
  10.1103/PhysRevLett.106.047006} {\bibfield  {journal} {\bibinfo  {journal}
  {Phys. Rev. Lett.}\ }\textbf {\bibinfo {volume} {106}},\ \bibinfo {pages}
  {047006} (\bibinfo {year} {2011})}\BibitemShut {NoStop}%
\bibitem [{\citenamefont {Gr\"uninger}\ \emph {et~al.}(2000)\citenamefont
  {Gr\"uninger}, \citenamefont {van~der Marel}, \citenamefont {Tsvetkov},\ and\
  \citenamefont {Erb}}]{erb_prl00}%
  \BibitemOpen
  \bibfield  {author} {\bibinfo {author} {\bibfnamefont {M.}~\bibnamefont
  {Gr\"uninger}}, \bibinfo {author} {\bibfnamefont {D.}~\bibnamefont {van~der
  Marel}}, \bibinfo {author} {\bibfnamefont {A.~A.}\ \bibnamefont {Tsvetkov}},
  \ and\ \bibinfo {author} {\bibfnamefont {A.}~\bibnamefont {Erb}},\ }\href
  {\doibase 10.1103/PhysRevLett.84.1575} {\bibfield  {journal} {\bibinfo
  {journal} {Phys. Rev. Lett.}\ }\textbf {\bibinfo {volume} {84}},\ \bibinfo
  {pages} {1575} (\bibinfo {year} {2000})}\BibitemShut {NoStop}%
\bibitem [{\citenamefont {Uykur}\ \emph {et~al.}(2014)\citenamefont {Uykur},
  \citenamefont {Tanaka}, \citenamefont {Masui}, \citenamefont {Miyasaka},\
  and\ \citenamefont {Tajima}}]{tajima_prl09}%
  \BibitemOpen
  \bibfield  {author} {\bibinfo {author} {\bibfnamefont {E.}~\bibnamefont
  {Uykur}}, \bibinfo {author} {\bibfnamefont {K.}~\bibnamefont {Tanaka}},
  \bibinfo {author} {\bibfnamefont {T.}~\bibnamefont {Masui}}, \bibinfo
  {author} {\bibfnamefont {S.}~\bibnamefont {Miyasaka}}, \ and\ \bibinfo
  {author} {\bibfnamefont {S.}~\bibnamefont {Tajima}},\ }\href {\doibase
  10.1103/PhysRevLett.112.127003} {\bibfield  {journal} {\bibinfo  {journal}
  {Phys. Rev. Lett.}\ }\textbf {\bibinfo {volume} {112}},\ \bibinfo {pages}
  {127003} (\bibinfo {year} {2014})}\BibitemShut {NoStop}%
\bibitem [{\citenamefont {Shibata}\ and\ \citenamefont
  {Yamada}(1998)}]{yamada_prl98}%
  \BibitemOpen
  \bibfield  {author} {\bibinfo {author} {\bibfnamefont {H.}~\bibnamefont
  {Shibata}}\ and\ \bibinfo {author} {\bibfnamefont {T.}~\bibnamefont
  {Yamada}},\ }\href {\doibase 10.1103/PhysRevLett.81.3519} {\bibfield
  {journal} {\bibinfo  {journal} {Phys. Rev. Lett.}\ }\textbf {\bibinfo
  {volume} {81}},\ \bibinfo {pages} {3519} (\bibinfo {year}
  {1998})}\BibitemShut {NoStop}%
\bibitem [{\citenamefont {Zhang}\ \emph {et~al.}(2020)\citenamefont {Zhang},
  \citenamefont {Wang}, \citenamefont {Xiang}, \citenamefont {Yao},
  \citenamefont {Liu}, \citenamefont {Shi}, \citenamefont {Lin}, \citenamefont
  {Dong}, \citenamefont {Wu},\ and\ \citenamefont {Wang}}]{wangNL_prx20}%
  \BibitemOpen
  \bibfield  {author} {\bibinfo {author} {\bibfnamefont {S.~J.}\ \bibnamefont
  {Zhang}}, \bibinfo {author} {\bibfnamefont {Z.~X.}\ \bibnamefont {Wang}},
  \bibinfo {author} {\bibfnamefont {H.}~\bibnamefont {Xiang}}, \bibinfo
  {author} {\bibfnamefont {X.}~\bibnamefont {Yao}}, \bibinfo {author}
  {\bibfnamefont {Q.~M.}\ \bibnamefont {Liu}}, \bibinfo {author} {\bibfnamefont
  {L.~Y.}\ \bibnamefont {Shi}}, \bibinfo {author} {\bibfnamefont
  {T.}~\bibnamefont {Lin}}, \bibinfo {author} {\bibfnamefont {T.}~\bibnamefont
  {Dong}}, \bibinfo {author} {\bibfnamefont {D.}~\bibnamefont {Wu}}, \ and\
  \bibinfo {author} {\bibfnamefont {N.~L.}\ \bibnamefont {Wang}},\ }\href
  {\doibase 10.1103/PhysRevX.10.011056} {\bibfield  {journal} {\bibinfo
  {journal} {Phys. Rev. X}\ }\textbf {\bibinfo {volume} {10}},\ \bibinfo
  {pages} {011056} (\bibinfo {year} {2020})}\BibitemShut {NoStop}%
\bibitem [{\citenamefont {Leggett}(1966)}]{leggett}%
  \BibitemOpen
  \bibfield  {author} {\bibinfo {author} {\bibfnamefont {A.~J.}\ \bibnamefont
  {Leggett}},\ }\href {\doibase 10.1143/PTP.36.901} {\bibfield  {journal}
  {\bibinfo  {journal} {Progress of Theoretical Physics}\ }\textbf {\bibinfo
  {volume} {36}},\ \bibinfo {pages} {901} (\bibinfo {year} {1966})},\ \Eprint
  {http://arxiv.org/abs/https://academic.oup.com/ptp/article-pdf/36/5/901/5256693/36-5-901.pdf}
  {https://academic.oup.com/ptp/article-pdf/36/5/901/5256693/36-5-901.pdf}
  \BibitemShut {NoStop}%
\bibitem [{\citenamefont {Hubbard}(1959)}]{hubbard}%
  \BibitemOpen
  \bibfield  {author} {\bibinfo {author} {\bibfnamefont {J.}~\bibnamefont
  {Hubbard}},\ }\href {\doibase 10.1103/PhysRevLett.3.77} {\bibfield  {journal}
  {\bibinfo  {journal} {Phys. Rev. Lett.}\ }\textbf {\bibinfo {volume} {3}},\
  \bibinfo {pages} {77} (\bibinfo {year} {1959})}\BibitemShut {NoStop}%
\bibitem [{\citenamefont {{Stratonovich}}(1957)}]{stratonovich}%
  \BibitemOpen
  \bibfield  {author} {\bibinfo {author} {\bibfnamefont {R.~L.}\ \bibnamefont
  {{Stratonovich}}},\ }\href@noop {} {\bibfield  {journal} {\bibinfo  {journal}
  {Soviet Physics Doklady}\ }\textbf {\bibinfo {volume} {2}},\ \bibinfo {pages}
  {416} (\bibinfo {year} {1957})}\BibitemShut {NoStop}%
\bibitem [{\citenamefont {Aitchison}\ \emph {et~al.}(1995)\citenamefont
  {Aitchison}, \citenamefont {Ao}, \citenamefont {Thouless},\ and\
  \citenamefont {Zhu}}]{aitchison_prb95}%
  \BibitemOpen
  \bibfield  {author} {\bibinfo {author} {\bibfnamefont {I.~J.~R.}\
  \bibnamefont {Aitchison}}, \bibinfo {author} {\bibfnamefont {P.}~\bibnamefont
  {Ao}}, \bibinfo {author} {\bibfnamefont {D.~J.}\ \bibnamefont {Thouless}}, \
  and\ \bibinfo {author} {\bibfnamefont {X.-M.}\ \bibnamefont {Zhu}},\ }\href
  {\doibase 10.1103/PhysRevB.51.6531} {\bibfield  {journal} {\bibinfo
  {journal} {Phys. Rev. B}\ }\textbf {\bibinfo {volume} {51}},\ \bibinfo
  {pages} {6531} (\bibinfo {year} {1995})}\BibitemShut {NoStop}%
\bibitem [{\citenamefont {De~Palo}\ \emph {et~al.}(1999)\citenamefont
  {De~Palo}, \citenamefont {Castellani}, \citenamefont {Di~Castro},\ and\
  \citenamefont {Chakraverty}}]{depalo_prb99}%
  \BibitemOpen
  \bibfield  {author} {\bibinfo {author} {\bibfnamefont {S.}~\bibnamefont
  {De~Palo}}, \bibinfo {author} {\bibfnamefont {C.}~\bibnamefont {Castellani}},
  \bibinfo {author} {\bibfnamefont {C.}~\bibnamefont {Di~Castro}}, \ and\
  \bibinfo {author} {\bibfnamefont {B.~K.}\ \bibnamefont {Chakraverty}},\
  }\href {\doibase 10.1103/PhysRevB.60.564} {\bibfield  {journal} {\bibinfo
  {journal} {Phys. Rev. B}\ }\textbf {\bibinfo {volume} {60}},\ \bibinfo
  {pages} {564} (\bibinfo {year} {1999})}\BibitemShut {NoStop}%
\bibitem [{\citenamefont {Paramekanti}\ \emph {et~al.}(2000)\citenamefont
  {Paramekanti}, \citenamefont {Randeria}, \citenamefont {Ramakrishnan},\ and\
  \citenamefont {Mandal}}]{randeria_prb00}%
  \BibitemOpen
  \bibfield  {author} {\bibinfo {author} {\bibfnamefont {A.}~\bibnamefont
  {Paramekanti}}, \bibinfo {author} {\bibfnamefont {M.}~\bibnamefont
  {Randeria}}, \bibinfo {author} {\bibfnamefont {T.~V.}\ \bibnamefont
  {Ramakrishnan}}, \ and\ \bibinfo {author} {\bibfnamefont {S.~S.}\
  \bibnamefont {Mandal}},\ }\href {\doibase 10.1103/PhysRevB.62.6786}
  {\bibfield  {journal} {\bibinfo  {journal} {Phys. Rev. B}\ }\textbf {\bibinfo
  {volume} {62}},\ \bibinfo {pages} {6786} (\bibinfo {year}
  {2000})}\BibitemShut {NoStop}%
\bibitem [{\citenamefont {Benfatto}\ \emph {et~al.}(2001)\citenamefont
  {Benfatto}, \citenamefont {Caprara}, \citenamefont {Castellani},
  \citenamefont {Paramekanti},\ and\ \citenamefont
  {Randeria}}]{benfatto_prb01}%
  \BibitemOpen
  \bibfield  {author} {\bibinfo {author} {\bibfnamefont {L.}~\bibnamefont
  {Benfatto}}, \bibinfo {author} {\bibfnamefont {S.}~\bibnamefont {Caprara}},
  \bibinfo {author} {\bibfnamefont {C.}~\bibnamefont {Castellani}}, \bibinfo
  {author} {\bibfnamefont {A.}~\bibnamefont {Paramekanti}}, \ and\ \bibinfo
  {author} {\bibfnamefont {M.}~\bibnamefont {Randeria}},\ }\href {\doibase
  10.1103/PhysRevB.63.174513} {\bibfield  {journal} {\bibinfo  {journal} {Phys.
  Rev. B}\ }\textbf {\bibinfo {volume} {63}},\ \bibinfo {pages} {174513}
  (\bibinfo {year} {2001})}\BibitemShut {NoStop}%
\bibitem [{\citenamefont {Benfatto}\ \emph {et~al.}(2004)\citenamefont
  {Benfatto}, \citenamefont {Toschi},\ and\ \citenamefont
  {Caprara}}]{benfatto_prb04}%
  \BibitemOpen
  \bibfield  {author} {\bibinfo {author} {\bibfnamefont {L.}~\bibnamefont
  {Benfatto}}, \bibinfo {author} {\bibfnamefont {A.}~\bibnamefont {Toschi}}, \
  and\ \bibinfo {author} {\bibfnamefont {S.}~\bibnamefont {Caprara}},\ }\href
  {\doibase 10.1103/PhysRevB.69.184510} {\bibfield  {journal} {\bibinfo
  {journal} {Phys. Rev. B}\ }\textbf {\bibinfo {volume} {69}},\ \bibinfo
  {pages} {184510} (\bibinfo {year} {2004})}\BibitemShut {NoStop}%
\bibitem [{\citenamefont {Sun}\ \emph {et~al.}(2020)\citenamefont {Sun},
  \citenamefont {Fogler}, \citenamefont {Basov},\ and\ \citenamefont
  {Millis}}]{millis_prr20}%
  \BibitemOpen
  \bibfield  {author} {\bibinfo {author} {\bibfnamefont {Z.}~\bibnamefont
  {Sun}}, \bibinfo {author} {\bibfnamefont {M.~M.}\ \bibnamefont {Fogler}},
  \bibinfo {author} {\bibfnamefont {D.~N.}\ \bibnamefont {Basov}}, \ and\
  \bibinfo {author} {\bibfnamefont {A.~J.}\ \bibnamefont {Millis}},\ }\href
  {\doibase 10.1103/PhysRevResearch.2.023413} {\bibfield  {journal} {\bibinfo
  {journal} {Phys. Rev. Research}\ }\textbf {\bibinfo {volume} {2}},\ \bibinfo
  {pages} {023413} (\bibinfo {year} {2020})}\BibitemShut {NoStop}%
\bibitem [{\citenamefont {Fertig}\ and\ \citenamefont
  {Das~Sarma}(1990)}]{dassarma_prl90}%
  \BibitemOpen
  \bibfield  {author} {\bibinfo {author} {\bibfnamefont {H.~A.}\ \bibnamefont
  {Fertig}}\ and\ \bibinfo {author} {\bibfnamefont {S.}~\bibnamefont
  {Das~Sarma}},\ }\href {\doibase 10.1103/PhysRevLett.65.1482} {\bibfield
  {journal} {\bibinfo  {journal} {Phys. Rev. Lett.}\ }\textbf {\bibinfo
  {volume} {65}},\ \bibinfo {pages} {1482} (\bibinfo {year}
  {1990})}\BibitemShut {NoStop}%
\bibitem [{\citenamefont {Fertig}\ and\ \citenamefont
  {Das~Sarma}(1991)}]{dassarma_prb91}%
  \BibitemOpen
  \bibfield  {author} {\bibinfo {author} {\bibfnamefont {H.~A.}\ \bibnamefont
  {Fertig}}\ and\ \bibinfo {author} {\bibfnamefont {S.}~\bibnamefont
  {Das~Sarma}},\ }\href {\doibase 10.1103/PhysRevB.44.4480} {\bibfield
  {journal} {\bibinfo  {journal} {Phys. Rev. B}\ }\textbf {\bibinfo {volume}
  {44}},\ \bibinfo {pages} {4480} (\bibinfo {year} {1991})}\BibitemShut
  {NoStop}%
\bibitem [{\citenamefont {Hwang}\ and\ \citenamefont
  {Das~Sarma}(1995)}]{dassarma_prb95}%
  \BibitemOpen
  \bibfield  {author} {\bibinfo {author} {\bibfnamefont {E.~H.}\ \bibnamefont
  {Hwang}}\ and\ \bibinfo {author} {\bibfnamefont {S.}~\bibnamefont
  {Das~Sarma}},\ }\href {\doibase 10.1103/PhysRevB.52.R7010} {\bibfield
  {journal} {\bibinfo  {journal} {Phys. Rev. B}\ }\textbf {\bibinfo {volume}
  {52}},\ \bibinfo {pages} {R7010} (\bibinfo {year} {1995})}\BibitemShut
  {NoStop}%
\bibitem [{\citenamefont {Schrieffer}(1988)}]{schrieffer}%
  \BibitemOpen
  \bibfield  {author} {\bibinfo {author} {\bibfnamefont {J.}~\bibnamefont
  {Schrieffer}},\ }\href {https://books.google.it/books?id=XchWnQAACAAJ} {\emph
  {\bibinfo {title} {Theory of Superconductivity}}},\ Frontiers in physics\
  (\bibinfo  {publisher} {Addison-Wesley},\ \bibinfo {year} {1988})\BibitemShut
  {NoStop}%
\bibitem [{\citenamefont {Anderson}(1958)}]{anderson_pr58}%
  \BibitemOpen
  \bibfield  {author} {\bibinfo {author} {\bibfnamefont {P.~W.}\ \bibnamefont
  {Anderson}},\ }\href {\doibase 10.1103/PhysRev.112.1900} {\bibfield
  {journal} {\bibinfo  {journal} {Phys. Rev.}\ }\textbf {\bibinfo {volume}
  {112}},\ \bibinfo {pages} {1900} (\bibinfo {year} {1958})}\BibitemShut
  {NoStop}%
\bibitem [{\citenamefont {Anderson}(1963)}]{anderson_pr63}%
  \BibitemOpen
  \bibfield  {author} {\bibinfo {author} {\bibfnamefont {P.~W.}\ \bibnamefont
  {Anderson}},\ }\href {\doibase 10.1103/PhysRev.130.439} {\bibfield  {journal}
  {\bibinfo  {journal} {Phys. Rev.}\ }\textbf {\bibinfo {volume} {130}},\
  \bibinfo {pages} {439} (\bibinfo {year} {1963})}\BibitemShut {NoStop}%
\bibitem [{\citenamefont {Keimer}\ \emph {et~al.}(2015)\citenamefont {Keimer},
  \citenamefont {Kivelson}, \citenamefont {Norman}, \citenamefont {Uchida},\
  and\ \citenamefont {Zaanen}}]{keimer_review15}%
  \BibitemOpen
  \bibfield  {author} {\bibinfo {author} {\bibfnamefont {B.}~\bibnamefont
  {Keimer}}, \bibinfo {author} {\bibfnamefont {S.~A.}\ \bibnamefont
  {Kivelson}}, \bibinfo {author} {\bibfnamefont {M.~R.}\ \bibnamefont
  {Norman}}, \bibinfo {author} {\bibfnamefont {S.}~\bibnamefont {Uchida}}, \
  and\ \bibinfo {author} {\bibfnamefont {J.}~\bibnamefont {Zaanen}},\ }\href
  {\doibase 10.1038/nature14165} {\bibfield  {journal} {\bibinfo  {journal}
  {Nature}\ }\textbf {\bibinfo {volume} {518}},\ \bibinfo {pages} {179}
  (\bibinfo {year} {2015})}\BibitemShut {NoStop}%
\bibitem [{\citenamefont {Shibauchi}\ \emph {et~al.}(1994)\citenamefont
  {Shibauchi}, \citenamefont {Kitano}, \citenamefont {Uchinokura},
  \citenamefont {Maeda}, \citenamefont {Kimura},\ and\ \citenamefont
  {Kishio}}]{shibauchi_prl94}%
  \BibitemOpen
  \bibfield  {author} {\bibinfo {author} {\bibfnamefont {T.}~\bibnamefont
  {Shibauchi}}, \bibinfo {author} {\bibfnamefont {H.}~\bibnamefont {Kitano}},
  \bibinfo {author} {\bibfnamefont {K.}~\bibnamefont {Uchinokura}}, \bibinfo
  {author} {\bibfnamefont {A.}~\bibnamefont {Maeda}}, \bibinfo {author}
  {\bibfnamefont {T.}~\bibnamefont {Kimura}}, \ and\ \bibinfo {author}
  {\bibfnamefont {K.}~\bibnamefont {Kishio}},\ }\href {\doibase
  10.1103/PhysRevLett.72.2263} {\bibfield  {journal} {\bibinfo  {journal}
  {Phys. Rev. Lett.}\ }\textbf {\bibinfo {volume} {72}},\ \bibinfo {pages}
  {2263} (\bibinfo {year} {1994})}\BibitemShut {NoStop}%
\bibitem [{\citenamefont {Panagopoulos}\ \emph {et~al.}(1996)\citenamefont
  {Panagopoulos}, \citenamefont {Cooper}, \citenamefont {Peacock},
  \citenamefont {Gameson}, \citenamefont {Edwards}, \citenamefont
  {Schmidbauer},\ and\ \citenamefont {Hodby}}]{panagopoulos_prb96}%
  \BibitemOpen
  \bibfield  {author} {\bibinfo {author} {\bibfnamefont {C.}~\bibnamefont
  {Panagopoulos}}, \bibinfo {author} {\bibfnamefont {J.~R.}\ \bibnamefont
  {Cooper}}, \bibinfo {author} {\bibfnamefont {G.~B.}\ \bibnamefont {Peacock}},
  \bibinfo {author} {\bibfnamefont {I.}~\bibnamefont {Gameson}}, \bibinfo
  {author} {\bibfnamefont {P.~P.}\ \bibnamefont {Edwards}}, \bibinfo {author}
  {\bibfnamefont {W.}~\bibnamefont {Schmidbauer}}, \ and\ \bibinfo {author}
  {\bibfnamefont {J.~W.}\ \bibnamefont {Hodby}},\ }\href {\doibase
  10.1103/PhysRevB.53.R2999} {\bibfield  {journal} {\bibinfo  {journal} {Phys.
  Rev. B}\ }\textbf {\bibinfo {volume} {53}},\ \bibinfo {pages} {R2999}
  (\bibinfo {year} {1996})}\BibitemShut {NoStop}%
\bibitem [{\citenamefont {Hosseini}\ \emph {et~al.}(2004)\citenamefont
  {Hosseini}, \citenamefont {Broun}, \citenamefont {Sheehy}, \citenamefont
  {Davis}, \citenamefont {Franz}, \citenamefont {Hardy}, \citenamefont
  {Liang},\ and\ \citenamefont {Bonn}}]{bonn_prl04}%
  \BibitemOpen
  \bibfield  {author} {\bibinfo {author} {\bibfnamefont {A.}~\bibnamefont
  {Hosseini}}, \bibinfo {author} {\bibfnamefont {D.~M.}\ \bibnamefont {Broun}},
  \bibinfo {author} {\bibfnamefont {D.~E.}\ \bibnamefont {Sheehy}}, \bibinfo
  {author} {\bibfnamefont {T.~P.}\ \bibnamefont {Davis}}, \bibinfo {author}
  {\bibfnamefont {M.}~\bibnamefont {Franz}}, \bibinfo {author} {\bibfnamefont
  {W.~N.}\ \bibnamefont {Hardy}}, \bibinfo {author} {\bibfnamefont
  {R.}~\bibnamefont {Liang}}, \ and\ \bibinfo {author} {\bibfnamefont {D.~A.}\
  \bibnamefont {Bonn}},\ }\href {\doibase 10.1103/PhysRevLett.93.107003}
  {\bibfield  {journal} {\bibinfo  {journal} {Phys. Rev. Lett.}\ }\textbf
  {\bibinfo {volume} {93}},\ \bibinfo {pages} {107003} (\bibinfo {year}
  {2004})}\BibitemShut {NoStop}%
\bibitem [{\citenamefont {Fazio}\ and\ \citenamefont {{van der
  Zant}}(2001)}]{fazio_review01}%
  \BibitemOpen
  \bibfield  {author} {\bibinfo {author} {\bibfnamefont {R.}~\bibnamefont
  {Fazio}}\ and\ \bibinfo {author} {\bibfnamefont {H.}~\bibnamefont {{van der
  Zant}}},\ }\href {\doibase https://doi.org/10.1016/S0370-1573(01)00022-9}
  {\bibfield  {journal} {\bibinfo  {journal} {Physics Reports}\ }\textbf
  {\bibinfo {volume} {355}},\ \bibinfo {pages} {235} (\bibinfo {year}
  {2001})}\BibitemShut {NoStop}%
\bibitem [{\citenamefont {Konsin}\ and\ \citenamefont
  {Sorkin}(1998)}]{konsin_prb98}%
  \BibitemOpen
  \bibfield  {author} {\bibinfo {author} {\bibfnamefont {P.}~\bibnamefont
  {Konsin}}\ and\ \bibinfo {author} {\bibfnamefont {B.}~\bibnamefont
  {Sorkin}},\ }\href {\doibase 10.1103/PhysRevB.58.5795} {\bibfield  {journal}
  {\bibinfo  {journal} {Phys. Rev. B}\ }\textbf {\bibinfo {volume} {58}},\
  \bibinfo {pages} {5795} (\bibinfo {year} {1998})}\BibitemShut {NoStop}%
\bibitem [{\citenamefont {Alpeggiani}\ and\ \citenamefont
  {Andreani}(2013)}]{alpeggiani_prb13}%
  \BibitemOpen
  \bibfield  {author} {\bibinfo {author} {\bibfnamefont {F.}~\bibnamefont
  {Alpeggiani}}\ and\ \bibinfo {author} {\bibfnamefont {L.~C.}\ \bibnamefont
  {Andreani}},\ }\href {\doibase 10.1103/PhysRevB.88.174513} {\bibfield
  {journal} {\bibinfo  {journal} {Phys. Rev. B}\ }\textbf {\bibinfo {volume}
  {88}},\ \bibinfo {pages} {174513} (\bibinfo {year} {2013})}\BibitemShut
  {NoStop}%
\bibitem [{\citenamefont {Hepting}\ \emph {et~al.}(2018)\citenamefont
  {Hepting}, \citenamefont {Chaix}, \citenamefont {Huang}, \citenamefont
  {Fumagalli}, \citenamefont {Peng}, \citenamefont {Moritz}, \citenamefont
  {Kummer}, \citenamefont {Brookes}, \citenamefont {Lee}, \citenamefont
  {Hashimoto}, \citenamefont {Sarkar}, \citenamefont {He}, \citenamefont
  {Rotundu}, \citenamefont {Lee}, \citenamefont {Greene}, \citenamefont
  {Braicovich}, \citenamefont {Ghiringhelli}, \citenamefont {Shen},
  \citenamefont {Devereaux},\ and\ \citenamefont {Lee}}]{lee_rixs_nature18}%
  \BibitemOpen
  \bibfield  {author} {\bibinfo {author} {\bibfnamefont {M.}~\bibnamefont
  {Hepting}}, \bibinfo {author} {\bibfnamefont {L.}~\bibnamefont {Chaix}},
  \bibinfo {author} {\bibfnamefont {E.~W.}\ \bibnamefont {Huang}}, \bibinfo
  {author} {\bibfnamefont {R.}~\bibnamefont {Fumagalli}}, \bibinfo {author}
  {\bibfnamefont {Y.~Y.}\ \bibnamefont {Peng}}, \bibinfo {author}
  {\bibfnamefont {B.}~\bibnamefont {Moritz}}, \bibinfo {author} {\bibfnamefont
  {K.}~\bibnamefont {Kummer}}, \bibinfo {author} {\bibfnamefont {N.~B.}\
  \bibnamefont {Brookes}}, \bibinfo {author} {\bibfnamefont {W.~C.}\
  \bibnamefont {Lee}}, \bibinfo {author} {\bibfnamefont {M.}~\bibnamefont
  {Hashimoto}}, \bibinfo {author} {\bibfnamefont {T.}~\bibnamefont {Sarkar}},
  \bibinfo {author} {\bibfnamefont {J.~F.}\ \bibnamefont {He}}, \bibinfo
  {author} {\bibfnamefont {C.~R.}\ \bibnamefont {Rotundu}}, \bibinfo {author}
  {\bibfnamefont {Y.~S.}\ \bibnamefont {Lee}}, \bibinfo {author} {\bibfnamefont
  {R.~L.}\ \bibnamefont {Greene}}, \bibinfo {author} {\bibfnamefont
  {L.}~\bibnamefont {Braicovich}}, \bibinfo {author} {\bibfnamefont
  {G.}~\bibnamefont {Ghiringhelli}}, \bibinfo {author} {\bibfnamefont {Z.~X.}\
  \bibnamefont {Shen}}, \bibinfo {author} {\bibfnamefont {T.~P.}\ \bibnamefont
  {Devereaux}}, \ and\ \bibinfo {author} {\bibfnamefont {W.~S.}\ \bibnamefont
  {Lee}},\ }\href {\doibase 10.1038/s41586-018-0648-3} {\bibfield  {journal}
  {\bibinfo  {journal} {Nature}\ }\textbf {\bibinfo {volume} {563}},\ \bibinfo
  {pages} {374} (\bibinfo {year} {2018})}\BibitemShut {NoStop}%
\bibitem [{\citenamefont {Lin}\ \emph {et~al.}(2020)\citenamefont {Lin},
  \citenamefont {Yuan}, \citenamefont {Jin}, \citenamefont {Yin}, \citenamefont
  {Li}, \citenamefont {Zhou}, \citenamefont {Lu}, \citenamefont {Dantz},
  \citenamefont {Schmitt}, \citenamefont {Ding}, \citenamefont {Guo},
  \citenamefont {Dean},\ and\ \citenamefont {Liu}}]{liu_rixs_npjqm20}%
  \BibitemOpen
  \bibfield  {author} {\bibinfo {author} {\bibfnamefont {J.}~\bibnamefont
  {Lin}}, \bibinfo {author} {\bibfnamefont {J.}~\bibnamefont {Yuan}}, \bibinfo
  {author} {\bibfnamefont {K.}~\bibnamefont {Jin}}, \bibinfo {author}
  {\bibfnamefont {Z.}~\bibnamefont {Yin}}, \bibinfo {author} {\bibfnamefont
  {G.}~\bibnamefont {Li}}, \bibinfo {author} {\bibfnamefont {K.-J.}\
  \bibnamefont {Zhou}}, \bibinfo {author} {\bibfnamefont {X.}~\bibnamefont
  {Lu}}, \bibinfo {author} {\bibfnamefont {M.}~\bibnamefont {Dantz}}, \bibinfo
  {author} {\bibfnamefont {T.}~\bibnamefont {Schmitt}}, \bibinfo {author}
  {\bibfnamefont {H.}~\bibnamefont {Ding}}, \bibinfo {author} {\bibfnamefont
  {H.}~\bibnamefont {Guo}}, \bibinfo {author} {\bibfnamefont {M.~P.~M.}\
  \bibnamefont {Dean}}, \ and\ \bibinfo {author} {\bibfnamefont
  {X.}~\bibnamefont {Liu}},\ }\href {\doibase 10.1038/s41535-019-0205-9}
  {\bibfield  {journal} {\bibinfo  {journal} {npj Quantum Materials}\ }\textbf
  {\bibinfo {volume} {5}},\ \bibinfo {pages} {4} (\bibinfo {year}
  {2020})}\BibitemShut {NoStop}%
\bibitem [{\citenamefont {Nag}\ \emph {et~al.}(2020)\citenamefont {Nag},
  \citenamefont {Zhu}, \citenamefont {Bejas}, \citenamefont {Li}, \citenamefont
  {Robarts}, \citenamefont {Yamase}, \citenamefont {Petsch}, \citenamefont
  {Song}, \citenamefont {Eisaki}, \citenamefont {Walters}, \citenamefont
  {Garc\'{\i}a-Fern\'andez}, \citenamefont {Greco}, \citenamefont {Hayden},\
  and\ \citenamefont {Zhou}}]{zhou_prl20}%
  \BibitemOpen
  \bibfield  {author} {\bibinfo {author} {\bibfnamefont {A.}~\bibnamefont
  {Nag}}, \bibinfo {author} {\bibfnamefont {M.}~\bibnamefont {Zhu}}, \bibinfo
  {author} {\bibfnamefont {M.}~\bibnamefont {Bejas}}, \bibinfo {author}
  {\bibfnamefont {J.}~\bibnamefont {Li}}, \bibinfo {author} {\bibfnamefont
  {H.~C.}\ \bibnamefont {Robarts}}, \bibinfo {author} {\bibfnamefont
  {H.}~\bibnamefont {Yamase}}, \bibinfo {author} {\bibfnamefont {A.~N.}\
  \bibnamefont {Petsch}}, \bibinfo {author} {\bibfnamefont {D.}~\bibnamefont
  {Song}}, \bibinfo {author} {\bibfnamefont {H.}~\bibnamefont {Eisaki}},
  \bibinfo {author} {\bibfnamefont {A.~C.}\ \bibnamefont {Walters}}, \bibinfo
  {author} {\bibfnamefont {M.}~\bibnamefont {Garc\'{\i}a-Fern\'andez}},
  \bibinfo {author} {\bibfnamefont {A.}~\bibnamefont {Greco}}, \bibinfo
  {author} {\bibfnamefont {S.~M.}\ \bibnamefont {Hayden}}, \ and\ \bibinfo
  {author} {\bibfnamefont {K.-J.}\ \bibnamefont {Zhou}},\ }\href {\doibase
  10.1103/PhysRevLett.125.257002} {\bibfield  {journal} {\bibinfo  {journal}
  {Phys. Rev. Lett.}\ }\textbf {\bibinfo {volume} {125}},\ \bibinfo {pages}
  {257002} (\bibinfo {year} {2020})}\BibitemShut {NoStop}%
\bibitem [{\citenamefont {Kaiser}\ \emph {et~al.}(2014)\citenamefont {Kaiser},
  \citenamefont {Hunt}, \citenamefont {Nicoletti}, \citenamefont {Hu},
  \citenamefont {Gierz}, \citenamefont {Liu}, \citenamefont {Le~Tacon},
  \citenamefont {Loew}, \citenamefont {Haug}, \citenamefont {Keimer},\ and\
  \citenamefont {Cavalleri}}]{kaiser_prb14}%
  \BibitemOpen
  \bibfield  {author} {\bibinfo {author} {\bibfnamefont {S.}~\bibnamefont
  {Kaiser}}, \bibinfo {author} {\bibfnamefont {C.~R.}\ \bibnamefont {Hunt}},
  \bibinfo {author} {\bibfnamefont {D.}~\bibnamefont {Nicoletti}}, \bibinfo
  {author} {\bibfnamefont {W.}~\bibnamefont {Hu}}, \bibinfo {author}
  {\bibfnamefont {I.}~\bibnamefont {Gierz}}, \bibinfo {author} {\bibfnamefont
  {H.~Y.}\ \bibnamefont {Liu}}, \bibinfo {author} {\bibfnamefont
  {M.}~\bibnamefont {Le~Tacon}}, \bibinfo {author} {\bibfnamefont
  {T.}~\bibnamefont {Loew}}, \bibinfo {author} {\bibfnamefont {D.}~\bibnamefont
  {Haug}}, \bibinfo {author} {\bibfnamefont {B.}~\bibnamefont {Keimer}}, \ and\
  \bibinfo {author} {\bibfnamefont {A.}~\bibnamefont {Cavalleri}},\ }\href
  {\doibase 10.1103/PhysRevB.89.184516} {\bibfield  {journal} {\bibinfo
  {journal} {Phys. Rev. B}\ }\textbf {\bibinfo {volume} {89}},\ \bibinfo
  {pages} {184516} (\bibinfo {year} {2014})}\BibitemShut {NoStop}%
\bibitem [{\citenamefont {Nozieres}\ and\ \citenamefont {Pines}(1999)}]{pines}%
  \BibitemOpen
  \bibfield  {author} {\bibinfo {author} {\bibfnamefont {P.}~\bibnamefont
  {Nozieres}}\ and\ \bibinfo {author} {\bibfnamefont {D.}~\bibnamefont
  {Pines}},\ }\href {https://books.google.it/books?id=q3wCwaV-gmUC} {\emph
  {\bibinfo {title} {Theory Of Quantum Liquids}}},\ Advanced Books Classics\
  (\bibinfo  {publisher} {Avalon Publishing, New York, NY},\ \bibinfo {year}
  {1999})\BibitemShut {NoStop}%
\bibitem [{\citenamefont {Pick}\ \emph {et~al.}(1970)\citenamefont {Pick},
  \citenamefont {Cohen},\ and\ \citenamefont {Martin}}]{pick_prb70}%
  \BibitemOpen
  \bibfield  {author} {\bibinfo {author} {\bibfnamefont {R.~M.}\ \bibnamefont
  {Pick}}, \bibinfo {author} {\bibfnamefont {M.~H.}\ \bibnamefont {Cohen}}, \
  and\ \bibinfo {author} {\bibfnamefont {R.~M.}\ \bibnamefont {Martin}},\
  }\href {\doibase 10.1103/PhysRevB.1.910} {\bibfield  {journal} {\bibinfo
  {journal} {Phys. Rev. B}\ }\textbf {\bibinfo {volume} {1}},\ \bibinfo {pages}
  {910} (\bibinfo {year} {1970})}\BibitemShut {NoStop}%
\bibitem [{\citenamefont {Belitz}\ \emph {et~al.}(1989)\citenamefont {Belitz},
  \citenamefont {De~Souza-Machado}, \citenamefont {Devereaux},\ and\
  \citenamefont {Hoard}}]{belitz_prb89}%
  \BibitemOpen
  \bibfield  {author} {\bibinfo {author} {\bibfnamefont {D.}~\bibnamefont
  {Belitz}}, \bibinfo {author} {\bibfnamefont {S.}~\bibnamefont
  {De~Souza-Machado}}, \bibinfo {author} {\bibfnamefont {T.~P.}\ \bibnamefont
  {Devereaux}}, \ and\ \bibinfo {author} {\bibfnamefont {D.~W.}\ \bibnamefont
  {Hoard}},\ }\href {\doibase 10.1103/PhysRevB.39.2072} {\bibfield  {journal}
  {\bibinfo  {journal} {Phys. Rev. B}\ }\textbf {\bibinfo {volume} {39}},\
  \bibinfo {pages} {2072} (\bibinfo {year} {1989})}\BibitemShut {NoStop}%
\bibitem [{\citenamefont {Cea}\ \emph {et~al.}(2014)\citenamefont {Cea},
  \citenamefont {Bucheli}, \citenamefont {Seibold}, \citenamefont {Benfatto},
  \citenamefont {Lorenzana},\ and\ \citenamefont {Castellani}}]{cea_prb14}%
  \BibitemOpen
  \bibfield  {author} {\bibinfo {author} {\bibfnamefont {T.}~\bibnamefont
  {Cea}}, \bibinfo {author} {\bibfnamefont {D.}~\bibnamefont {Bucheli}},
  \bibinfo {author} {\bibfnamefont {G.}~\bibnamefont {Seibold}}, \bibinfo
  {author} {\bibfnamefont {L.}~\bibnamefont {Benfatto}}, \bibinfo {author}
  {\bibfnamefont {J.}~\bibnamefont {Lorenzana}}, \ and\ \bibinfo {author}
  {\bibfnamefont {C.}~\bibnamefont {Castellani}},\ }\href {\doibase
  10.1103/PhysRevB.89.174506} {\bibfield  {journal} {\bibinfo  {journal} {Phys.
  Rev. B}\ }\textbf {\bibinfo {volume} {89}},\ \bibinfo {pages} {174506}
  (\bibinfo {year} {2014})}\BibitemShut {NoStop}%
\bibitem [{\citenamefont {Seibold}\ \emph {et~al.}(2015)\citenamefont
  {Seibold}, \citenamefont {Benfatto}, \citenamefont {Castellani},\ and\
  \citenamefont {Lorenzana}}]{seibold_prb15}%
  \BibitemOpen
  \bibfield  {author} {\bibinfo {author} {\bibfnamefont {G.}~\bibnamefont
  {Seibold}}, \bibinfo {author} {\bibfnamefont {L.}~\bibnamefont {Benfatto}},
  \bibinfo {author} {\bibfnamefont {C.}~\bibnamefont {Castellani}}, \ and\
  \bibinfo {author} {\bibfnamefont {J.}~\bibnamefont {Lorenzana}},\ }\href
  {\doibase 10.1103/PhysRevB.92.064512} {\bibfield  {journal} {\bibinfo
  {journal} {Phys. Rev. B}\ }\textbf {\bibinfo {volume} {92}},\ \bibinfo
  {pages} {064512} (\bibinfo {year} {2015})}\BibitemShut {NoStop}%
\bibitem [{\citenamefont {Fetter}\ and\ \citenamefont
  {Walecka}(1971)}]{fetter}%
  \BibitemOpen
  \bibfield  {author} {\bibinfo {author} {\bibfnamefont {A.~L.}\ \bibnamefont
  {Fetter}}\ and\ \bibinfo {author} {\bibfnamefont {J.~D.}\ \bibnamefont
  {Walecka}},\ }\href@noop {} {\emph {\bibinfo {title} {Quantum Theory of
  Many-Particle Systems}}}\ (\bibinfo  {publisher} {McGraw-Hill},\ \bibinfo
  {address} {Boston},\ \bibinfo {year} {1971})\BibitemShut {NoStop}%
\bibitem [{\citenamefont {Homann}\ \emph {et~al.}(2020)\citenamefont {Homann},
  \citenamefont {Cosme},\ and\ \citenamefont {Mathey}}]{homann_prl20}%
  \BibitemOpen
  \bibfield  {author} {\bibinfo {author} {\bibfnamefont {G.}~\bibnamefont
  {Homann}}, \bibinfo {author} {\bibfnamefont {J.~G.}\ \bibnamefont {Cosme}}, \
  and\ \bibinfo {author} {\bibfnamefont {L.}~\bibnamefont {Mathey}},\
  }\href@noop {} {\bibfield  {journal} {\bibinfo  {journal} {Physical Review
  Research}\ }\textbf {\bibinfo {volume} {2}},\ \bibinfo {pages} {043214}
  (\bibinfo {year} {2020})}\BibitemShut {NoStop}%
\bibitem [{\citenamefont {Homann}\ \emph {et~al.}(2022)\citenamefont {Homann},
  \citenamefont {Cosme},\ and\ \citenamefont {Mathey}}]{homann_22}%
  \BibitemOpen
  \bibfield  {author} {\bibinfo {author} {\bibfnamefont {G.}~\bibnamefont
  {Homann}}, \bibinfo {author} {\bibfnamefont {J.~G.}\ \bibnamefont {Cosme}}, \
  and\ \bibinfo {author} {\bibfnamefont {L.}~\bibnamefont {Mathey}},\ }\href
  {\doibase 10.1088/1367-2630/ac9b83} {\bibfield  {journal} {\bibinfo
  {journal} {New Journal of Physics}\ }\textbf {\bibinfo {volume} {24}},\
  \bibinfo {pages} {113007} (\bibinfo {year} {2022})}\BibitemShut {NoStop}%
\bibitem [{\citenamefont {Homann}\ \emph {et~al.}(2021)\citenamefont {Homann},
  \citenamefont {Cosme}, \citenamefont {Okamoto},\ and\ \citenamefont
  {Mathey}}]{homann_prb21}%
  \BibitemOpen
  \bibfield  {author} {\bibinfo {author} {\bibfnamefont {G.}~\bibnamefont
  {Homann}}, \bibinfo {author} {\bibfnamefont {J.~G.}\ \bibnamefont {Cosme}},
  \bibinfo {author} {\bibfnamefont {J.}~\bibnamefont {Okamoto}}, \ and\
  \bibinfo {author} {\bibfnamefont {L.}~\bibnamefont {Mathey}},\ }\href@noop {}
  {\bibfield  {journal} {\bibinfo  {journal} {Physical Review B}\ }\textbf
  {\bibinfo {volume} {103}},\ \bibinfo {pages} {224503} (\bibinfo {year}
  {2021})}\BibitemShut {NoStop}%
\bibitem [{\citenamefont {Koyama}\ and\ \citenamefont
  {Tachiki}(1996)}]{koyama_prb96}%
  \BibitemOpen
  \bibfield  {author} {\bibinfo {author} {\bibfnamefont {T.}~\bibnamefont
  {Koyama}}\ and\ \bibinfo {author} {\bibfnamefont {M.}~\bibnamefont
  {Tachiki}},\ }\href {\doibase 10.1103/PhysRevB.54.16183} {\bibfield
  {journal} {\bibinfo  {journal} {Phys. Rev. B}\ }\textbf {\bibinfo {volume}
  {54}},\ \bibinfo {pages} {16183} (\bibinfo {year} {1996})}\BibitemShut
  {NoStop}%
\bibitem [{\citenamefont {Koyama}(2002)}]{koyama02}%
  \BibitemOpen
  \bibfield  {author} {\bibinfo {author} {\bibfnamefont {T.}~\bibnamefont
  {Koyama}},\ }\href {\doibase 10.1143/JPSJ.71.2986} {\bibfield  {journal}
  {\bibinfo  {journal} {Journal of the Physical Society of Japan}\ }\textbf
  {\bibinfo {volume} {71}},\ \bibinfo {pages} {2986} (\bibinfo {year}
  {2002})},\ \Eprint
  {http://arxiv.org/abs/https://doi.org/10.1143/JPSJ.71.2986}
  {https://doi.org/10.1143/JPSJ.71.2986} \BibitemShut {NoStop}%
\bibitem [{\citenamefont {Okamoto}\ \emph {et~al.}(2016)\citenamefont
  {Okamoto}, \citenamefont {Cavalleri},\ and\ \citenamefont
  {Mathey}}]{cavalleri_prl16}%
  \BibitemOpen
  \bibfield  {author} {\bibinfo {author} {\bibfnamefont {J.-i.}\ \bibnamefont
  {Okamoto}}, \bibinfo {author} {\bibfnamefont {A.}~\bibnamefont {Cavalleri}},
  \ and\ \bibinfo {author} {\bibfnamefont {L.}~\bibnamefont {Mathey}},\ }\href
  {\doibase 10.1103/PhysRevLett.117.227001} {\bibfield  {journal} {\bibinfo
  {journal} {Phys. Rev. Lett.}\ }\textbf {\bibinfo {volume} {117}},\ \bibinfo
  {pages} {227001} (\bibinfo {year} {2016})}\BibitemShut {NoStop}%
\bibitem [{\citenamefont {Okamoto}\ \emph {et~al.}(2017)\citenamefont
  {Okamoto}, \citenamefont {Hu}, \citenamefont {Cavalleri},\ and\ \citenamefont
  {Mathey}}]{cavalleri_prb17}%
  \BibitemOpen
  \bibfield  {author} {\bibinfo {author} {\bibfnamefont {J.-i.}\ \bibnamefont
  {Okamoto}}, \bibinfo {author} {\bibfnamefont {W.}~\bibnamefont {Hu}},
  \bibinfo {author} {\bibfnamefont {A.}~\bibnamefont {Cavalleri}}, \ and\
  \bibinfo {author} {\bibfnamefont {L.}~\bibnamefont {Mathey}},\ }\href
  {\doibase 10.1103/PhysRevB.96.144505} {\bibfield  {journal} {\bibinfo
  {journal} {Phys. Rev. B}\ }\textbf {\bibinfo {volume} {96}},\ \bibinfo
  {pages} {144505} (\bibinfo {year} {2017})}\BibitemShut {NoStop}%
\end{thebibliography}%

\end{document}